%% file: ComplexityCFTD.tex
\documentclass[prl,aps,a4paper,twocolumn,superscriptaddress,nofootinbib,longbibliography ]{revtex4-2}

\usepackage[utf8x]{inputenc}
\usepackage{amssymb}
\usepackage{amsmath}
\usepackage{graphicx}
\usepackage{bbm}
\usepackage{psfrag}
\usepackage{latexsym}
\usepackage{hyperref}
\usepackage{color}
\usepackage{physics}
\usepackage{changepage}
\usepackage{caption}
\usepackage{subcaption}

\newcommand{\be}{\begin{equation}}
\newcommand{\ee}{\end{equation}}
\newcommand{\bea}{\begin{eqnarray}}
\newcommand{\eea}{\end{eqnarray}}
\newcommand{\ba}{\begin{eqnarray}}
\newcommand{\ea}{\end{eqnarray}}

\newcommand{\beq}{\begin{equation}}
\newcommand{\eeq}{\end{equation}}
\newcommand{\beqa}{\begin{eqnarray}}
\newcommand{\eeqa}{\end{eqnarray}}
\newcommand{\beqar}{\begin{eqnarray*}}
\newcommand{\eeqar}{\end{eqnarray*}}

\newcommand{\del}{\partial}

\newcommand{\Real}{{\mathbbm R}}

\newcommand{\nn}{{\mathcal N}}

\usepackage[normalem]{ulem}
\usepackage{soul}

\usepackage{etoolbox}
\preto\subequations{\ifhmode\unskip\fi}

\definecolor{purple}{rgb}{0.5,0.15,0.5}

\newcommand{\eg}{{\it e.g.,}\ }
\newcommand{\ie}{{\it i.e.,}\ }
\newcommand{\atanh}{\mathrm{tanh^{-1}}}

\makeatletter
\newcommand*{\textlabel}[2]{%
  \edef\@currentlabel{#1}% Set target label
  \phantomsection% Correct hyper reference link
  #1\label{#2}% Print and store label
}
\makeatother

\begin{document}

\author{Nicolas Chagnet}
\email{chagnet@lorentz.leidenuniv.nl}
\affiliation{\emph{Instituut-Lorentz, Universiteit Leiden, P.O. Box 9506, 2300 RA Leiden, The Netherlands}}

\author{Shira Chapman}
\email{schapman@bgu.ac.il}
\affiliation{\emph{Department of Physics, Ben-Gurion University of the Negev, Beer Sheva 84105, Israel}}

\author{Jan de Boer}
\email{J.deBoer@uva.nl}
\affiliation{\emph{Institute for Theoretical Physics, University of Amsterdam, Science Park 904, Postbus 94485, 1090 GL Amsterdam, The Netherlands}}

\author{Claire Zukowski}
\email{c.e.zukowski@uva.nl}
\affiliation{\emph{Institute for Theoretical Physics, University of Amsterdam, Science Park 904, Postbus 94485,
1090 GL Amsterdam, The Netherlands}}

\title{Complexity for Conformal Field Theories in General Dimensions}

\begin{abstract}
We study circuit complexity for conformal field theory states in arbitrary dimensions. Our circuits start from a primary state and move along a unitary representation of the Lorentzian conformal group.
Different choices of distance functions can be understood
in terms of the geometry of coadjoint orbits of the conformal group.
We explicitly relate our circuits to timelike geodesics in anti-de Sitter space and the complexity metric to distances between these geodesics.  We extend our method to circuits in other symmetry groups using a group theoretic generalization of the notion of coherent states.
\end{abstract}

\maketitle

\noindent \emph{\textlabel{1}{sec:intro}.~Introduction.--}
\addcontentsline{toc}{section}{1. Introduction}
\input{sections/introduction}

\hspace{10pt}

\noindent \emph{\textlabel{3}{sec:gdc}.~Complexity in General Dimensions.--}
\addcontentsline{toc}{section}{3. Complexity in General Dimensions}
\input{sections/generalD}

\hspace{10pt}

\noindent \emph{\textlabel{4}{sec:coad}.~Geometric Action and Coadjoint Orbits.-- }
\addcontentsline{toc}{section}{4. Geometric Action and Coadjoint Orbits}
\input{sections/coadjoint}

\hspace{10pt}

\noindent \emph{\textlabel{5}{sec:csg}.~Coherent State Generalization.--}
\addcontentsline{toc}{section}{5. Coherent State Generalization}
\input{sections/coherentstate}

\hspace{10pt}

\noindent \emph{\textlabel{6}{sec:holo}.~Holography.--}
\addcontentsline{toc}{section}{6. Holography}
\input{sections/Holography}

\hspace{10pt}

\noindent \emph{\textlabel{7}{sec:sum}.~Summary and Outlook.--}
\addcontentsline{toc}{section}{7. Summary and Outlook}

\input{sections/discussion}

\begin{acknowledgments}
\section*{Acknowledgments}
We would like to thank Igal Arav, Costas Bachas, Ning Bao, Alexandre Belin, Adam Chapman, Bartlomiej Czech, Lorenzo Di Pietro, Ben Freivogel, Vladimir Gritsev, Austin Joyce, Kurt Hinterbichler, Marco Meineri, Yaron Oz, Giuseppe Policastro and Jan Zaanen, for valuable comments and discussions.  We are particularly grateful to Bartlomiej Czech for suggesting to look for a relation between the FS metric and minimal and maximal distances between timelike geodesics. 
This work is supported by the Delta ITP consortium, a program of the Netherlands Organisation for Scientific Research (NWO) that is funded by the Dutch Ministry of Education, Culture and Science (OCW). 
The work of SC is supported by the Israel Science
Foundation (grant No. 1417/21). 
SC acknowledges the support of Carole and Marcus Weinstein through the BGU Presidential Faculty Recruitment Fund.  

\end{acknowledgments}

\vspace{10 pt}

\bibliography{referencesCCTFD}
\appendix{}

\titlepage

\pagebreak

\setcounter{page}{1}

\begin{center}
{\LARGE Supplemental Material}
\addcontentsline{toc}{chapter}{Supplemental Material}
\end{center}

\input{sections/AppendixLoreEuc}

\input{sections/AppendixHigherD}
\input{sections/complexityfunctional}

\input{sections/canonicalvariables}

\input{sections/bounds}

\input{sections/AppendixD2}

\input{sections/AppendixSpin}
\input{sections/AppendixFundam}

\input{sections/cartankilling}
\input{sections/holography_appendix}

\end{document}

%% file: sections/introduction.tex
The peculiarity of quantum systems is rooted in their entanglement pattern. Hence, there is increasing interest in studying measures characterizing entanglement in quantum states. The most famous of these measures is the entanglement entropy, which estimates the knowledge a given subsystem has about the full quantum state. In recent years, it became apparent that entanglement entropy is not enough to capture the full information about quantum correlations in a state. As a consequence, a new measure from quantum information became prominent in studies of quantum states. This measure, known as \emph{quantum computational complexity} (QCC), estimates how hard it is to construct a given state from a set of elementary operations \cite{watrous2008quantum,Aaronson:2016vto,nielsen2002quantum}. QCC is also of clear interest in recent efforts to construct quantum computers.

QCC has attracted a lot of attention in high energy theory due to its proposed relation to black holes \cite{Susskind:2018pmk,Susskind:2014moa}. This relation was explicitly formulated within the holographic (or AdS/CFT) correspondence \cite{Aharony:1999ti}. 
It turns out that the growth of black hole interiors behaves in a very similar way to the growth of complexity during Hamiltonian evolution in quantum systems, see, \eg \cite{Susskind:2014rva,Brown:2015lvg,brown2016holographic,Carmi:2017jqz,Stanford:2014jda,Chapman:2018dem,Chapman:2018lsv}.
These ideas suggest a promising avenue to address puzzles related to black hole spacetimes and their interior geometry.

However, the lack of a complete framework for studying QCC within quantum field theory (QFT) has been a stumbling block towards rigorously establishing  the connection between black hole interiors and QCC. Significant progress was made for free and weakly coupled QFTs \cite{Jefferson:2017sdb,chapman2018,Khan:2018rzm,Hackl:2018ptj,chapman2019,Chapman:2019clq,Bhattacharyya:2018bbv} and for strongly coupled  two dimensional conformal field theories (CFTs)  \cite{caputa2019,caputa2019,Erdmenger:2020sup,flory2020,flory2020a,Bueno:2019ajd}.
Yet, no results exist at present for circuit complexity in CFTs in $d>2$ and further, its precise connection with holography has not been established in any dimension. The importance of studying complexity in $d>2$ becomes evident when noting that holographic complexity behaves very differently in $d=2$ and in $d>2$, in particular when studying the complexity of formation of thermofield double states \cite{chapman2017} or its sensitivity to defects \cite{Chapman:2018bqj,Sato:2019kik}.
The goal of this letter is to bridge these gaps by studying complexity of CFTs in $d>2$ and further by establishing a rigorous connection between complexity and geometry in holography.

We employ the symmetry generators to construct circuits in unitary representations of the Lorentzian conformal group and present explicit results for state-dependent distance functions along these circuits. Our circuits live in a phase space which is a coadjoint orbit of the conformal group and the various cost functions take the form of simple geometric notions on these orbits.
Using symmetry generators to construct circuits restricts the circuits to move in the space of generalized coherent states. We use this fact to generalize our results to general symmetry groups. We illustrate our methods by focusing on circuits starting from a scalar primary state whose coadjoint orbit can be identified with the coset space $SO(d,2)/(SO(2)\times SO(d))$, but our techniques are also applicable to more general spinning states. We derive bounds on the complexity and its rate of change.

We explicitly relate our unitary circuits to timelike geodesics in anti-de Sitter spacetimes.  We find that the line element in the complexity metric admits a very simple interpretation as the average of the minimal and maximal squared distances between two nearby geodesics. This provide a novel bulk description for complexity which is rigorously derived from the CFT and opens new possibilities for testing the holographic complexity proposals.

This paper is organized as follows: in \S \ref{sec:prel}, we introduce the relevant complexity distance functions. In \S \ref{sec:gdc} we present the result for the complexity of CFT states in general dimensions. In \S \ref{sec:coad}-\ref{sec:csg}, we connect our results to the notions of coadjoint orbits and generalized coherent states. In \S \ref{sec:holo} we connect our results to holography. We conclude in \S \ref{sec:sum} with a summary and outlook.

\hspace{10pt}

\noindent \emph{\textlabel{2}{sec:prel}. Preliminaries.--}
\addcontentsline{toc}{section}{2. Preliminaries}
Explicitly, QCC is defined as the minimal number of gates required to reach a desired \emph{target state}, starting from a (typically simpler) \emph{reference state}. For several applications, it is advantageous to focus on continuous notions of complexity rather than a discrete gate counting. Such ideas were put forward by Nielsen \cite{nielsen2006quantum,nielsen2005,dowling2008geometry} who translated the problem of studying minimal gate complexity to that of studying geodesics on the space of unitary transformations. In a very similar way, we can study notions of continuous complexity using geodesics through the space of quantum states.

Continuous complexity is defined using a cost function $\mathcal F(\sigma)$, with circuit parameter $\sigma$.
The complexity is the minimal cost among all possible trajectories moving from the reference state to the target state: 
$\mathcal C \equiv \min \int \dd \sigma \, \mathcal F(\sigma)$.
Past attempts to study state complexity in CFTs (\eg \cite{caputa2019}) focused on two cost functions:
the $\mathcal F_1$ cost function and the Fubini-Study (FS) norm defined as
\begin{subequations}\label{eq:costfunctions}
	\begin{align}
		&\hspace{-10pt}\mathcal F_1(\sigma) \dd\sigma = \left| \braket{\psi}{\partial_\sigma \psi} \right|  \dd\sigma =\left| \ev{U^\dagger \dd U}{\psi_R} \right| ,\label{def:1norm}\\
\begin{split}
&\hspace{-10pt} \mathcal F_{FS}(\sigma) \dd \sigma
=\sqrt{\ev{\dd U^\dagger \dd U}{\psi_R} - \left|\ev{U^\dagger \dd U}{\psi_R}\right|^2},\label{eq:FSmetricgeneral2}
\end{split}
	\end{align}
\end{subequations}
where $\ket{\psi(\sigma)} \equiv U(\sigma) \ket{\psi_R}$ are the states along the unitary circuit, $\ket{\psi_R}$ is the reference state and $\dd s_{FS}^2=\mathcal F_{FS}^2(\sigma) \dd \sigma^2$ is the well known FS-metric. Our analysis in the next section demonstrates that the $\mathcal F_1$ cost function assigns zero cost to certain gates and has therefore disadvantages as a complexity measure.

The FS-metric along straight-line trajectories $e^{itH}|\psi_R\rangle$ is proportional to the variance $\Delta E = \sqrt{\ev{H^2} - |\ev{H}|^2}$. We can interpret $H$ as the Hamiltonian and $t$ as the time. This variance was shown by \cite{Anandan:1990fq} to bound the time required to reach an orthogonal state $\tau_{\text{orth.}} \geq \pi \hbar/(2 \Delta E)$ on compact spaces. Inspired by these bounds on orthogonality time, Lloyd conjectured a bound on the rate of computation \cite{lloyd2000ultimate} (see also \cite{Brown:2015lvg}). 
Unlike \cite{Anandan:1990fq}, our state manifold is non-compact and our states never become orthogonal. Nonetheless, we will derive bounds on the complexity and its rate of change by other means. Deriving bounds on the state overlap in our setup is an interesting question for future study.

%% file: sections/generalD.tex
Consider the Euclidean conformal algebra in $d \geq 2$ 
 with  $D,P_\mu, K_\mu, L_{\mu\nu}$ the Euclidean conformal generators (used to construct unitary representations of the Lorentzian conformal group \cite{SupMat}A) satisfying 
\begin{equation}
	\label{eq:radialquant}
	\begin{aligned}
		D^\dagger & = D~, & \quad K_\mu^\dagger & = P_\mu~, & \quad L_{\mu\nu}^\dagger & = - L_{\mu\nu}~,\\
	\end{aligned}
\end{equation}
in radial quantization.

As the reference state, we consider a scalar primary state $\ket{\psi_R} = \ket{\Delta}$ satisfying
$D \ket \Delta = \Delta \ket \Delta$ and $K_\mu \ket \Delta =L_{\mu\nu} \ket \Delta = 0$
and focus on circuits generated by the unitary
\begin{equation}
	\label{eq:unitaryDef}
 \hspace{-3pt}	U(\sigma) \equiv e^{i \alpha(\sigma)\cdot P} e^{i \gamma_D(\sigma) D} \left(\prod_{\mu < \nu} e^{i \lambda_{\mu\nu}(\sigma) L_{\mu\nu}}\right)  e^{i \beta(\sigma) \cdot K}\,,
\end{equation}
with  $\sigma$  a circuit parameter and $\alpha_\mu, \beta_\mu, \gamma_D$ and $\lambda_{\mu\nu}$ a-priori complex parameters, further constrained by the restriction that $U(\sigma)$ be unitary.
The circuits take the form $\ket{\alpha(\sigma)} \equiv U(\sigma) \ket \Delta \equiv \nn(\sigma) e^{i \alpha(\sigma) \cdot P} \ket \Delta$ 
where $\nn(\sigma) \equiv \exp(i \gamma_D(\sigma) \Delta)$ is a normalization factor and $\gamma_D(\sigma) \equiv \gamma_D^{R}(\sigma)+i \gamma_D^I(\sigma)$, with $R/I$ indicating the real/imaginary part. Unitarity of $U(\sigma)$  implies $\gamma_D^{I}(\sigma)= -\frac{1}{2} \log A(\alpha, \alpha^*)$ (see \cite{SupMat}B) where
\begin{equation}\label{domain}
A(\alpha, \alpha^*)  \equiv 1 - 2 \, \alpha \cdot \alpha^* + \alpha^2 \alpha^{*2}>0~,
\end{equation}
and requiring a positive spectrum for the Hamiltonian $D$ along the circuit implies $\alpha^* \cdot \alpha<1$ 
(equivalently $\alpha^2 \alpha^{*2}<1$).

Substituting $\ket{\alpha(\sigma)}$ into the cost-functions \eqref{def:1norm}-\eqref{eq:FSmetricgeneral2} and using the  expectation values of $\{P_\mu, K_\mu, K_\mu P_\nu\}$ (see \cite{SupMat}B), 
we find for the $\mathcal{F}_1$ cost function
\begin{equation}\label{eq:higherd1norm}
\hspace{-2pt}	\frac{\mathcal{F}_1}{\Delta} =  \left|\dfrac{\dot \alpha \cdot \alpha^* -\dot \alpha^* \cdot \alpha+  \alpha^2 \,(\dot \alpha^* \cdot \alpha^*)-\alpha^{*2} (\dot \alpha \cdot \alpha)}{A(\alpha, \alpha^*)} + i \dot \gamma_D^R \right| ,
\end{equation}
while for the FS-metric we obtain
\begin{equation}
	\label{eq:FSGen}
	\dv{s_{FS}^2}{\sigma^2}= 2 \Delta \left[\dfrac{\dot \alpha \cdot \dot \alpha^{*} - 2|\dot \alpha \cdot \alpha|^2}{A(\alpha, \alpha^*)} + 2\dfrac{\left|\dot \alpha \cdot \alpha^* - \alpha^{*2} \, \alpha \cdot \dot \alpha \right|^2}{A(\alpha, \alpha^*)^2}\right] \, .
\end{equation}

The FS-metric \eqref{eq:FSGen} is a positive-definite Einstein-K\"{a}hler metric on the complex manifold of states with $d$ complex coordinates $\alpha$ bounded inside the domain \eqref{domain}. It satisfies 
$\dd s_{FS}^2 = \partial_\alpha \partial_{\alpha^*}  K(\alpha,\alpha^*) \, \dd \alpha\,  \dd \alpha^*$, 
where the associated K\"{a}hler potential is defined as $K(\alpha, \alpha^*) =  -\Delta \, \log A(\alpha,\alpha^*)$. 
Denoting collectively the indices of $\alpha$ and $\alpha^*$ by capital Latin letters, one finds that $R_{AB}=-\frac{2d}{\Delta}g_{AB}$ and $R=-\frac{4 d^2}{\Delta}$ and that all sectional curvatures are negative.  This means that geodesics will deviate from each other.

In fact, \eqref{eq:FSGen} is a natural metric on the following quotient space of the conformal group
\begin{equation}\label{eq:coset}
	\mathcal M = \dfrac{\text{SO}(d,2)}{\text{SO}(2)\times\text{SO}(d)}~,
\end{equation}
which can also be identified with the space of timelike geodesics in AdS${}_{d+1}$~\cite{gibbons2000, Andrianopoli:2005de}, see \S\ref{sec:holo}.  This is similar to the relation between the metric on kinematic space and spacelike geodesics in AdS${}_{d+1}$,  \cite{Czech:2015qta, deBoer:2015kda, Czech:2016xec, deBoer:2016pqk} where the relevant orbit is $SO(d,2)/SO(1,1)\times SO(1,d-1)$ \cite{penna2019}.
While some of the above observations are well known in the context of geometry of Lie groups \cite{gallier2019differential,helgason}, here they find a novel role in the context of circuit complexity.

Since the coset space \eqref{eq:coset} is a negatively curved symmetric space, its geodesics passing through $\ket{\psi_R}$ take the form \cite{Barrett1983}
\begin{equation}
\label{eq:singleExponential}
\ket{\psi(\sigma)} = \exp \left[i \sigma (\tilde \alpha P_\mu + \tilde \alpha^* K_\mu)\right] \ket{\psi_R},
\end{equation}
and do not reconnect, \ie \eqref{eq:coset} has no conjugate points~\cite{helgason}.
Here, we parametrized our geodesics in terms of the straight-line-trajectory-parameter $\tilde \alpha$ rather than $\alpha$. Explicitly, in terms of the $\alpha$ parametrization, the complexity of a target state $\ket{\alpha(\sigma=1)}\equiv |\alpha_T\rangle$ is
\begin{equation}
\begin{split}
	\label{eq:generalComplexityGeodesics22}
	\mathcal C[\tilde \alpha] &\,= \sqrt{2 \Delta ~ \tilde \alpha^* \cdot \tilde \alpha} ~,
	\\
%	\label{eq:atatsFormula}
 	2\tilde \alpha \cdot \tilde \alpha^* &\, =  \left[\left(\atanh\Omega^S_T \right)^2 + \left(\atanh\Omega^A_T \right)^2 \right]~,
\end{split}
\end{equation}
where $\Omega^\pm_T \equiv \Omega^{S}_T\pm \Omega^{A}_T \equiv \sqrt{2~ \alpha_T \cdot \alpha_T^* \pm 2 |\alpha_T^2|}$ 
(see \cite{SupMat}C--D). 
Earlier, we chose to parametrize the states with $\alpha(\sigma)$ rather than $\tilde\alpha$ since this facilitates the evaluation of correlation functions in the state and therefore provides its more natural characterization. We will see later that the relation to holography is also done using the parameter $\alpha$.
The complexity \eqref{eq:generalComplexityGeodesics22} can be bounded by employing the inequalities around~\eqref{domain}
\begin{equation}\label{eq:compbound1}
 \dfrac{\Delta}{E_T + \Delta}\sqrt{ (E_T-\Delta)} \leq \mathcal C[\alpha_T] \leq \sqrt{E_T - \Delta}
\end{equation}
where $E_T \equiv \ev{D}{\alpha_T} = \Delta (1 - \alpha_T^2 \alpha_T^{*2})/A(\alpha_T, \alpha_T^*)$  is the energy of the target state in radial quantization  
(see \cite{SupMat}E).

A substantial difference between the $\mathcal F_1$ cost function and the FS metric is that the former depends on $\gamma_D^R$ which induces an overall phase in the states through which our circuits pass. In fact, the $\mathcal{F}_1$ cost function \eqref{eq:higherd1norm} without absolute values vanishes on-shell except for its part associated with the overall phase $\gamma_D^R$ and is simply proportional to the Berry gauge field, cf.~\cite{Oblak:2017ect,Akal:2019hxa,Erdmenger:2020sup}.

We close by observing that the FS distance along time evolved states $e^{i \tau D} \ket{\alpha_0}$ satisfies a Lloyd-like bound \cite{lloyd2000ultimate}
\begin{equation}
   \dv{s_{\mathrm{FS}}}{\tau} \leq \dfrac{E}{\sqrt{\Delta}} \leq \sqrt{\dfrac{2}{d - 2}} E ~.
\end{equation}
where $E\equiv\ev{D}{\alpha_0}$ is the energy, $\ket{\alpha_0}$ an arbitrary initial state and we used the unitarity bound $\Delta \geq d/2 - 1$  \cite{rychkov2017}.

We compare our results to the existing literature for $d = 2$ CFTs in \cite{SupMat}F. In that case, holomorphic factorization allows us to also treat spinning states \cite{SupMat}G.

%% file: sections/coadjoint.tex
Our results for the cost functions \eqref{eq:higherd1norm}-\eqref{eq:FSGen} can be understood in terms of the geometry of coadjoint orbits, see, \eg \cite{witten1988,kirillov2004}. A similar connection was pointed out in two dimensions in \cite{caputa2019,Erdmenger:2020sup}.

Let us start by briefly describing the coadjoint orbit method in representation theory. Consider a Lie group $G$ with Lie algebra $\mathfrak{g}$, a dual space $\mathfrak{g}^*$ consisting of linear maps on $\mathfrak{g}$, and a pairing $\left<\cdot,\cdot\right>$ between the Lie algebra and dual space.
For matrix groups, the adjoint action of $u \in G$ on $X \in \mathfrak g$ is defined as $\mbox{Ad}_u(X) = u X u^{-1}$. 
At the level of the algebra, the adjoint action is simply the commutator $\mbox{ad}_Y(X)
= [Y, X]$ where $X,Y\in \mathfrak{g}$. 
The Maurer-Cartan (MC) form on the full group is $\Theta \equiv u^{-1} \dd u$ where $u\in G$
and it satisfies $\dd \Theta = - \Theta \wedge \Theta$.

The coadjoint action on the dual space is defined implicitly by
\begin{equation} \label{eq:mcDef}
\hspace{-6pt}\left<\mbox{Ad}_u^* \xi, X\right> = \left<\xi, \mbox{Ad}_{u^{-1}} X\right>~,\quad \xi\in \mathfrak{g}^*~,~ X\in\mathfrak{g}~, ~u \in G~,
\end{equation}
from which one can build the coadjoint orbit $\mathcal O_\lambda \equiv \{\mathrm{Ad}^*_u \lambda \vert u \in G\} \subset \mathfrak g^*$ of a given dual algebra element $\lambda \in \mathfrak g^*$. $\mathcal{O}_\lambda$ can be identified with the coset space $G/H_\lambda$, where the subgroup  $H_\lambda =\text{Stab}(\lambda) \equiv \{u \in G ~\vert~ \mathrm{Ad}^*_u \lambda=\lambda \}$ is the stabilizer 
and the corresponding algebra is $\mathfrak h_\lambda\equiv \text{stab}(\lambda)$.

Each coadjoint orbit corresponds to a symplectic manifold with a local pre-symplectic form $\mathcal A_\lambda$ and the Kirillov-Kostant symplectic form $\omega_\lambda$ defined as
\begin{equation}\label{Alambda}
\mathcal A_\lambda = \left<\lambda, \Theta \right>~, \indent \omega_\lambda = \left<\lambda, \dd\Theta \right>~.
\end{equation}
The geometric action associated to the coadjoint orbit is
$S_\lambda = \int \mathcal A_\lambda$ \cite{alekseev1988, alekseev2018}.

The symplectic form $\omega_\lambda$ is compatible with a complex structure $J_\lambda$ satisfying $J_\lambda^2=-1$ if
$\omega_\lambda(J_\lambda x,J_\lambda y)=\omega_\lambda(x,y)$.
In this case it is possible to define a K\"ahler metric $\dd s^2_{G/H_\lambda}(x,y) =  \omega_\lambda(x,J_\lambda  y)$ on the
coadjoint orbit $\mathcal{O}_\lambda$.

In \cite{SupMat}H,  we apply the above definitions in the fundamental (matrix) representation of the conformal algebra $\mathfrak{so}(d,2)$ with  representative $\lambda$ taken to be proportional to the dilatation matrix with stabilizer group $\mathfrak{h_\lambda}=\mathfrak{so}(2)\times\mathfrak{so}(d)$ and orbit corresponding to the quotient space $G/H_\lambda$ from Eq.~\eqref{eq:coset}. This 
yields an agreement with eqs.~\eqref{eq:higherd1norm}-\eqref{eq:FSGen}, \ie 
\begin{equation}
	\label{eq:resultLink2}
 	\begin{aligned}
 		\mathcal F_1 \, \dd \sigma & = \left| \mathcal A_\lambda \right|~, \qquad \dd s_{FS}^2 & = \dd s^2_{G/H_\lambda} ~.
 	\end{aligned}
\end{equation}
As alluded to above, $\mathcal A_\lambda$ can also be interpreted as a Berry gauge field, and the Berry curvature is simply the symplectic form $\omega_\lambda$.
Circuits starting from spinning primary states in $d>2$ amount to a different choice of representative to match with the relevant reduced stabilizer group.

%% file: sections/coherentstate.tex
The equivalence of the FS-metric and the $\mathcal{F}_1$ cost function with their geometric counterparts on the coadjoint orbit is also valid within infinite dimensional Hilbert spaces obtained via geometric quantization of the orbits of arbitrary Lie groups \cite{caputa2019,Bueno:2019ajd,Taylor:1993zp}.  This can be understood using a group theoretical generalization of the notion of coherent states, see \eg \cite{perelomov1972,gilmore,RevModPhys.54.407,Provost:1980nc}.
The existence of these states is intrinsically connected to the representation theory of the symmetry in question.  In this section we explain how the  coadjoint orbit perspective leads to the complexity functionals of \eqref{eq:higherd1norm}-\eqref{eq:FSGen} for general Lie groups.

As before, we consider some real Lie group $G$ with Lie algebra $\mathfrak{g}$. The corresponding complex algebra admits a decomposition
$\mathfrak{g}_\mathbb{C}=\mathfrak{n}_{+}+\mathfrak{h}_\mathbb{C}+\mathfrak{n}_{-}$ with a real structure (a dagger) which maps $\mathfrak{h}_\mathbb{C}$ to itself and $\mathfrak{n}_{+}$ to $\mathfrak{n}_{-}$. For a detailed account of this decomposition, see \cite{SupMat}I. The generators of the real Lie algebra are anti-Hermitian. We denote the real subalgebra of $\mathfrak{h}_\mathbb{C}$ by $\mathfrak{h}$ and its associated Lie group $H$.
We also assume that $[\mathfrak{n}_{+} ,\mathfrak{n}_{+}] \subset \mathfrak{n}_{+}$ and similarly for $\mathfrak{n}_{-}$ and that $[\mathfrak{h}_\mathbb{C} ,\mathfrak{n}_{\pm}] \subset \mathfrak{n}_{\pm}$. We take a basis of raising operators $E_{\alpha}$ for $\mathfrak{n}_{+}$ and lowering operators $E_{-\alpha}$ for $\mathfrak{n}_{-}$ with $E_{\alpha}^{\dagger}=E_{-\alpha}$ and a basis $h_i$ for $\mathfrak{h}$.

We consider a unitary highest weight representation generated by a one-dimensional base state $|\psi_R\rangle$ satisfying $\mathcal{D}(E_{\alpha})|\psi_R\rangle=0$ and $\mathcal{D}(h_i)|\psi_R\rangle=\chi_i|\psi_R\rangle$ with $\chi_i$  constants and where $\mathcal{D}$ is the representation on the Hilbert space.
In other words, the base state is invariant up to a phase under the action of the stabilizer subgroup $H\subset G$. This includes the possibility of spinning highest weight representations, cf.~\cite{Rowe1985,Rowe1988,Rowe:2012yi,Bartlett2002}, in which case the stabilizer subgroup will be smaller compared to the spinless case.

We act on our base state with a unitary transformation 
$U=\exp\left( \sum_{\alpha}(\lambda_{\alpha} E_{\alpha} - \lambda^*_\alpha E_{-\alpha}) + \sum_i x_i h_i\right)$
in order to produce generalized coherent states
\begin{equation}\label{eq:gpthycoherent}
|u\rangle \equiv U |\psi_R\rangle =  \mathcal{N}_H(z,z^*,x) \exp{z_\alpha \mathcal{D}(E_{-\alpha})}|\psi_R\rangle~,
\end{equation}
with $\mathcal{N}_H$ a normalization factor (including possibly an overall phase). $x$ are real coordinates on the stabilizer and $z$, $z^*$ are holomorphic coordinates on the orbit. The relation between the coordinates which appear in $U$ and the coordinates $z$ can be quite complicated in general.
Of course, multiplying $U$ from the right by an element of $H$ does not modify $|u\rangle$ (up to an overall phase) and therefore $U$ can be thought of as an element of $\mathcal{D}(G/H)$.

Generalized coherent states can be understood in terms of coadjoint orbits. Consider the dual element
\begin{equation}
 \label{eq:representative}
 \lambda(\mathcal{O}) = i \,  \mbox{Tr}\left[\left|\psi_R\right>\left<\psi_R\right| \mathcal{D}(\mathcal{O})\right]~,
\end{equation}
where the trace is taken in the infinite dimensional representation space.
The coadjoint action \eqref{eq:mcDef} on $\lambda$ is simply
$\left<\mbox{Ad}^*_U \lambda, \mathcal{O}\right> = i\, \mbox{Tr}\left(\left|\psi_R\right>\left<\psi_R\right| U^{-1} \mathcal{D}(\mathcal{O}) U\right)=i\ev{\mathcal{D}(\mathcal O)}{u}$, which indeed remains unmodified by the stabilizing elements $U \in \mathcal{D}(H)$. Thus we can view $\lambda$ as a representative that selects the orbit $G/H$.

The MC form of the unitary $U$ in Eq.~\eqref{eq:gpthycoherent} can be decomposed as $\Theta\equiv U^\dagger \dd U \equiv \Theta^-+\Theta^{(H)}+\Theta^{+}$ with $\Theta_{\pm}\in \mathfrak{n}_{\pm}$, $\Theta^{(H)}\in h_\mathbb{C}$. When acting with it on the base state we obtain 
\begin{equation}
\begin{split}
&\hspace{-3pt} \Theta|\psi_R\rangle = U^{-1}\left[ \frac{\dd\mathcal{N}_H}{\mathcal{N}_H} U
+  \mathcal{N}_H \dd(e^{z_{\alpha} \mathcal{D}(E_{-\alpha})}) \right]|\psi_R\rangle~.
\end{split}
\end{equation}
So $\Theta_-|\psi\rangle = (U^{-1} \mathcal{N}_H \dd e^{z_{\alpha} \mathcal{D}(E_{-\alpha})})_-|\psi\rangle$ and this
only depends on $\dd z_{\alpha}$ and not on $\dd z^*_{\alpha}$. Therefore $\Theta^-|\psi_R\rangle = \Theta^-_\mu \dd z^\mu|\psi_R\rangle$ and by conjugation $\langle\psi_R| \Theta^+ = \langle\psi_R| \Theta^+_\mu \dd z^{*}{}^\mu$. Notice also that $\Theta^\dagger = - \Theta$
and therefore
the FS-metric \eqref{eq:FSmetricgeneral2}  becomes $\dd s_{FS}^2 = -\langle \psi_R | \Theta^+_\mu \Theta^-_\nu |\psi_R\rangle \dd z^*{}^\mu  \dd z^\nu$.

The metric has a manifest complex structure $J$ compatible with the dagger which maps $z$ to $-i z$ and $z^*$ to $i z^*$. Together, the metric and the complex structure define a closed $2$-form according to $\omega(X,Y) = -g (X, JY)$, \ie
\begin{equation}
\begin{split}\label{omegaGenCo}
\omega & = -i \langle \psi_R | \Theta^+_\mu \Theta^-_\nu |\psi_R\rangle\,  \dd z^*{}^\mu \wedge \dd z^\nu \\
&= - i \langle \psi_R | \Theta \wedge \Theta |\psi_R\rangle = i \langle \psi_R | \dd\Theta |\psi_R\rangle~.
\end{split}
\end{equation} 
We recognize this as the Kirillov-Kostant symplectic form \eqref{Alambda} through the representative \eqref{eq:representative}.

Finally, the geometric action of the coadjoint orbit associated with the representative \eqref{eq:representative} relates to the $\mathcal{F}_1$ cost function \eqref{def:1norm}
\begin{equation}
\begin{split}
\mathcal{F}_1 \dd \sigma &= |\left<\psi_R\right| U^\dagger \dd U \left|\psi_R\right>|
= |\left< \lambda, \Theta\right>| = |\mathcal{A}_\lambda|~.\label{eq:cosetrep}
\end{split}
\end{equation}

For the specific case of the conformal algebra considered in \S\ref{sec:gdc}, we can take as base states the scalar primary states, $\ket{\psi_R}=\left|\Delta\right>$. The stabilizing subalgebra is $\mathfrak{h}=\mathfrak{so}(2)\times\mathfrak{so}(d)$, generated by $D$ and $L_{\mu\nu}$. The raising operators $\mathfrak{n}_+=\{K_\mu\}$ annihilate highest weight states and  the lowering operators are their conjugates $\mathfrak{n}_-=\{P_\mu\}$. Together these parametrize the coset \eqref{eq:coset}.

%% file: sections/Holography.tex
The symplectic geometry we found equally describes the space of timelike geodesics in AdS, and this allows us to rigorously derive a bulk description of complexity. Explicitly,  our circuits \eqref{eq:unitaryDef} starting from a scalar primary are mapped to the following particle trajectory  in embedding coordinates in $AdS_{d+1}$ of curvature radius $R$  (following \cite{Dorn:2005jt}'s conventions)
\begin{equation}
\label{eq:geodesicCorrespondence}
\begin{split}
X_0& \,=r(t) \cos(t/R), \quad X_{0'} = r(t) \sin(t/R), \\
X_\mu&\, = \frac{E_0 \, r(t)}{E \,A(\alpha,\alpha^*)} \left(\alpha_\mu  B^*(t,\alpha^*)+
\alpha^*_\mu B(t,\alpha)\right)
\end{split}
\end{equation}
where
\begin{equation}\label{eq:geodesicCorrespondence2}
\begin{split}
&\hspace{-10pt}r(t)= \frac{R\, E}{E_0} \frac{\sqrt{A(\alpha,\alpha^*)} }{|B(t,\alpha)|}, \qquad E=E_0\frac{(1-\alpha^2  \alpha^*{}^2)}{A(\alpha,\alpha^*)}\\
&\hspace{40pt}B(t,\alpha) = e^{ i t/R} \alpha^2-e^{- i t/R}~.
\end{split}
\end{equation}
Here, $\alpha$ parametrizes the phase space of the geodesics, and $A(\alpha,\alpha^*) > 0$ and $\alpha^2 \alpha^{*2} < 1$. $E$ is the energy of the massive particle, which is minimal at rest and equal to $E_0 = m R ~ (1 + O(1/mR))$, with $m$ the mass of the particle. The phase space is identical to that of the CFT${}_d$ with the identification $\Delta = E_0$.
Time evolution $e^{i \tau D} \ket \alpha = \ket{\alpha e^{i \tau}}$ amounts to translating the geodesic in time in $\mathrm{AdS}_{d+1}$ and fixed radius geodesics correspond to $\alpha^2 = 0$. 
The complexity \eqref{eq:generalComplexityGeodesics22}
is expressed in terms of the energy $E$ and the angular momentum $J$ 
 of the massive particle through  
$	\Omega_T^{S/A} = \sqrt{\dfrac{E \pm J - \Delta}{E \pm J + \Delta}}$ (see \cite{SupMat}J). 
For a circuit of circular geodesics starting at the origin and ending at a radius $r_T = R^2/\delta$ close to the boundary,  the complexity diverges as $\mathcal C[\delta] \sim \sqrt{\Delta} \log [2 R / \delta]$.

\begin{figure}[h]
\centering
\begin{subfigure}[b]{.45\linewidth}
\includegraphics[width=\linewidth]{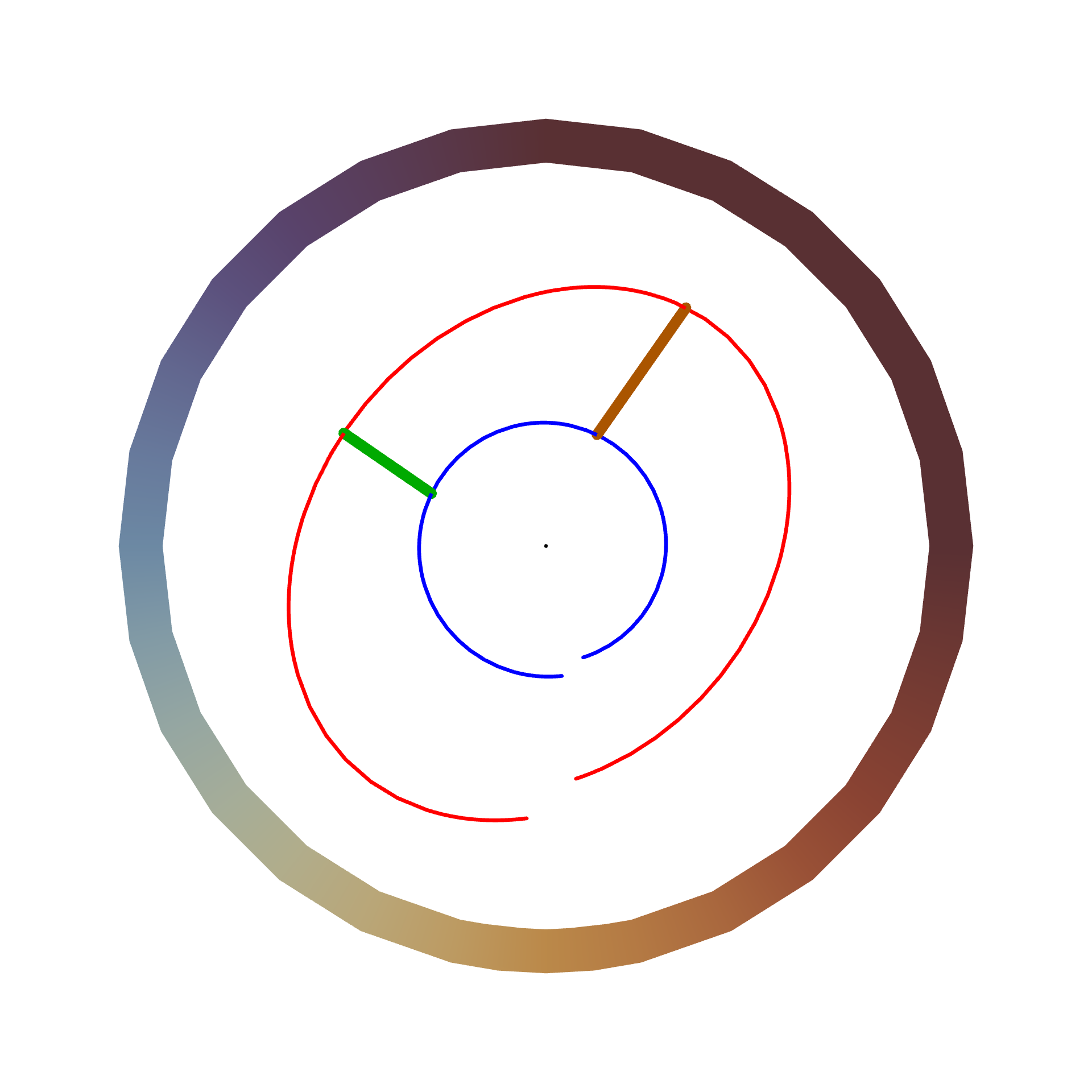}
\label{fig:top}
\end{subfigure}
\begin{subfigure}[b]{.45\linewidth}
\includegraphics[scale=0.17]{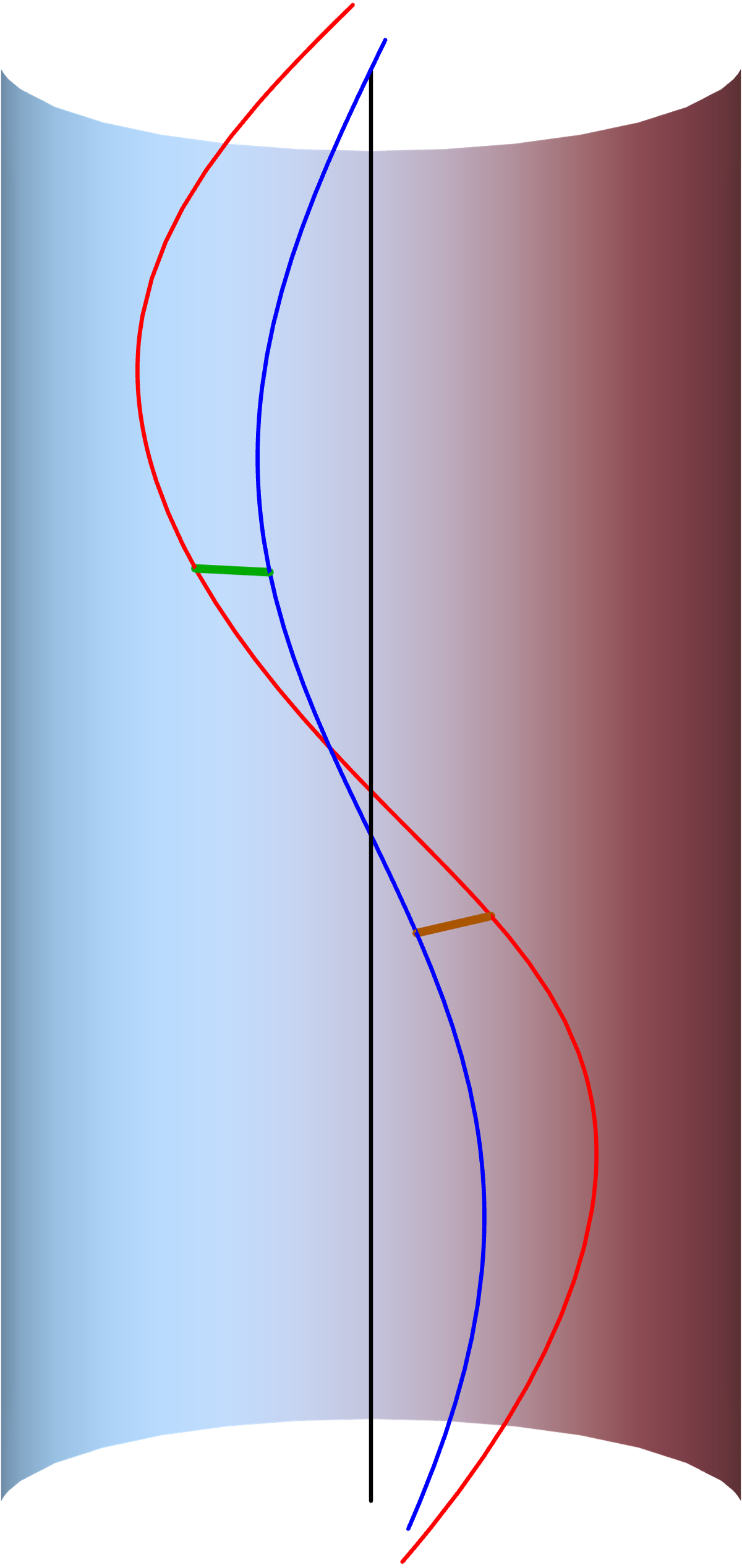}
\label{fig:front}
\end{subfigure}

\begin{subfigure}[b]{\linewidth}
\includegraphics[width=\linewidth]{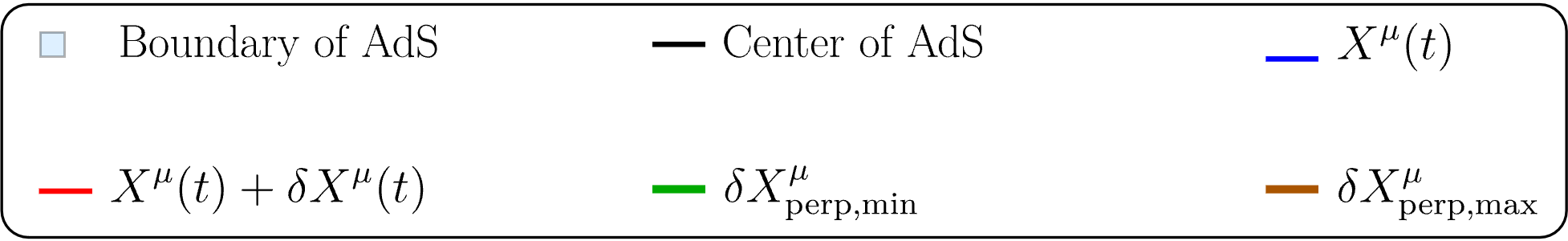}
\end{subfigure}
\caption{Illustration of two nearby timelike geodesics in $\mathrm{AdS}_3$ (blue, red) corresponding to two boundary circuits  and the minimal (green) and maximal (brown) perpendicular distance between them. The infinitesimal variation was exaggerated to improve the visualization.}
\label{fig:geo}
\end{figure}

The FS-metric over the space of circuits receives a surprisingly simple interpretation in terms of the maximal and minimal perpendicular distance between two infinitesimally nearby geodesics (as illustrated in figure \ref{fig:geo}, see \cite{SupMat}J)
\begin{equation}
\label{eq:geometricArgumentProp}
\dd s^2_{\mathrm{FS}} = \dfrac{\Delta}{2 R^2} \left(\delta X^2_{\mathrm{perp,min}} + \delta X^2_{\mathrm{perp,max}} \right)~.
\end{equation}

%% file: sections/discussion.tex
We studied the circuit complexity of trajectories associated to unitary representations of the conformal group in general dimensions. We considered primary states as reference states. Boundary states which are disentangled \cite{Miyaji:2014mca} could be an interesting alternative.
Our gates, consisting of global conformal transformations, are non-local similarly to the gates relevant for holographic complexity \cite{Fu:2018kcp}. We explained how our results  can be understood using the geometry of coadjoint orbits.
We presented general proofs relating the FS-metric and $\mathcal F_1$ cost function to a coadjoint orbit metric and geometric action in the context of generalized coherent states. These proofs are also applicable to circuits starting from spinning primaries and to other symmetry groups.

Our complexity geometry does not provide a notion of distance between any two states in the CFT Hilbert space. 
It is an important question for the future to describe the complexity for circuits moving across different conformal families. Furthermore,  considering more general states formed by non-local insertions could reveal  the role of OPE coefficients in studying complexity.

Considering the complexity of mixed states in CFT, \eg thermal states or subregions of the vacuum is another important question. For example, coherent states can be used as a starting point for the ensemble approach to mixed state complexity \cite{Agon:2018zso}.  It is also interesting to explore the complexity of states with a conformal timelike defect/boundary and compare to holography \cite{Chapman:2018bqj,Sato:2019kik,Braccia:2019xxi}.

The path-integral approach to  complexity  \cite{caputa2017,Camargo:2019isp,Milsted:2018yur,Milsted:2018san,Czech:2017ryf,Chandra:2021kdv} involves the two-dimensional Liouville action and central charge. Hence, it relates to circuits going beyond the global conformal group. It is therefore compelling to study the $d>2$ complexity of circuits constructed from general smearings of the stress tensor and tie the result to a higher-dimensional Liouville action \cite{caputa2017,Levy:2018bdc}.

Our complexity geometry is highly symmetric. It is interesting to break some of the symmetry by adding penalty factors -- effectively favoring certain directions through the manifold of conformal unitaries.  
Our  $\mathcal{F}_1$ cost function  \eqref{def:1norm} (also considered in \cite{caputa2017}) vanishes along certain non-trivial trajectories and differs from the $\mathcal{F}_1$ norm used when studying the complexity of Gaussian states (\eg \cite{Jefferson:2017sdb}). The difference is reminiscent of exchanging the order of the absolute value in the complexity definition and the sum over circuit generators. We intend to compare these two different definitions in the future.

We rigorously derived a bulk description of our circuits as trajectories between timelike geodesics in AdS. We could  connect this picture to the holographic complexity proposals for instance by exploring the influence of massive particles on the action. It is valuable to study generalizations of our circuit-geodesic duality in other spacetimes and for more than one (possibly spinning) particle. Further, it is important  to explore the relation of our bulk picture to the phase space of Euclidean sources  \cite{Marolf:2017kvq,Belin:2018fxe,Belin:2018bpg,Miyaji:2015woj} and hence possibly to the complexity=volume proposal (see also \cite{Abt:2018ywl}). Another compelling possibility is to connect our bulk picture to a parallel transport problem of timelike geodesics. similarly to what was done for spacelike geodesics \cite{Czech:2017zfq, Czech:2019vih} in the context of kinematic space \cite{Czech:2015qta, deBoer:2015kda, Czech:2016xec, deBoer:2016pqk}.

Complexity provides us with a new measure of entanglement in CFTs and it is interesting to probe its potential in diagnosing  phase transitions. Some inspiration can be drawn from \cite{PhysRevLett.99.095701,PhysRevB.88.064304,SciPostPhys.2.3.021,PhysRevB.102.174304,PhysRevResearch.2.013323}. We hope to come back to this question in the future.

%% file: sections/AppendixLoreEuc.tex
\section{A. Relating the Euclidean and Lorentzian Conformal Generators}\label{app:EucliLoreConf}

The Euclidean conformal generators
\begin{align}
 \begin{split}\label{EuclideanConformalAlgebra}
  & \hspace{-5pt} {[D, P_\mu]  =  P_\mu~, \atop  	~~~[D, K_\mu] = - K_\mu~,} \quad
 {[L_{\mu\nu}, P_\rho]  =  \delta_{\nu\rho} P_\mu - \delta_{\mu\rho} P_\nu~, \atop [L_{\mu\nu}, K_\rho]  =   \delta_{\nu\rho} K_\mu - \delta_{\mu\rho} K_\nu~,}
 \\
 & ~~[K_\mu, P_\nu]  =  2\left(\delta_{\mu\nu} D - L_{\mu\nu}\right)~,
% 	\\
% \hspace{-8pt} 	[L_{\mu\nu}, P_\rho] & =  \delta_{\nu\rho} P_\mu - \delta_{\mu\rho} P_\nu, \quad
%  [L_{\mu\nu}, K_\rho]  =   \delta_{\nu\rho} K_\mu - \delta_{\mu\rho} K_\nu,
 \\
 & ~~ 	[L_{\mu\nu}, L_{\rho\sigma}]  =  -L_{\mu\rho} \delta_{\nu\sigma} + L_{\mu\sigma} \delta_{\nu\rho}
 -( \mu \leftrightarrow \nu)~,
 \end{split}
 \end{align}
% \eqref{EuclideanConformalAlgebra} 
 used in our analysis obey an $SO(d+1,1)$ algebra. This might seem confusing since we are interested in studying unitary circuits of the Lorentzian conformal group $SO(d,2)$. However, the choice of Hermiticity conditions \eqref{eq:radialquant} ensures that we are building unitary circuits of the Lorentzian conformal group.
We present below the explicit relation between the Euclidean conformal generators on $\mathbb{R}^d$ and the Lorentzian conformal generators on $\mathbb{R}^{1,d-1}$. The Euclidean generators are then used to construct unitary representations of the Lorentzian conformal group and in fact play a role analogous to that of the ladder operators $J_{\pm}$ in the quantum mechanical treatment of angular momentum. Although this idea is very commonly used in defining the Hilbert space in CFTs with respect to constant time slices on the Lorentzian cylinder, it is not always explicitly addressed. We found that it was most clearly explained in \cite{Dalimil,Minwalla1998}, see also \cite{luscher1975global}. Since here the generators play an essential role in constructing the circuits for defining complexity, we found it worthwhile to present some of the details of this construction here for completeness.

\begin{figure}[h]
  \centering \includegraphics[width=0.6\textwidth]{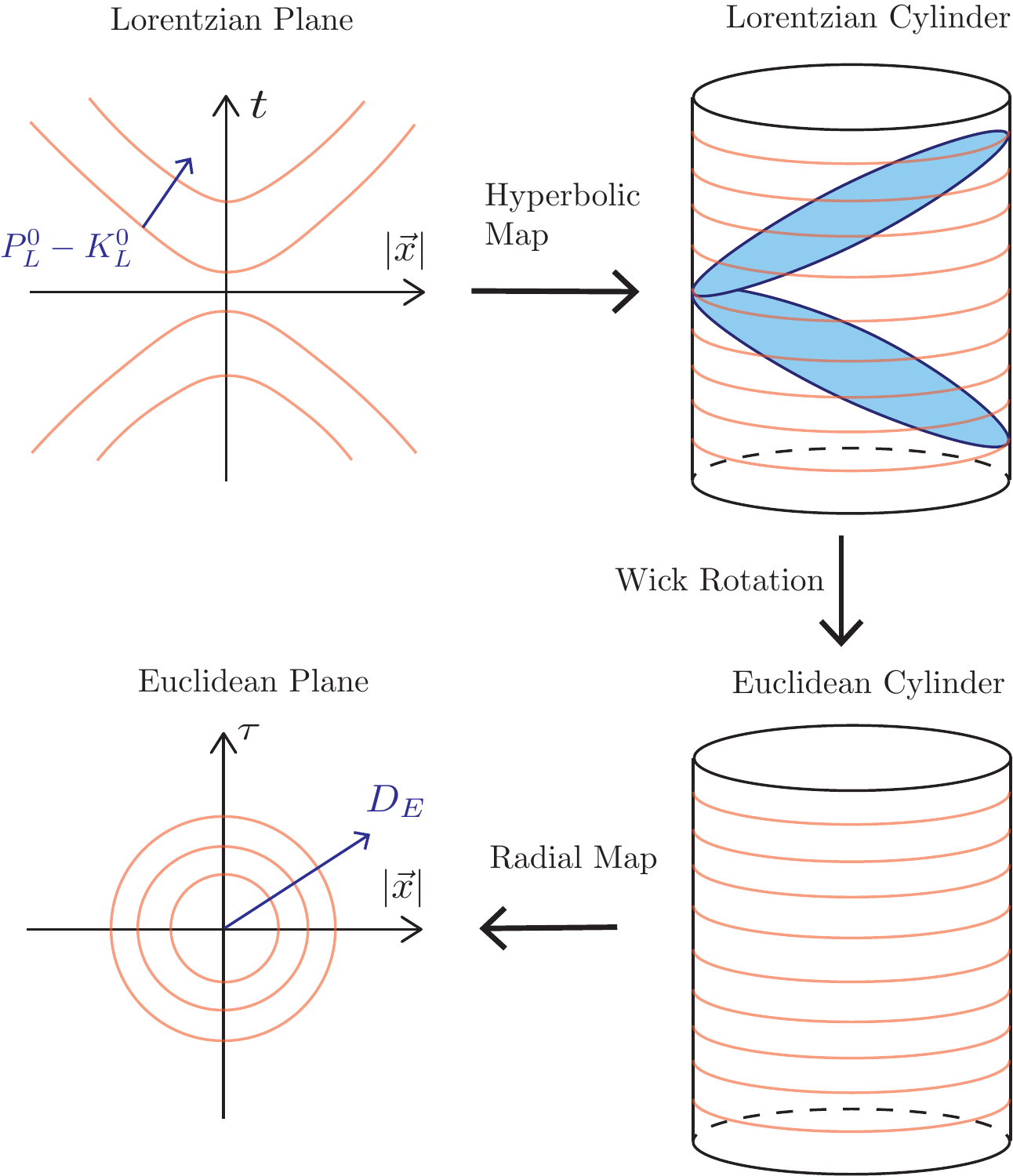}
  \caption{Illustration of the process of mapping the Lorentzian generators to the Euclidean generators. We start with quantizing the theory with respect to constant $P^0-K^0$ slices on the Lorentzian plane. Those are mapped to constant time slices on the Lorentzian cylinder. The Wick rotation maps those to constant time slices on the Euclidean cylinder. Finally we map the Euclidean cylinder to the Euclidean plane via a radial map.}
  \label{fig:LE2}
\end{figure}

The idea is to construct the Hilbert space with respect to constant time slices on the Lorentzian cylinder, using the cylinder translation generator as the Hamiltonian in quantizing the theory. This means that as a starting point for our circuits, we will consider states which are eigenstates of this generator. Since we are constructing unitary Lorentzian representations, all the Lorentzian generators are taken to be anti-Hermitian. The generator of time translations on the cylinder is mapped via a hyperbolic map $\tan (t_{cyl}) = \frac{2t}{1+|\vec x|^2-t^2}$ to the generator $P^0_L-K^0_L$ on the Lorentzian plane, which plays the role of the Hamiltonian when quantizing the theory on the Lorentzian plane. The relevant time slices are illustrated in figure \ref{fig:LE2}. Here and throughout the following, we will use a subscript $L$ to denote Lorentzian generators  on $\mathbb{R}^{1,d-1}$ and $E$ to denote Euclidean generators on $\mathbb{R}^d$. To go to the Euclidean picture one has to Wick rotate with respect to the time direction on the Lorentzian cylinder $\tau_{cyl} = i t_{cyl}$, in such a way as to obtain a Euclidean cylinder, and finally map radially to the Euclidean plane $\tau_{cyl} = \log(r)$. The Hermiticity conditions of the Euclidean generators follow from this mapping procedure and are as given in Eq.~\eqref{eq:radialquant}.

The explicit relations between the generators are given by \cite{Minwalla1998}:
\begin{subequations}
\begin{equation}
\begin{split}
L_{pq}^E = \begin{cases}
 L^L_{pq} &  \quad{1\leq p\leq d-1 \quad \text{and} \quad 1\leq q\leq d-1}~,\\
-\frac{1}{2} (P_p^L+K_p^L) & \quad {q=d \quad \text{and} \quad 1\leq p\leq d-1}~,\\
\frac{1}{2} (P_q^L+K_q^L) & \quad {p=d \quad \text{and} \quad 1\leq q\leq d-1}~, \\
0 &  \quad{p=d \quad \text{and} \quad q=d}~, \\
\end{cases}
\end{split}
\end{equation}
\begin{equation}
\begin{split}
P_{p}^E = \begin{cases}
\frac{1}{2} (P^L_p-K^L_p)- i L^L_{p0}\quad {1\leq p\leq d-1}~,\\
 D^L-\frac{i}{2} (P_0^L+K_0^L) \quad \ \, p=d~,\\
\end{cases}
\end{split}
\end{equation}
\begin{equation}
\begin{split}
K_{p}^E = \begin{cases}
-\frac{1}{2} (P^L_p-K^L_p)- i L^L_{p0}\quad {1\leq p\leq d-1}~,\\
-D^L-\frac{i}{2} (P_0^L+K_0^L) \quad \ \, p=d~,\\
\end{cases}
\end{split}
\end{equation}
\begin{equation}
\begin{split}
D^E = \frac{i}{2}(P^L_0-K^L_0)~,
\end{split}
\end{equation}
\end{subequations}
where the index $p$ takes values between $1$ and $d$. The Lorentzian generators are the usual anti-Hermitian ones. In terms of the differential representation for example:
\begin{equation}
\begin{split}
L_{\mu\nu}^L = & \, x_\mu \del_\nu-x_\nu \del_\mu~,\\
P_\mu^L = &\, - \del_\mu~, \\
K^L_\mu = &\, x^2 \del_\mu - 2 x_\mu x^\nu \del_\nu~, \\
D^L = & \, - x^\mu \del_\mu~,
\end{split}
\end{equation}
where $\mu=0\ldots d-1$. It then follows straightforwardly that the Euclidean generators satisfy the algebra \eqref{EuclideanConformalAlgebra} and the  Hermiticity conditions \eqref{eq:radialquant}.

%% file: sections/AppendixHigherD.tex
\section{B. Expectation Values of the Conformal Generators in \texorpdfstring{$d\geq 2$}{d >= 2}}
Here, we present additional details about the derivation of the expectation values of the symmetry generators in the state $|\alpha(\sigma)\rangle$.  We start by considering the following conjugation relations
\begin{equation}
	g(\alpha M, N) \equiv e^{-i \alpha M} N e^{i \alpha M}\, ,\label{eq:recombfunctions}
\end{equation}
where $M$ and $N$ are two generators of a given algebra. The conjugation functions obey a system of differential equations
\begin{equation}
	\dv{}{\alpha} g(\alpha M, N) = -i e^{-i \alpha M} [M, N] e^{i \alpha M}~, \quad g(0, N) = N~,
\end{equation}
which can then be solved to obtain the conjugation relations for the generators of the algebra.
Specifying to the case of the conformal group \eqref{EuclideanConformalAlgebra} in $d\geq 2$ we obtain
\begin{subequations}
	\label{eq:commutationRelationsHigherD}
	\begin{align}
		g(\alpha \cdot P, D) & = D + i \alpha \cdot P \, , \label{conjPD}\\
		g(\alpha \cdot K, D) & = D - i \alpha \cdot K \, ,\label{conjKD}\\
		g(\alpha \cdot P, L_{\mu\nu}) & = L_{\mu\nu} - i \alpha_\mu P_\nu + i \alpha_\nu P_\mu \, ,\label{conjPL}\\
		g(\alpha \cdot K, L_{\mu\nu}) & = L_{\mu\nu} - i \alpha_\mu K_\nu + i \alpha_\nu K_\mu \, ,\label{conjKL}\\
		g(\alpha \cdot P, K_\mu) & = K_\mu + 2 i \alpha_\mu D - 2 i \alpha^\rho L_{\mu\rho} - \left(2 \alpha_\mu \alpha^\rho - \alpha^2 \delta_\mu^\rho\right) P_\rho \, ,\label{conjPK}\\
		g(\alpha \cdot K, P_\mu) & = P_\mu - 2 i \alpha_\mu D - 2 i \alpha^\rho L_{\mu\rho} - (2 \alpha_\mu \alpha^\rho - \alpha^2 \delta_\mu^\rho) K_\rho \, . \label{conjKP}
	\end{align}
\end{subequations}

Next, we explain how to evaluate the expectation values required to get Eq. \eqref{eq:higherd1norm}-\eqref{eq:FSGen}.  The one point functions are evaluated as follows:
\begin{equation}
\ev{K_\mu}{\alpha} = 2i \alpha_\mu \Delta- \left(2 \alpha_\mu \alpha^\rho - \alpha^2 \delta_\mu^\rho\right)\langle\alpha| P_\rho|\alpha\rangle~,
\end{equation}
where we have used the conjugation \eqref{conjPK} to derive this equality. We further make the ansatz $\ev{K_\mu}{\alpha}=(\ev{P_\mu}{\alpha})^*=A \alpha_\mu+ B^*\alpha^*_{\mu}$ with $A$ and $B$ two complex coefficients. Solving for the coefficients leads to
\begin{equation}
	\label{PEV}
	\ev{P_\mu}{\alpha}  = \left(\ev{K_\mu}{\alpha}\, \right)^* = -2 i \Delta \dfrac{\alpha^*_\mu - \alpha^{*2} \alpha_\mu}{A(\alpha,\alpha^*) }  \, ,
\end{equation}
where $A(\alpha,\alpha^*)$ was defined in Eq.~\eqref{domain}. 
Next, we evaluate the expectation value
\begin{equation}
\ev{K_\mu P_\nu}{\alpha} = \ev{[K_\mu, P_\nu]}{\alpha}+ \ev{P_\nu K_\mu}{\alpha} =2 \delta_{\mu\nu}\ev{D}{\alpha} -2 \ev{ L_{\mu\nu}}{\alpha}+ \ev{P_\nu K_\mu}{\alpha}~,
\end{equation}
where in the last equality, we have used the algebra \eqref{EuclideanConformalAlgebra}.
The expectation values $\ev{D}{\alpha}$ and $\ev{L_{\mu\nu}}{\alpha}$ can be related to those calculated in Eq.~\eqref{PEV} by using the conjugation relations \eqref{conjPD} and \eqref{conjPL}, respectively. The conjugations \eqref{conjPK} and \eqref{conjKP} can then be used to relate the expectation value $\ev{P_\nu K_\mu}{\alpha}$ to the unknown $\ev{K_\mu P_\nu}{\alpha}$. Finally we solve the entire relation by using the ansatz $\ev{K_\mu P_\nu}{\alpha}=A \delta_{\mu\nu} + B \alpha_\mu \alpha_\nu + B^* \alpha^*_\mu \alpha^*_\nu +C \alpha^*_\mu \alpha_\nu +D \alpha_\mu \alpha^*_\nu$ with $A, C$ and $D$ real coefficients and $B$ a complex coefficient. In making this ansatz we have taken into account that $\ev{K_\mu P_\nu}{\alpha}$ is invariant under complex conjugation and the exchange of its indices. Finally, solving for the coefficients leads to
\begin{equation}\label{KPEV}
		\ev{K_\mu P_\nu}{\alpha} = \dfrac{2 \Delta}{A(\alpha,\alpha^*)^2}  \left[ \delta_{\mu\nu}A(\alpha,\alpha^*) - 2 (\Delta + 1) \left(\alpha^{*2} \, \alpha_\mu \alpha_\nu + \alpha^{2} \, \alpha^*_\mu \alpha^*_\nu - \alpha_\mu \alpha_\nu^*\right)- 2\left( 1 - 2 \alpha \cdot \alpha^* - \Delta \alpha^2 \alpha^{*2}\right) \alpha^*_\mu \alpha_\nu\right]  \, .
\end{equation}
The expectation value of the dilatation generator is derived in a very similar manner 
\begin{equation}
\langle \alpha |D| \alpha \rangle = \Delta \frac{1-\alpha^2 \alpha^*{}^2}{A(\alpha,\alpha^*)}\, .
\end{equation}
Positivity of this expectation value together with the condition $A(\alpha,\alpha^*)>0$ implies $\alpha\cdot\alpha^*<1$ (or equivalently $\alpha^2 \alpha^{*2}<1$).  
Similarly, we derive the expectation value of $L_{\mu\nu}$ through
\begin{equation}
	\ev{L_{\mu\nu}}{\alpha} = i \alpha_\nu \ev{P_\mu}{\alpha} - i \alpha_\mu \ev{P_\nu}{\alpha} = 2 \Delta \dfrac{\alpha^*_\mu \alpha_\nu - \alpha_\mu \alpha^*_\nu}{A(\alpha,\alpha^*)}~,
\end{equation}
such that its contraction yields
\begin{equation}
	\dfrac{1}{2} \ev{i L_{\mu\nu}}{\alpha} \ev{i L_{\mu\nu}}{\alpha} = 4 \Delta^2 \dfrac{(\alpha\cdot\alpha^*)^2 - \alpha^2 \alpha^{*2}}{A(\alpha,\alpha^*)^2}~.
\end{equation}
When we introduce the bulk picture, $\ev{D}{\alpha}$ and $\frac{1}{2}\ev{iL_{\mu\nu}}{\alpha} \ev{iL_{\mu\nu}}{\alpha}$ will be the energy and the squared angular momentum of the particle, respectively.

{\bf Constraints from unitarity:} Using the above one- and two-point functions we can now explain how the requirement that Eq.~\eqref{eq:unitaryDef} encodes a unitary transformation leads to the constraint
\begin{equation}\label{finalofappendixB}
	\gamma_D^I = - \dfrac{1}{2} \log A(\alpha,\alpha^*)~,
\end{equation}
where $A(\alpha,\alpha^*)$ was defined in Eq.~\eqref{domain}. In our derivation below, we use similar techniques to those often used in the context of coherent states, see, e.g., \cite{Zhang:1990fy}. The requirement that $U$ in Eq.~\eqref{eq:unitaryDef} be unitary implies
\begin{equation}
	\begin{aligned}\label{eqtosol}
		1 & = \braket{\alpha}
 = e^{- 2 \gamma_D^I \Delta} \ev{e^{- i \alpha^* \cdot K} e^{i \alpha \cdot P}}{\Delta} \equiv e^{- 2 \gamma_D^I \Delta} \braket{\hat \alpha}\, ,
	\end{aligned}
\end{equation}
where in the last equality, we have defined the un-normalized coherent state
\begin{equation}
	\ket{\hat \alpha} \equiv e^{i \alpha \cdot P} \ket \Delta \, .
\end{equation}
In order to evaluate the overlap $\braket{\hat \alpha}$, let us apply successive derivatives to this expression
\begin{equation}
	\begin{aligned}
		\partial_{\alpha^{*\mu}} \log \braket{\hat \alpha} & = \dfrac{\ev{- i K_\mu}{\hat \alpha}}{\braket{\hat \alpha}} = \ev{- i K_\mu}{ \alpha}~, \\
		\partial_{\alpha^{\mu}} \log \braket{\hat \alpha} & = \dfrac{\ev{ i P_\mu}{\hat \alpha}}{\braket{\hat \alpha}} = \ev{ i P_\mu}{ \alpha}~.
	\end{aligned}
\end{equation}
Using the explicit expectation values in Eq.~\eqref{PEV} and integrating these equations, we find
\begin{equation}
	\log \braket{\hat \alpha} = -\Delta \log A(\alpha, \alpha^*)+ c_1 \,, 
\end{equation}
with $c_1$ an arbitrary constant. The trivial solution for $\alpha=\alpha^*=0$ can then be used to fix $c_1=0$.
Finally, substituting the overlap $\braket{\hat \alpha}$ into Eq.~\eqref{eqtosol}, we obtain Eq.~\eqref{finalofappendixB}.

%% file: sections/complexityfunctional.tex
\section{C. Geodesic Trajectories in the Complexity Metric}
%H.

As mentioned in \S\ref{sec:gdc}, the geodesics on the coset space \eqref{eq:coset} take the form 
\begin{equation}\label{app:endstate}
	\ket{\alpha(\sigma)} \equiv \mathcal N(\sigma) e^{i \alpha(\sigma) \cdot P} \ket \Delta \equiv e^{i \sigma \left(\tilde \alpha \cdot P + \tilde \alpha^* \cdot K\right)} \ket \Delta \equiv e^{i \sigma X} \ket \Delta~,
\end{equation}
where the relation between $\alpha(\sigma)$ and $\{\tilde \alpha,\sigma\}$ is yet to be determined. We choose a convention in which the trajectory will reach its end point (the target state) at $\sigma=1$ where $\alpha(\sigma=1)\equiv\alpha_T$. We will explain in (\cite{SupMat}D) how to relate $\alpha_T$ and $\tilde \alpha$. For the moment, we will only need the relation
\begin{equation}\label{eq:TargetComplexityRel}
   2\,\tilde \alpha \cdot \tilde \alpha^* =\left( \atanh\Omega^S_T\right)^2 + \left(\atanh\Omega^A_T\right)^2 
\end{equation}
where $\Omega^{S/A}_T = (\Omega^+_T \pm \Omega^-_T)/2$ and $\Omega^{\pm}_T = \sqrt 2 \sqrt{\alpha_T\cdot \alpha^*_T \pm \sqrt{\alpha^2_T \alpha^*_T{}^2}}$.
A useful simplification is
$
\left(\Omega^{S/A}\right)^2=
 \alpha_T \cdot \alpha_T^* \pm  \sqrt{(\alpha_T \cdot \alpha_T^*)^2 - \alpha_T^2 \alpha_T^{*2}}$.
The Fubini-Study metric associated to the trajectory \eqref{app:endstate} is 
\begin{equation}
	\begin{aligned}
		\dv{s^2}{\sigma^2} & = \ev{X^\dagger X}{\Delta} - \left| \ev{X}{\Delta} \right|^2 = \tilde \alpha^{* \mu} \tilde \alpha^\nu \ev{K_\mu P_\nu}{\Delta} = 2 \Delta ~ \tilde \alpha \cdot \tilde \alpha^*~,
	\end{aligned}
\end{equation}
so the complexity of the state $|\alpha_T\rangle$ is given by
\begin{equation}
	\label{eq:generalComplexityGeodesics}
	\mathcal C[\tilde \alpha] = \sqrt{2 \Delta ~ \tilde \alpha^* \cdot \tilde \alpha} ~,
\end{equation}
where it remains to substitute the relation \eqref{eq:TargetComplexityRel}. We can see from this result that $\mathcal C$ admits no non-trivial null directions since $\mathcal C = 0$ would require $\Omega^+_T = \Omega_T^- = 0$ which in turn implies that $\alpha_T \cdot \alpha_T^* = \alpha_T^2 = 0$. Therefore, $\alpha_T = 0$.
In the bulk of the paper we insisted on parametrizing the state in terms of $\alpha(\sigma)$ rather than $\tilde\alpha$ since, as we saw in \cite{SupMat}B, this facilitates the evaluation of  correlation functions in the state. We therefore regard it as a more explicit characterization of the state throughout the paper. We will also see later that the description in terms of $\alpha(\sigma)$ lends itself to a more natural holographic interpretation.

As an example, let us consider the one-dimensional case for which we have
\begin{equation}\label{zetaaa}
    \ket{\alpha(\sigma)} = e^{i \alpha(\sigma) P} e^{i \gamma(\sigma) D} e^{i \beta(\sigma) K} \ket{\Delta} = e^{i \sigma (\tilde \alpha P + \tilde \alpha^* K)} \ket{\Delta}~.
\end{equation}
In this case the relation between $\alpha(\sigma)$ and $\{\tilde\alpha,\sigma\}$ is straightforward to derive and we obtain 
\begin{equation}\label{eq:recomb1d}
\alpha(\sigma) = \dfrac{\tilde \alpha}{|\tilde \alpha|} \tanh \left( \sigma | \tilde \alpha|\right)
\end{equation} 
(see, \eg \S 11.3.3 of \cite{klimovchumakov} or appendix C of \cite{chapman2018} for this recombination formula). It is possible to verify that this function solves the affinely parametrized geodesic equations $\alpha''(\sigma)= \frac{2 \alpha^*(\sigma) \alpha'(\sigma)^2}{\alpha(\sigma) \alpha^*(\sigma)-1}$ (and its complex conjugate) for the $d=1$ metric 
 $\dd s^2_{FS} = 2 \Delta \left[\dfrac{\dd \alpha \dd \alpha^*}{\left(1 - |\alpha|^2\right)^2}  \right]$ cf. Eq. \eqref{eq:FSGen}.
The target state obtained at $\sigma=1$, \ie $\alpha(1) = \alpha_T$ satisfies $\tanh(|\tilde \alpha|) = |\alpha_T|$. This leads to the complexity
 \begin{equation}
    \mathcal C[\alpha_T] = \sqrt{2 \Delta} ~ \atanh(|\alpha_T|) ~,
 \end{equation}
 which is indeed consistent with Eqs.~\eqref{eq:TargetComplexityRel}, \eqref{eq:generalComplexityGeodesics}.

 In two dimensions, we can use holomorphic factorization to decompose the metric as $\dd s^2 = \dd s^2_{H} + \dd s^2_{AH}$ where H stands for holomorphic and AH stands for  anti-holomorphic, see Eq. \eqref{TheMetricqueue}  below. The geodesic trajectories take the form of a direct product of two elements of the form \eqref{zetaaa} with parameters  $\{ \zeta(\sigma),\tilde \zeta\}$ for the holomorphic part and $\{  \bar \zeta(\sigma),\tilde{\bar \zeta}\}$ for the anti-holomorphic part (cf. Eq. \eqref{eq:2dDiffeom} below). The metric along these (straight-line trajectories) reads
 \begin{equation}
 	\dd s^2 = 2 \dd \sigma^2 \left[h |\tilde \zeta|^2 + \bar h |\tilde{\bar \zeta}|^2\right] ~.
 \end{equation}
Due to the holomorphic factorization we have that
 \begin{equation}
 	\zeta (\sigma)= \dfrac{\tilde \zeta}{|\tilde \zeta|} \tanh \left( \sigma | \tilde \zeta|\right) ~, \quad
 	\bar \zeta(\sigma)= \dfrac{\tilde{\bar \zeta}}{|\tilde{\bar \zeta}|} \tanh \left( \sigma |\tilde{\bar \zeta}|\right) ~.
 \end{equation}
 Using the same logic as before, we can compute the complexity for the end state $\{\zeta_T,    \bar \zeta_T\}$ and obtain 
 \begin{equation}
 	\mathcal C[\zeta_T, \bar\zeta_T] = \sqrt{2 h ~ (\atanh|\zeta_T|)^2 + 2 \bar h ~ \left(\atanh|\bar \zeta_T|\right)^2}~.
 \end{equation}
Note that this equation is consistent with the result Eqs.~\eqref{eq:TargetComplexityRel}, \eqref{eq:generalComplexityGeodesics} in general dimensions (using the relation \eqref{eq:2dsplitCopies} below).

%% file: sections/canonicalvariables.tex
\section{D. Canonical Variables and Recombination Formula}
%I

In this part of the supplementary material, we will explain how to establish the relation $\alpha(\tilde \alpha)$ in general dimensions, see \cite{SupMat}C. This method will be based on reference \cite{feinsilver2001canonical} which explains how to use the matrix representation of the algebra to find the coset representative of an arbitrary group element. 

We will work in the fundamental representation of the conformal group corresponding to 
the matrix algebra  $\mathfrak{so}(d,2)$ spanned by
\begin{equation}
	\label{eq:basisFund}
	(M_{AB})^C {}_D \equiv \delta_{A}{}^{C}\, g_{B D} - \delta_{B}{}^{C}\, g_{A D}~,
\end{equation}
where $g = \mathrm{Diag} (-, -, +, \dots, +)$ is the flat metric over $\mathbb R^{d,2}$, $\delta_{AB}$ is the Kronecker delta, and $A,B,C,D \in \{-1,0,\dots,d-1,d\}$. These matrices obey the commutation relations
\begin{equation}
	[M_{AB},M_{CD}] =  - g_{BD} M_{AC}  + g_{BC} M_{AD}   +  g_{AD}  M_{BC} - g_{AC} M_{BD}   \, ,
\end{equation}
and the orthogonality condition
\begin{equation}
	g M^T = - M g \, .
\end{equation}
The Euclidean conformal generators are given in the fundamental representation (denoted by $R$) by 
\begin{gather}\label{RgenFund}
	R(D) \equiv -i M_{-1,0}~, \quad R(L_{\mu\nu}) \equiv M_{\mu,\nu}~, \quad
	R(P_\mu) \equiv M_{-1,\mu} - i M_{0, \mu}~, \quad R(K_\mu) \equiv -(M_{-1,\mu} + i M_{0, \mu})~,
\end{gather}
where $\mu,\nu \in \{1, \dots, d\}$. Note that this representation obeys the algebra \eqref{EuclideanConformalAlgebra}, and is consistent with the Hermiticity conditions \eqref{eq:radialquant} through
\begin{equation}
R(\mathcal{O}^\dagger) = g^{-1} R(\mathcal{O})^\dagger g
\end{equation}
for any operator $\mathcal{O}$ in the conformal algebra.

We can define a vacuum vector $\Omega \in \mathbb{R}^{d,2}$ as
\begin{equation}
	\Omega = (1, i, 0 \dots)~, \quad R(D) \cdot \Omega = - \Omega~, \quad R(K_{\mu}) \cdot \Omega = 0~, \quad R(L_{\mu\nu}) \cdot \Omega = 0~.
\end{equation}
Note that the $D$ eigenvalue for the vector $\Omega$ above is not generic. However, in the following equation, the result of exponentiating $D$ is absorbed in the normalization constant $\mathcal{N}$ and hence changing it amounts to changing the overall phase of the coherent state. This phase eventually cancels out in Eq. \eqref{helloworld} and hence this choice does not influence our results. The action of a coherent element on the vacuum vector takes the form
\begin{equation}
	\mathcal N e^{i \alpha \cdot R(P)} \cdot \Omega = \mathcal N \begin{bmatrix} 1 - \alpha^2\\ i(1 + \alpha^2)\\ 2 i \alpha_1\\ \vdots\\ 2 i \alpha_d \end{bmatrix} ~, 
\end{equation}
which means that we can find the coset representative $\alpha$ associated to an arbitrary group element $g$ from the vector $v = g \cdot \Omega$ through
\begin{equation}\label{helloworld}
	\alpha_\mu = - i \dfrac{v_{\mu}}{v_{-1} - i v_{0}} ~ .
\end{equation}
Choosing now a geodesic path $g(\sigma) = e^{i \sigma \left(\tilde \alpha \cdot R(P) + \tilde \alpha^* \cdot R(K) \right)}$, we can use this method to find its associated coherent state parametrization $\alpha(\tilde \alpha,\sigma)$.

In $d = 1$, this method confirms the recombination formula  \eqref{eq:recomb1d}.  
In that case, the vector $v$ associated with the group element $g(\sigma)$ is
\begin{equation}
	v = \begin{bmatrix}
		\cosh\left(|\tilde\alpha|\sigma\right)^2 -\frac{\tilde\alpha}{\tilde \alpha^*} \sinh\left(|\tilde\alpha| \sigma\right)^2
		\\
\frac{i}{2} \left(\left(1+\frac{\tilde\alpha}{\tilde \alpha^*}\right) \cosh (2 \sigma  |\tilde \alpha|)+1-\frac{\tilde\alpha}{\tilde \alpha^*}\right)
\\
		i \sqrt{\frac{\tilde\alpha}{\tilde \alpha^*}} \sinh(2 \sigma |\tilde\alpha|)
	\end{bmatrix}~,
\end{equation}
and $\alpha(\tilde \alpha,\sigma)$ is given by
\begin{equation}\label{eq:another1d}
	\alpha(\tilde \alpha,\sigma) = \dfrac{\tilde \alpha}{|\tilde \alpha|} \tanh\left( |\tilde \alpha| \sigma \right)~.
\end{equation}
This equation can be inverted as follows
\begin{equation}
	\tilde \alpha(\alpha) = \dfrac{ \alpha}{|\alpha|} \atanh(|\alpha|)/\sigma ~.
\end{equation}

In $d\geq2$, we find that 
\begin{equation}
	\label{eq:recombFormula}
	\alpha_\mu = \dfrac{\sinh(\tilde \omega_+ \sigma)}{\cosh(\tilde \omega_+ \sigma) + \cosh(\tilde \omega_- \sigma)} \dfrac{\tilde \alpha^*_\mu e^{i \tilde \beta} + \tilde \alpha_\mu}{ \tilde \omega_+} - \dfrac{\sinh(\tilde \omega_- \sigma)}{\cosh(\tilde \omega_+ \sigma) + \cosh(\tilde \omega_- \sigma)} \dfrac{\tilde \alpha^*_\mu e^{i \tilde \beta} - \tilde \alpha_\mu}{ \tilde \omega_-}
\end{equation}
where $\tilde \omega_\pm = \sqrt{2}\sqrt{\tilde \alpha \cdot \tilde \alpha^* \pm \sqrt{\tilde\alpha^2~\tilde\alpha^{*2}} }$ 
 and $e^{2 i \tilde \beta} = \tilde \alpha^2 / \tilde \alpha^{*2}$. This method is applicable for any dimension $d$. 
As a consistency check, we note that in 
 $d = 2$, the above formula is consistent with holomorphic factorization, \ie using  it together with the relation \eqref{eq:2dsplitCopies} below, we simply obtain a double copy of the relation \eqref{eq:another1d}
\begin{equation}
	\zeta = \dfrac{\tilde \zeta}{|\tilde \zeta|} \tanh(\sigma |\tilde \zeta|)~, \quad \bar \zeta = \dfrac{\tilde{\bar \zeta}}{|\tilde{\bar \zeta}|} \tanh(\sigma |\tilde{\bar \zeta}|)~.
\end{equation}

 The inversion (finding $\tilde\alpha$ in terms of $\alpha$) can be translated to the problem of solving three equations (for $\alpha \cdot \alpha$, $\alpha^* \cdot \alpha^*$ and $\alpha \cdot \alpha^*$) in terms of three variables ($\tilde\alpha \cdot \tilde\alpha$, $\tilde\alpha^* \cdot \tilde\alpha^*$ and $\tilde\alpha \cdot \tilde\alpha^*$) using the above relations. This system of equations can be solved analytically. To start, note that 
\begin{equation}
	\alpha(\sigma)^2 = \sqrt{\dfrac{\tilde \alpha^2}{\tilde \alpha^{*2}}} \dfrac{\cosh\left(\tilde \omega_+ \sigma\right) - \cosh\left(\tilde \omega_- \sigma\right)}{\cosh\left(\tilde \omega_+ \sigma\right) + \cosh\left(\tilde \omega_- \sigma\right)}~, \quad \alpha(\sigma) \cdot \alpha(\sigma)^* = \dfrac{\sinh\left(\tilde \omega_+ \sigma\right)^2 + \sinh\left(\tilde \omega_- \sigma\right)^2}{\left(\cosh\left(\tilde \omega_+ \sigma\right) + \cosh\left(\tilde \omega_- \sigma\right)\right)^2}~,
\end{equation} 
and ${\alpha^2}/{\alpha^{*2}}={\tilde \alpha^2}/{\tilde\alpha^{*2}}$.
It is useful to consider the following combinations: 
\begin{equation}
	\Omega_\pm(\sigma) = \dfrac{2 \sinh\left(\tilde \omega_\pm \sigma\right)}{\cosh\left(\tilde \omega_+ \sigma\right) + \cosh\left(\tilde \omega_- \sigma\right)} \text{ where } \Omega_\pm(\sigma) \equiv \sqrt{2} \sqrt{\alpha(\sigma) \cdot \alpha^*(\sigma) \pm \sqrt{\alpha(\sigma)^2 \alpha^*(\sigma)^{2}}}~,
\end{equation}
which can be inverted as
\begin{equation}
\tilde\omega_{\pm} = \frac{1}{\sigma} \,\,
\tanh ^{-1}\left(\frac{4 \,\Omega_{\pm}(\sigma)}{4\pm(\Omega_{+}^2(\sigma)-\Omega_{-}^2(\sigma))}\right)
= 
\frac{1}{\sigma} \,\,
\left(\tanh ^{-1}\left[\Omega^S(\sigma)\right]\pm 
\tanh ^{-1}\left[\Omega^A(\sigma)\right]\right)
\end{equation}
where we have introduced the variables $\Omega^{S/A}(\sigma) \equiv (\Omega^+(\sigma) \pm \Omega^-(\sigma))/2$
and used the identity $\atanh(x) = \dfrac{1}{2} \log \left[\dfrac{1 + x}{1 - x}\right]$. 
Noting that $\tilde \alpha \cdot \tilde \alpha^* = (\tilde \omega_+^2 + \tilde \omega_-^2)/4$ and using the previous formula leads to 
\begin{equation}
	\tilde \alpha \cdot \tilde \alpha^* = \dfrac{1}{2 \sigma^2} \left[\left(\atanh \Omega^S(\sigma) \right)^2 + \left(\atanh\Omega^A(\sigma)\right)^2 \right].
\end{equation}
This is true for all $\sigma$ and so in particular for $\sigma = 1$  we find
\begin{equation}
	2 ~\tilde \alpha \cdot \tilde \alpha^* = \left(\atanh\left[\dfrac{\Omega^+_T + \Omega^-_T}{2}\right]\right)^2 + \left(\atanh\left[\dfrac{\Omega^+_T - \Omega^-_T}{2}\right]\right)^2~,
\end{equation}
with $\Omega^\pm_T \equiv \sqrt{2} \sqrt{\alpha_T \cdot \alpha^*_T \pm \sqrt{\alpha_T^2 \alpha^*_T{}^{2}}}$ where we have defined $\alpha(\sigma=1)\equiv \alpha_T$.
Similarly using $\tilde \alpha^2 =  \sqrt{\alpha^2/\alpha^{*2}}\left(\tilde\omega_+^2-\tilde\omega_-^2\right)/4$ we obtain
\begin{equation}
	\tilde \alpha^2 = \frac{1}{\sigma^2 } \sqrt{\frac{\alpha^2}{\alpha^{*2}}}\,\, \atanh[\Omega^A(\sigma)]\,\, \atanh[\Omega^S(\sigma)] ~.
\end{equation}
%we can compute 
Using the above, the full inversion of \eqref{eq:recombFormula} is straightforwardly obtained
% inverted formula
\begin{equation}
	\sigma \tilde \alpha_\mu = \atanh\left(\Omega^S(\sigma)\right)
	\dfrac{\Omega^S(\sigma) \alpha_\mu(\sigma) - \Omega^A(\sigma) \sqrt{\frac{\alpha^2}{\alpha^{*2}}} \alpha^*_\mu(\sigma)}{\Omega^+(\sigma) \Omega^-(\sigma)} - \atanh\left(\Omega^A(\sigma)\right)\dfrac{\Omega^A(\sigma) \alpha_\mu(\sigma) - \Omega^S(\sigma) \sqrt{\frac{\alpha^2}{\alpha^{*2}}} \alpha^*_\mu(\sigma)}{\Omega^+(\sigma) \Omega^-(\sigma)}~.
\end{equation}

%% file: sections/bounds.tex
\section{E. Bounds on Complexity and its Time Evolution}
%K

Here, we derive bounds on the complexity and its rate of change.

\subsection{Bounds on Complexity}
We consider bounds on the complexity \eqref{eq:generalComplexityGeodesics22} of a final state $\alpha_T$ with energy $E_T$.  
We will use the following bounds on $\log$
\begin{equation}
	\dfrac{y-1}{y+1} \leq \log(y) \leq \dfrac{y-1}{\sqrt{y}}~, \quad y \geq 1~,
\end{equation}
which lead to the following bounds on $\atanh(x)^2$% = \log\left(\dfrac{1+x}{1-x}\right)^2/4$
\begin{equation}
	\dfrac{x^2}{4} \leq \atanh(x)^2 \leq \dfrac{x^2}{1 - x^2}, \qquad -1<x<1~.
\end{equation}
We therefore have
\begin{equation}
	\begin{aligned}
		\dfrac{\sqrt{\Delta}}{2} \sqrt{(\Omega^{S}_T)^2 + (\Omega^{A}_T)^2} & \leq \mathcal C[\alpha_T] \leq \sqrt{\Delta} \sqrt{\dfrac{(\Omega^{S}_T)^2}{1 - (\Omega^{S}_T)^2} + \dfrac{(\Omega^{A}_T)^2}{1 - (\Omega^{A}_T)^2}}\\
		\text{\ie ~~}  \sqrt{\frac{\Delta}{2} \, \alpha_T \cdot \alpha^*_T} & \leq \mathcal C[\alpha_T] \leq \sqrt{E_T - \Delta}\\
	\end{aligned}
\end{equation} 
where we used $\left(\Omega^{S/A}\right)^2= \alpha_T \cdot \alpha_T^* \pm  \sqrt{(\alpha_T \cdot \alpha_T^*)^2 - \alpha_T^2 \alpha_T^{*2}}=\dfrac{E_T \pm J_T - \Delta}{E_T \pm J_T + \Delta}$ with $E_T \equiv \ev{D}{\alpha_T} = \Delta (1 - \alpha_T^2 \alpha_T^{*2})/A(\alpha_T, \alpha_T^*)$ the energy and $J_T \equiv \sqrt{\langle\alpha_T| i L_{\mu\nu}| \alpha_T\rangle \langle \alpha_T| i L_{\mu\nu}| \alpha_T\rangle/2} =  2\Delta \sqrt{(\alpha_T \cdot \alpha_T^*)^2 - \alpha_T^2 \alpha_T^{*2}}/A(\alpha, \alpha^*)$ the angular momentum. 
The lower bound can be expressed in terms of $E_T$ by using the inequalities %$0 \leq J_T \leq E_T - \Delta$ and
\begin{equation}
0 \leq J_T \leq E_T - \Delta \qquad
\Rightarrow
\qquad
	\alpha_T \cdot \alpha_T^* = \dfrac{E_T^2 - J_T^2 - \Delta^2}{(E_T + \Delta)^2 - J_T^2} \geq \dfrac{2 \Delta(E_T-\Delta)}{\left(E_T + \Delta\right)^2}~,
\end{equation}
leading to
\begin{equation}
	 \dfrac{\Delta}{E_T + \Delta}\sqrt{ (E_T-\Delta)} \leq \mathcal C[\alpha_T] \leq \sqrt{E_T - \Delta}~.
\end{equation}

The above bound uses the inequality $J_T + \Delta \leq E_T $ which can be proven using the definitions of $E_T$ and $J_T$ above (these will actually have a nice interpretation in holography, see \eqref{eq:EJalphasInv} below). For this note that 
\begin{equation}
   \left(E_T - \Delta\right)^2 - J_T^2 =  \dfrac{4 \Delta^2 \alpha^2 \alpha^{*2}}{A(\alpha, \alpha^*)} > 0~.
\end{equation} 
We then see indeed that $ J_T + \Delta \leq E_T$ when $ \Delta \leq E_T$ which in turn follows from positivity of the spectrum, see comment below Eq.~\eqref{domain}.

\subsection{Complexity of Time Evolved States}
Let us now consider the time evolution of a coherent state   $e^{i \tau D} \ket{\alpha_0} = \ket{\alpha_0 e^{i \tau}}$, for which the FS-metric reduces to
\begin{equation}
   \dd s_{FS}^2=2 \Delta  \dd \tau^2 \left(\frac{\alpha_0 \cdot \alpha_0^*}{A(\alpha_0,\alpha_0^*)}+2\frac{(\alpha_0 \cdot \alpha_0^*)^2-\alpha_0^2 \alpha_0^*{}^2}{{A(\alpha_0,\alpha_0^*)^2}}\right).
\end{equation}
This can be simplified and bounded as follows
\begin{equation}
   \begin{aligned}
      \dv{s_{FS}^2}{\tau^2} & = 2 \Delta \dfrac{\alpha_0 \cdot \alpha_0^* + (\alpha_0 \cdot \alpha_0^* - 2) \alpha_0^2 \alpha_0^{*2}}{A(\alpha_0,\alpha_0^*)^2}~,\\
      & = \dfrac{\ev{D}{\alpha_0}^2 }{\Delta}\dfrac{2 \alpha_0^2 \alpha_0^{*2} (\alpha_0 \cdot \alpha_0^* - 2) + 2 \alpha_0 \cdot \alpha_0^*}{\left(1 - \alpha_0^2 \alpha_0^{*2}\right)^2}\\
      & = \dfrac{\ev{D}{\alpha_0}^2}{\Delta} \left(1 - \dfrac{(1 + \alpha_0^2 \alpha_0^{*2}) A(\alpha_0, \alpha_0^*)}{\left(1 - \alpha_0^2 \alpha_0^{*2}\right)^2}\right)~\\
      & \leq \dfrac{\ev{D}{\alpha_0}^2}{\Delta} \equiv \dfrac{E^2}{\Delta} ~.
   \end{aligned}
\end{equation}
Unitarity further constrains $\Delta$ for $d \geq 3$ \cite{rychkov2017}
\begin{equation}
   \Delta \geq d/2 - 1.
\end{equation}
Therefore, the rate of change of the FS-distance along this trajectory in the space of states is bounded by 
\begin{equation}
   \dv{s_{FS}}{\tau} \leq \frac{E}{\sqrt{\Delta}} \leq E \sqrt{\dfrac{2}{d - 2}} ~.    
\end{equation}

%% file: sections/AppendixD2.tex
\section{F. Comparison with Previous Results in \texorpdfstring{$d=2$}{d = 2}}
%C

In this part of the supplementary material, we will focus on conformal circuits in two dimensions. In this case, the global conformal group can be extended to (two copies of) the full Virasoro group. The discussion is often phrased in terms of holomorphic and anti-holomorphic coordinates $z  = x + i \tau$ ($\bar z  = x - i \tau$).
The global conformal algebra $\mathfrak{so}(2,2)\simeq\mathfrak{sl}(2,\mathbb{R})\times \mathfrak{sl}(2,\mathbb{R})$ is generated by holomorphic generators $L_{-1}, L_0, L_1$ satisfying
\begin{equation}\label{algebraL}
[L_{\pm1},L_0]=\pm L_{\pm 1}~, \quad [L_1,L_{-1}]=2L_0~,
\end{equation}
and anti-holomorphic generators $\bar L_{-1}, \bar L_0, \bar L_1$ satisfying similar relations. In radial quantization, the generators satisfy the Hermiticity conditions $L_{1}^{\dagger}=L_{-1}$ (and $\bar L_{1}^{\dagger}=\bar L_{-1}$). Working in this language will allow us to compare our results with the previous complexity literature in $2d$ CFTs \cite{caputa2019, Erdmenger:2020sup, flory2020, flory2020a}.

For the reference state we again select a spinless highest weight state $\ket{\psi_R} = \ket{h, \bar h=h} \equiv |\bf h\rangle$, satisfying
$\bar L_0 \ket{\bf{h}} = L_0 \ket{\bf{h}} = h \ket{\bf{h}}$, $\bar L_1 \ket{\bf{h}}=L_1 \ket{\bf{h}} = 0$.
As expected, the cost functions \eqref{eq:costfunctions} factorize into holomorphic and anti-holomorphic parts, which can be treated separately. We will focus on holomorphic unitary circuits within the $2d$ global conformal group
\begin{equation}
	\label{eq:2dDiffeom}
	\begin{aligned}
	U(\sigma) & \equiv e^{i \zeta(\sigma) L_{-1}} e^{i \gamma(\sigma) L_0} e^{i \zeta_1(\sigma) L_1}~ ,\\
	\ket{\zeta} & \equiv U(\sigma) \ket{\bf{h}} = \nn(\sigma) e^{i \zeta(\sigma) L_{-1}} \ket{\bf{h}}~,
	\end{aligned}
\end{equation}
where $\gamma \equiv \gamma_R + i \gamma_I$ is decomposed into its real and imaginary parts and unitarity restricts $\gamma_I = - \log(1 - |\zeta|^2)$ with $|\zeta|^2<1$ and $\zeta_1 = \zeta^* e^{i \gamma_R}$. These relations can be derived using an explicit recombination formula, see \S 11.3.3 of \cite{klimovchumakov}.
Following the same logic as in \S\ref{sec:gdc} (see the expectation values of the conformal generators below) we can derive the $\mathcal F_1$ cost \eqref{def:1norm}
\begin{equation}
	\label{eq:2d1norm}
	\begin{aligned}
	\dfrac{\mathcal{F}_1}{h}
	& = \left| \dfrac{ \dot \zeta \zeta^*-\dot \zeta^* \zeta}{1 - |\zeta|^2}~ + i \dot \gamma_R + (\zeta \leftrightarrow \bar{\zeta})\right|~,
	\end{aligned}
\end{equation}
and the Fubini-Study metric \eqref{eq:FSmetricgeneral2}
\begin{equation}
	\dd s^2_{FS} = 2 h \left[\dfrac{\dd \zeta \dd \zeta^*}{\left(1 - |\zeta|^2\right)^2} + \dfrac{\dd \bar\zeta \dd \bar \zeta^*}{\left(1 - |\bar \zeta|^2\right)^2} \right] \, ,\label{eq:2dFSmetric}
\end{equation}
which corresponds to the hyperbolic geometry on (two copies of) the Poincar\'{e} unit disk.
These cost functions are obtained from those in the \S\ref{sec:gdc} by the substitutions $\alpha^\mu\del_\mu  = \left(\frac{\zeta-\bar \zeta}{2i}\right)  \del_\tau+\left(\frac{\zeta+ \bar \zeta}{2}\right) \del_x$ and $\gamma^R_D = (\gamma_R+\bar \gamma_R)/2$. It is worth noting that holomorphic factorization implies that in the case with spin $s=h-\bar h\neq 0$ in $2d$ the result is straightforwardly obtained by replacing the coefficient in front of the anti-holomorphic part of the metric by $\bar h$.

There are a number of existing results in the literature for the circuit complexity of the Virasoro group on the cylinder~\cite{caputa2019, Erdmenger:2020sup, flory2020, flory2020a}. However, in those papers the circuit complexity is given in terms of the diffeomorphism $f(z=e^{i\theta}) \in \text{Diff} (S^1)$ associated to the holomorphic part of the circuit.
We find it insightful to explicitly relate our approach to the previous literature for $d=2$. In addition to providing a consistency check for our results, this also provides a clear interpretation of the circuits and gates associated with the diffeomorphisms of \cite{caputa2019, Erdmenger:2020sup, flory2020, flory2020a}.
Restricting to the holomorphic copy, \cite{caputa2019} have shown that the $\mathcal F_1$ cost is given by
\begin{equation}
	\mathcal{F}_{1} =  \biggr| \int_0^{2 \pi} \frac{\dd \theta}{2\pi} \, \dfrac{\partial_\sigma f(\sigma, \theta)}{\partial_\theta f(\sigma, \theta)} \left[ - \tilde h + \dfrac{c}{12}\{f, \theta\}\right] \biggr| \, ,
	\label{eq:2DF1}
\end{equation}
where $c$ is the central charge, $\tilde h\equiv h-c/24$ is the shifted eigenvalue of the generator $L_0$ on the cylinder and $\{f, \theta\}$ is the Schwarzian derivative. Denoting $\varepsilon_i = \dfrac{\partial_\sigma f(\sigma, \theta_i)}{\partial_{\theta_i} f(\sigma, \theta_i)}$, the FS-metric is~\cite{flory2020,flory2020a}
\begin{equation}
\frac{ds^2_{FS}}{d\sigma^2}  = \int_0^{2 \pi}  \frac{\dd\theta_1}{2\pi}  \frac{\dd\theta_2}{2\pi} \, \varepsilon_1 \varepsilon_2 \left[ \dfrac{c}{32 \sin^4 \left[(\theta_1 - \theta_2)/2\right]} - \dfrac{h}{2 \sin^2 \left[(\theta_1 - \theta_2)/2\right]}\right] \label{eq:2DFSFH} \, .
\end{equation}
As explained in section \S3.3 of \cite{flory2020a}, this expression must be regularized and thus we have
\begin{equation}
	\label{eq:2DFSFHReg}
	\dv{s^2_{FS}}{\sigma^2} = \int_0^{2 \pi} \dd \theta_1 \dd \theta_2 \log \left[\sin(\dfrac{\theta_1 - \theta_2}{2})^2\right] \left[- \dfrac{c}{24} \partial_{\theta_1}^2 \varepsilon_1 \partial_{\theta_2}^2 \varepsilon_2 - \tilde h \, \partial_{\theta_1} \varepsilon_1 \partial_{\theta_2}\varepsilon_2\right] \, .
\end{equation}

The diffeomorphism associated with the circuit \eqref{eq:2dDiffeom} is a M\"{o}bius transformation on the circle parametrized by the coordinate $\theta$, \ie  $z=e^{i\theta}$
\begin{equation}
	\label{eq:moebiusTransformation}
	f(\sigma) \colon e^{i \theta} \to \dfrac{A(\sigma) e^{i \theta} + B(\sigma)}{B^*(\sigma) e^{i \theta} + A^*(\sigma)} \text{ with } |A|^2 - |B|^2 = 1 \, ,
\end{equation}
which maps the unit circle in the complex plane to itself.
Acting with our unitary \eqref{eq:2dDiffeom} expressed in terms of the differential generators $L_n = -z^{n+1} \del_z$ ($n=0,\pm 1$) straightforwardly leads to
\begin{equation}
	\label{eq:diff2D}
	f(\sigma) \colon e^{i \theta} \to \dfrac{e^{i \theta} \left(e^{i \gamma/2} + \zeta \zeta_1 e^{-i \gamma/2}\right) + i \zeta e^{-i \gamma/2}}{-i\zeta_1 e^{i \theta} e^{-i \gamma/2} + e^{-i \gamma/2}} ~,
\end{equation}
where the flip in the signs of the parameters is due to the usual active/passive transformation conversion.
Constraining this transformation to be M\"{o}bius yields
\begin{equation}
	\label{eq:2DunitaryConditions}
	\zeta_1 = \zeta^* e^{i \gamma_R}~, \quad \gamma_I = - \log(1 - |\zeta|^2)~.
\end{equation}
Note that these are precisely the relations we obtained by requiring that the circuit \eqref{eq:2dDiffeom} is a unitary.
Finally, substituting the relations \eqref{eq:2DunitaryConditions} into the unitary \eqref{eq:diff2D} we obtain
\begin{equation}
	\label{eq:moebiusTransfFinal}
	f(\sigma,e^{i \theta}) = \dfrac{ e^{i \gamma^*/2} e^{i \theta}+ i \zeta(\sigma) e^{- i \gamma /2}}{-i \zeta^*(\sigma) e^{ i \gamma^* / 2}e^{i \theta} + e^{-i \gamma/2}} \, .
\end{equation}
As expected for a global transformation in $d = 2$ we have $\{f, z\} = 0$, while the mapping to the cylinder creates a non-zero Schwarzian $\{f, \theta\} = 1/2$.
Substituting this diffeomorphism into Eqs.~\eqref{eq:2DF1} and \eqref{eq:2DFSFHReg}, we immediately recover our results \eqref{eq:2d1norm}-\eqref{eq:2dFSmetric} upon the addition of the second copy.

\section{Expectation Values of the Conformal Generators in \texorpdfstring{$d=2$}{d=2}}
We here derive the conjugation relations and expectation values of the conformal generators in the special case of $d = 2$ required for the derivation of Eqs.~\eqref{eq:2d1norm}-\eqref{eq:2dFSmetric}. For the conjugation relations \eqref{eq:recombfunctions} we obtain using the algebra \eqref{algebraL}
\begin{subequations}\label{eq:commutationRelations2d}
	\begin{align}
		g(\zeta L_1, L_0) & = L_0 - i \zeta L_1 \, ,\label{L1L0Comrel}\\
		g(\zeta L_{-1}, L_0) & = L_0 + i \zeta L_{-1} \, ,\label{Lm1L0Comrel}\\
		g(\zeta L_1, L_{-1}) & = L_{-1} - 2 i \zeta L_0 - \zeta^2 L_1 \, ,\label{L1Lm1Comrel}\\
		g(\zeta L_{-1}, L_1) & = L_1 + 2 i \zeta L_0 - \zeta^2 L_{-1} ~.\label{Lm1L1Comrel}
	\end{align}
\end{subequations}

Next, we explain how to evaluate the one- and two-point functions of the $2d$ conformal generators
 in the states \eqref{eq:2dDiffeom} along the circuit.
The one point functions are evaluated as follows:
\begin{align}\label{relation2323}
 \ev{L_1}{\zeta}
		& = 2 i \zeta h - \zeta^2  \ev{L_{-1}}{\zeta}~,
\end{align}
where in this equality we used the relation \eqref{Lm1L1Comrel}. The Hermiticity relations for the radial quantization further imply $\ev{L_1}{\zeta}=(\ev{L_{-1}}{\zeta})^*$. This then allows us to solve the relation \eqref{relation2323} and obtain the one-point functions of the conformal generators
	\begin{align}
&		\ev{L_{-1}}{\zeta}  =(\ev{L_{1}}{\zeta}\,)^* =  - 2 i h \dfrac{\zeta^*}{1 - |\zeta|^2}~.\label{one2d}
	\end{align}
Finally, the two point function can be computed as follows
\begin{equation}
\ev{L_1 L_{-1}}{\zeta} = \ev{[L_1, L_{-1}]}{\zeta}+ \ev{L_{-1} L_1 }{\zeta}  =2 \ev{L_0}{\zeta} + \ev{L_{-1} L_1 }{\zeta}~,
\end{equation}
where in the last equality, we have used the algebra \eqref{algebraL}.
The expectation value $\ev{L_0}{\zeta}$ can be related to those calculated in Eq.~\eqref{one2d} by using the conjugation relation \eqref{Lm1L0Comrel}. The conjugations \eqref{L1Lm1Comrel} and \eqref{Lm1L1Comrel} can then be used to relate the expectation value $\ev{L_{-1} L_1 }{\zeta}$ to the unknown $\ev{L_1 L_{-1}}{\zeta}$. Note that $\ev{L_1 L_{-1}}{\zeta}$ should be real due to the Hermiticity conditions. Finally we solve the entire relation, which leads to
\begin{equation}
		\ev{L_1 L_{-1}}{\zeta}  = 2h \, \dfrac{1 + 2 h |\zeta|^2}{(1 - |\zeta|^2)^2}~\label{two2d}.
\end{equation}

%% file: sections/AppendixSpin.tex
\section{G. Comments about Spinning States}
%E

In this part of the supplementary material we make use of holomorphic factorization in $2d$ to present extensions of the results \eqref{eq:2d1norm}-\eqref{eq:2dFSmetric} for the case with non-zero spin $s=h-\bar h$. We translate these results to the higher dimensional language in terms of the circuit parameter $\alpha^\mu$ in an attempt to reveal the structure of the result for the case with spin. However, let us emphasize that the results of this section are only valid in two dimensions and that this case is special in that spinning representations are one dimensional (\ie they consist of a single component state $|h,\bar h \rangle\equiv |\Delta,s\rangle$, with the scaling dimension $\Delta=h+\bar h$). We leave a detailed analysis of spinning states in general dimensions for future work.

Let us begin with the generalized version of Eqs.~\eqref{eq:2d1norm}-\eqref{eq:2dFSmetric} when $h \neq \bar h$
\begin{equation}\label{TheMetricqueue}
	\mathcal{F}_1
	 = \left| h \left(\dfrac{ \dot \zeta \zeta^*-\dot \zeta^* \zeta}{1 - |\zeta|^2}~ + i  \dot \gamma_R\right)
+\bar h \left(\dfrac{ \dot {\bar\zeta} \bar\zeta^*-\dot {\bar\zeta}^* \bar\zeta}{1 - |\bar\zeta|^2}~ + i \dot {\bar \gamma}_R \right)\right|~,
\qquad
	\dd s^2_{FS} = 2 h \left[\dfrac{\dd \zeta \dd \zeta^*}{\left(1 - |\zeta|^2\right)^2}  \right] +2 \bar h \left[ \dfrac{\dd \bar\zeta \dd \bar \zeta^*}{\left(1 - |\bar \zeta|^2\right)^2} \right]\, .
\end{equation}
Using the relation between the conformal generators
\begin{equation}
\begin{split}
L_{-1} = \frac{1}{2} (P_x-iP_\tau)~, \qquad L_{0} = \frac{1}{2}(D-i L_{\tau x})~, \qquad L_1 = \frac{1}{2} (K_x+iK_\tau)~,\\
\bar L_{-1} = \frac{1}{2} (P_x+iP_\tau)~, \qquad \bar L_{0} = \frac{1}{2}(D+i L_{\tau x})~, \qquad \bar L_1 = \frac{1}{2} (K_x-iK_\tau)~,\\
\end{split}
\end{equation}
allows us to identify the relation between the circuit parameters in two-dimensions \eqref{eq:2dDiffeom} and those in higher dimensions \eqref{eq:unitaryDef}. We obtain
\begin{equation}
\label{eq:2dsplitCopies}
\begin{split}
\alpha_x =\frac{\zeta+\bar\zeta}{2}~, \qquad \alpha_{\tau} = \frac{\zeta-\bar\zeta}{2i}~, \qquad
\gamma_D= \frac{\gamma+\bar \gamma}{2}~,\qquad \lambda_{\tau x}= \frac{\gamma-\bar \gamma}{2i}~,\qquad
\beta_x =\frac{\zeta_1+\bar\zeta_1}{2}~, \qquad \beta_{\tau} = -\frac{\zeta_1-\bar\zeta_1}{2i}~,
\end{split}
\end{equation}
which yields for the various costs
\begin{equation}
\begin{split}
\mathcal{F}_1
	 =&\, \biggr|\Delta  \left(\frac{\dot{\alpha}\cdot\alpha^*-\dot{\alpha}^*\cdot\alpha  +\alpha^2 (\dot{\alpha}^* \cdot \alpha^*)-\alpha^*{}^2 (\dot{\alpha} \cdot \alpha)}{A(\alpha, \alpha^*)}+i \dot \gamma_D^R\right)
\\
&+is ~ \frac{\dot\alpha \cdot M \cdot \alpha^*-\alpha\cdot M \cdot \dot\alpha^* +\alpha^2\, (\alpha^* \cdot M \cdot \dot\alpha^*)- \alpha^*{}^2\, (\dot \alpha \cdot M \cdot \alpha)}{A(\alpha, \alpha^*)}- i s \, \dot\lambda_{\tau x}^I \biggr|
\end{split}
\end{equation}
and
\begin{equation}
\begin{split}
\dd s^2_{FS} =& \, 2 \Delta \left[\dfrac{\dot \alpha \cdot \dot \alpha^{*} - 2|\dot \alpha \cdot \alpha|^2}{A(\alpha, \alpha^*)} + 2\dfrac{\left|\dot \alpha \cdot \alpha^* - \alpha^{*2} \, \alpha \cdot \dot \alpha \right|^2}{A(\alpha, \alpha^*)^2}\right]
\\
&+2 i s ~ \frac{ 2 (1-|\alpha|^2) (\dot\alpha\cdot \dot\alpha^*) (\alpha \cdot M \cdot\alpha^*)+\dot\alpha\cdot M \cdot \dot\alpha^* \left[2(1-|\alpha|^2)^2 -A(\alpha, \alpha^*)\right]}{A(\alpha, \alpha^*)^2}
\end{split}
\end{equation}
where $A(\alpha, \alpha^*)$ was defined in Eq.~\eqref{domain}, $M \equiv{\begin{pmatrix}0&1\\-1&0\end{pmatrix}}$, $\alpha = (\alpha_\tau, \alpha_x)$  and we have used superscripts to denote the real part of $\gamma_D$ and the imaginary part of $\lambda_{\tau x}$ (which remain unfixed by the unitarity constraint). Similar to the result in section \ref{sec:gdc}, the curvature of this metric is $R = -16 \dfrac{\Delta}{\Delta^2 - s^2} =-\frac{4}{h}-\frac{4}{\bar h} =  R_h + R_{\bar h} $ where $R_h,R_{\bar h}$ are the curvatures of the holomorphic and anti-holomorphic parts, as expected from the holomorphic factorization.
Of course, setting $s=0$ recovers Eqs.~\eqref{eq:higherd1norm}-\eqref{eq:FSGen}. 

%% file: sections/appendixFundam.tex
\section{H. Metric and Geometric Action in the Fundamental Representation of the Conformal Group}
\label{sec:fundRep}
%F

Here we will evaluate various quantities related to coadjoint orbits in the fundamental representation of the conformal group (see \cite{SupMat}D for our conventions of the fundamental representation of the conformal algebra  $\mathfrak{so}(d,2)$). 
In this case, the Lie algebra and the dual space are isomorphic since the algebra admits a non-degenerate bilinear form $(X, Y) \equiv \dfrac{1}{2} \Tr\left[X\cdot Y\right]$. 
Therefore each algebra element can be identified with a dual algebra element according to $\langle  \lambda , \cdot\rangle \equiv ( \lambda, \cdot)$. 
It is then straightforward to build $R(U)$, the matrix representation associated to the unitary $U$ in Eq.~\eqref{eq:unitaryDef}. The field theory unitarity conditions is imposed by requiring that we have
\begin{equation}
	\label{eq:fundUnitarity}
	R(U^\dagger) = g^{-1} R(U)^\dagger g= R(U)^{-1}~.
\end{equation}
This condition fixes the parameters $\{\beta^\mu, \gamma_D^I, \lambda^R_{\mu\nu}\}$ in the definition of $U$ in Eq.~\eqref{eq:unitaryDef} as a function of the remaining parameters $\{\alpha, \gamma_D^R, \lambda_{\mu\nu}^I\}$ where the superscripts $R$ and $I$ indicate the real and imaginary parts,  respectively. In particular, one of those constraints $\gamma_D^I= -\frac{1}{2} \log \left(1 - 2 \, \alpha \cdot \alpha^* + \alpha^2 \alpha^{*2}\right)$ should be familiar from the discussion in section \ref{sec:gdc}. We were able to solve the constraints explicitly for $d=1$ and $d=2$ and order by order in a perturbative expansion in $|\alpha|$ in $d>2$, however the explicit expressions are cumbersome and not particularly illuminating, and so we do not include them here. 
One can then compute the MC form associated to the trajectory implemented by the unitary $U$ in terms of the coordinates $\{\alpha, \gamma_D^R, \lambda_{\mu\nu}^I\}$ by evaluating
\begin{equation}
	R(\Theta) \equiv R(U)^{-1} \dd R(U)~.
\end{equation}

To make the connection to our previous results we consider a representative in the dual space $\lambda(\mathcal{O}) \equiv (i \Delta  R(D), \mathcal{O})  = \frac{1}{2}\Tr( \Delta M_{-1,0} \cdot \mathcal{O})$. This element of the dual algebra is of course identified via the bilinear form with the algebra element $\lambda = i \Delta R(D)$ (here we slightly abuse the notation since we give the same name to the dual algebra element and the corresponding algebra element, but since the bilinear form is non-degenerate indeed the two can be identified). The stabilizer algebra is $\mathfrak h_\lambda  = \mathfrak{so}(2)\times \mathfrak{so}(d)$ and it naturally leads to an orbit which can be identified with the coset space in Eq.~\eqref{eq:coset}. In the $d = 2$ spinning case, where $\bar h \neq h$ (see \cite{SupMat}G), the relevant representative is identified with the algebra element $\lambda =i \Delta R(D) + s R(L_{\tau x})$ where $\Delta=h+\bar h$ and $s = h-\bar h$.

In order to evaluate the metric along the orbit associated with the representative \eqref{eq:representative}, we compute the symplectic form
\begin{equation}
	\omega = \dfrac{1}{2} \Tr \left[\lambda \cdot \dd R(\Theta) \right]
\end{equation}
and then obtain the components
\begin{equation}
    \omega  = \omega_{\bar \mu \nu} \dd \alpha^{*\bar \mu} \wedge \dd \alpha^\nu \, ,
\end{equation}
where we have introduced barred indices to formally distinguish between $\alpha$ and $\alpha^*$.
The wedge product is canonically identified with the tensor product through $x \wedge y \equiv \dfrac{1}{2} (x \otimes y - y \otimes x)$ hence
\begin{equation}\label{sym124}
	\begin{aligned}
		\omega & = \dfrac{1}{2} \omega_{\bar \mu\nu} \left( \dd \alpha^{* \bar \mu} \otimes \dd \alpha^\nu -
\dd \alpha^\nu \otimes \dd \alpha^{* \bar \mu}\right)~.
	\end{aligned}
\end{equation}
The metric is then obtained by contracting with the complex structure $J$ given by
\begin{equation}
	J^\mu{}_\rho \dd \alpha^\rho = - i \dd \alpha^\mu~, \quad J^{\bar \nu}{}_{\bar \rho} \dd \alpha^{* \bar \rho} = i \dd \alpha^{* \bar \nu} \, ,
\end{equation}
such that
\begin{equation}
	\begin{aligned}
		\dd s^{2}_{G/H_\lambda} & = \dfrac{1}{2} \omega_{\bar \mu\rho} \dd \alpha^{* \bar \mu} \otimes  J^\rho{}_\nu \dd \alpha^\nu - \dfrac{1}{2} \omega_{\bar \rho\nu} \dd \alpha^\nu \otimes J^{\bar \rho}{}_{\bar \mu} \dd \alpha^{* \bar \mu}
	\end{aligned}
\end{equation}
leading to
\begin{equation}
	\label{eq:appMetricCoset}
\dd s^{2}_{G/H_\lambda} = -i \omega_{\bar \mu \nu} \dd \alpha^{*\bar \mu} \dd \alpha^\nu \, .
\end{equation}
\noindent
Finally, the pre-symplectic form \eqref{Alambda} is given by
\begin{equation}\label{prese1234}
	\mathcal{A_\lambda} = \dfrac{1}{2} \Tr \left[\lambda  \cdot R(\Theta) \right]  \, .
\end{equation}
Our results for the metric on the coset space coincide with the Fubini-Study metric in Eq.~\eqref{eq:FSGen} and those for the pre-symplectic potential becomes the $\mathcal{F}_1$ cost function in Eq.~\eqref{eq:higherd1norm} up to an absolute value, \ie
\begin{equation}
	\label{eq:resultLink}
 	\begin{aligned}
 		\mathcal F_1 \, \dd \sigma & = \left| \mathcal A_\lambda \right|~, \qquad \dd s_{FS}^2 & = \dd s^2_{G/H_\lambda} ~.
 	\end{aligned}
\end{equation}
We have verified this analytically in $ d = 2$  and  in a perturbative expansion  in $d > 2$.

\vspace{10pt}

{\bf Explicit Example -- SO(2,1):} Let us demonstrate how all this works for the simplest example of the $\mathfrak{so}(2,1)$ algebra.
This algebra is locally isomorphic to $\mathfrak{sl}(2,\mathbb{R})$, which represents (for example) the holomorphic copy in two dimensions. The relevant generators are
\begin{equation}
P_{1}\simeq L_{-1} = \begin{bmatrix}
		0 & 0 & 1\\
		0 & 0 & - i\\
		1 & -i & 0
	\end{bmatrix}
\quad
D\simeq L_0 = \begin{bmatrix}
		0 & i & 0\\
		-i & 0 & 0\\
		0 & 0 & 0
	\end{bmatrix}
\quad
K_{1} \simeq L_1 = \begin{bmatrix}
		0 & 0 & -1\\
		0 & 0 & -i\\
		-1 & -i & 0
	\end{bmatrix}  \, .
\end{equation}
The unitary is given by
\begin{equation}
	R(U) \equiv e^{i \zeta L_{-1}} e^{i (\gamma_R + i \gamma_I) L_0} e^{i \zeta_1 L_1} \, .
\end{equation}
Requiring the unitarity condition \eqref{eq:fundUnitarity} imposes
\begin{equation}
	\gamma_I = - \log(1 - |\zeta|^2)~, \quad \zeta_1 = e^{i \gamma_R} \zeta^* \, .
\end{equation}
The MC form is
\begin{equation}
\Theta=L_{-1}\frac{\left(i e^{-i \gamma_R} \dd\zeta\right) }{1-|\zeta|^2} +L_0 \left(\frac{\zeta^* \dd\zeta -\zeta  \dd\zeta^*}{1-|\zeta|^2}+i \, \dd\gamma_R\right)
+L_1\frac{ \left(i e^{i \gamma_R} \dd\zeta^*\right)}{1-|\zeta|^2}.
\end{equation}
The representative is $\lambda = i h L_0$ whose stabilizer is spanned by the generator $L_0$. The exterior derivative of the MC form is given by
\begin{equation}
	\dd \Theta = \dfrac{i}{(1- |\zeta|^2)^2} \left[ \zeta^* e^{i\gamma_R} L_1 - \zeta e^{-i \gamma_R} L_{-1} + 2 i L_0\right] \dd \zeta \wedge \dd \zeta^* + \dfrac{\dd \gamma_R}{1- |\zeta|^2} \wedge \left[ e^{-i \gamma_R} \dd \zeta L_{-1} -e^{i \gamma_R} \dd \zeta^* L_1\right].
\end{equation}
The metric on the coset space \eqref{eq:appMetricCoset} and the pre-symplectic form \eqref{prese1234} read
\begin{equation}
	\dd s^2_{G/H_\lambda} =  2h \dfrac{\dd \zeta \dd \zeta^*}{\left(1 - |\zeta|^2\right)^2}~, \qquad\qquad \mathcal{A_\lambda} = i h \left( \dfrac{\zeta^* \dd \zeta - \zeta \dd \zeta^*}{1 - |\zeta|^2}+i \dd \gamma_R\right)   \, ,
\end{equation}
which matches our results in Eqs.~\eqref{eq:2d1norm}-\eqref{eq:2dFSmetric} (up to an absolute value for $\mathcal{F}_1$) when focusing on a single $\mathfrak{sl}(2,\mathbb{R})$ copy.

%% file: sections/cartankilling.tex
\section{I. Root Space Decomposition}\label{app:FSmetric}
%G

The assumptions used in section \ref{sec:csg} to equate the Fubini-Study metric with the metric compatible with the symplectic form on a coadjoint orbit follow   naturally from the structure of a root space decomposition (for a summary of how this works for other coherent state symmetry groups, which use a slightly different root space decomposition, see \cite{Zhang:1990fy}). For the conformal algebra, the relevant decomposition is known as the minimal Bruhat decomposition~\cite{Dobrev}, see~\cite{Penedones:2015aga,Yamazaki:2016vqi} for applications to conformal field theory in the context of parabolic Verma modules. In this part of the supplementary material we will both motivate the assumptions of the proof and explain how to construct the representations of interest using a more mathematical language.

Consider a unitarily represented semisimple group $G$ with Lie algebra $\mathfrak{g}$, with $\mathcal{D}$ a highest-weight representation on the Hilbert space. For our purposes we have in mind the Euclidean conformal group, but for now we keep the group arbitrary. The group theoretic generalization of a coherent state is often defined in terms of a base state $\ket{\psi_R}$ left invariant up to a phase by a subgroup $H\subset G$, or equivalently one that is an eigenstate of the corresponding subalgebra $\mathfrak{h}$,
\be x \ket{\psi_R} = \chi \ket{\psi_R}~, \indent \forall x \in \mathcal{D}(\mathfrak{h})~. \label{eq:eigeneqn}\ee
In the case of vector coherent states~\cite{Rowe1985,Rowe1988,Rowe:2012yi,Bartlett2002} (relevant to circuits constructed from a spinning primary in the conformal algebra in $d>2$), one often considers instead a collection of base states that transform into each other under the action of a subgroup, much like primaries with spin transform among each other under the action of the $L_{\mu\nu}$'s. We will explain how the subalgebra $\mathfrak{h}$ relates to a portion of a root space decomposition, and also indicate how this works for the vector coherent states.

Any semisimple algebra admits some Cartan decomposition $\mathfrak{g} = \mathfrak{s} + \mathfrak{t}$ where $[\mathfrak{s},\mathfrak{s}]\subset \mathfrak{s}$, $[\mathfrak{s},\mathfrak{t}]\subset \mathfrak{t}$ and $[\mathfrak{t},\mathfrak{t}]\subset \mathfrak{s}$. Let $\mathfrak{a} \subset \mathfrak{t}$ be a maximal abelian subalgebra for $\mathfrak{t}$, and $\mathfrak{m}\subset \mathfrak{s}$ be the centralizer of $\mathfrak{a}$ in $\mathfrak{s}$, in other words the elements $X\in \mathfrak{s}$ such that $[X,h] = 0$ for all $h\in \mathfrak{a}$. The adjoint action with respect to $\mathfrak{a}$ can be diagonalized, with the eigenspaces known as the restricted root spaces $\mathfrak{g}_\alpha$:
\be \mathfrak{g}_\alpha =  \{ X \in \mathfrak{g} : [h,X] = \alpha(h) X~~ \forall h\in \mathfrak{a}\}~.\label{eq:rootspaces}\ee
The linear functionals $\alpha(h)$ are the roots. A root is called positive with respect to a given basis of the dual space if its coefficients in the expansion over this basis are positive. We denote the set of all roots as $\Phi$ and the set of positive and negative roots as $\Phi^\pm$. Unlike for the standard semisimple Lie algebra root space decomposition, these roots are defined with respect to an abelian algebra $\mathfrak{a}$ that is not maximal.

The minimal Bruhat decomposition is the resulting root space decomposition:
\be \mathfrak{g} = \mathfrak{n}^- \oplus \mathfrak{m} \oplus \mathfrak{a} \oplus \mathfrak{n}^+~, \indent \mathfrak{n}^\pm = \bigoplus_{\alpha \in \Phi^\pm} \mathfrak{g}_\alpha~.\label{eq:Bruhat}\ee
Here $\mathfrak{g}_0\equiv\mathfrak{m}\oplus \mathfrak{a}$ is the centralizer of $\mathfrak{a}$, since $\mathfrak{m}$ is the centralizer of $\mathfrak{a}$ in $\mathfrak{s}$ and $\mathfrak{a}$ is maximal in $\mathfrak{t}$. $\mathfrak{n}^\pm$ are the positive and negative root spaces.

Pick a basis $E_{\alpha,p} \in \mathfrak{g}_\alpha$. The label $p$ here accounts for any root degeneracy, which is possible given that the abelian algebra $\mathfrak{a}$ used in the root space decomposition~\eqref{eq:Bruhat} is not the full maximal abelian algebra (the Cartan subalgebra) for $\mathfrak{g}$. By Eq.~\eqref{eq:rootspaces} these satisfy
\be [h_i, E_{\alpha, p}] = \alpha_i E_{\alpha,p}~,\indent \forall h_i \in \mathfrak{a}~.\label{eq:rooteigencondition}\ee
For an inner product with $h_i^\dagger = h_i$, applying the dagger to Eq.~\eqref{eq:rooteigencondition} gives $E_{\alpha,p}^\dagger = E_{-\alpha,p}$.

Using the Jacobi relation applied to Eq.~\eqref{eq:rooteigencondition}, we can prove inclusions for the commutation relations for the root spaces,
\begin{align}
[\mathfrak{g}_\alpha, \mathfrak{g}_{\beta}] &\subset \mathfrak{g}_{\alpha+\beta}~, \indent \alpha+\beta \in \Phi~ \mbox{ or } \alpha+\beta = 0~, \label{eq:rootcommutations}
\end{align}
and for the part of the centralizer not in the abelian algebra $\mathfrak{a}$,
\begin{align}
[\mathfrak{m}, \mathfrak{g}_\alpha] &\subset \mathfrak{g}_\alpha~, \indent [\mathfrak{m},\mathfrak{m}] \subset \mathfrak{m}~.\label{eq:rootcommutations2}
\end{align}
For an ordinary root space decomposition, the commutator of a root vector with its Hermitian conjugate would take values in the Cartan subalgebra. Notice that here, it instead takes values in the centralizer $\mathfrak{g}_0$ of the abelian algebra $\mathfrak{a}$ defining the root decomposition.

Consider a highest-weight representation for this root decomposition. This consists of states $\ket{\lambda}$ labeled by their eigenvalues $\lambda$ under the $h_i$,
\be h_i \ket{\lambda} = \lambda_i \ket{\lambda}~,\label{eq:highestweight}\ee
where a highest-weight state $\ket{\lambda_0}$ is annihilated by all the raising operators,
\be E_{\alpha,p} \ket{\lambda_0} = 0~,\quad \forall \alpha, p~.\label{eq:roothighestweight}\ee
By Eq.~\eqref{eq:rooteigencondition}, $E_{\alpha,p}$ raises the eigenvalue under $h_i$ by $\alpha_i$, thus it can be interpreted as a ladder operator. Likewise, $E_{\alpha,p}^\dagger =  E_{-\alpha,p}$ is a ladder operator that lowers the eigenvalue. We build the representation by applying the lowering operators successively starting from the highest weight state. Note that here, in order to match with the standard CFT literature, we use opposite conventions to those used in the definition of coherent states for the Heisenberg group in quantum mechanics where one typically starts with a lowest weight state.

Now consider an element $x\in \mathfrak{m}$ (in the centralizer but not in $\mathfrak{a}$). Then $[\mathfrak m, \mathfrak g_\alpha] \subset \mathfrak g_\alpha$ by Eq.~\eqref{eq:rootcommutations2}, and acting the commutator on $\ket{\lambda_0}$ and applying the highest weight condition \eqref{eq:roothighestweight} gives $\mathfrak g_\alpha x \ket{\lambda_0} = 0$. So $x \ket{\lambda_0}$ must be a highest-weight state. If there is a \emph{single} state satisfying Eq.~\eqref{eq:roothighestweight}, then $x\ket{\lambda_0}$ is proportional to $\ket{\lambda_0}$. Thus the eigenvalue condition~\eqref{eq:highestweight} extends from $\mathfrak{a}$ to $\mathfrak{g}_0$,
\be x\ket{\lambda_0} = \chi \ket{\lambda_0}~, \indent \forall x \in \mathfrak{g}_0~.\label{eq:extension}\ee
Recalling Eq.~\eqref{eq:eigeneqn}, this means that the highest weight state of a root space representation using the Bruhat decomposition is a natural candidate for our base state in the coherent state construction, with an invariance group $H\subset G$ whose algebra $\mathfrak{h}$ is just the stabilizer $\mathfrak{g}_0$ of $\mathfrak{a}$:
\be \ket{\psi_R} = \ket{\lambda_0}, \indent \mathfrak{h} = \mathfrak{g}_0~. \ee
The orthogonal complements to $\mathfrak{h}$ are $\mathfrak{n}^\pm$. These consist of raising and lowering operators that are related by Hermitian conjugation and build the representation starting from $\ket{\lambda_0}$.

In the spinning case, a highest-weight representation can still be built from a preferred highest weight state, with a highest weight condition \eqref{eq:roothighestweight} that includes ladder operators in $\mathfrak{m}$.
Thus, $\ket{\lambda_0}$ satisfying only Eq.~\eqref{eq:roothighestweight} is not unique and $x\ket{\lambda_0}$ defines a subspace of states. This subset of states will participate in the generalization of the eigenvalue condition \eqref{eq:eigeneqn} to vector coherent states. Imposing Eq.~\eqref{eq:extension} applied to the preferred highest weight state results in an invariance subalgebra that is smaller than the centralizer.

We now return to consider how this structure ties to the proof in section \ref{sec:csg}. We saw that we could identify $\mathfrak{g}_0 \supseteq \mathfrak{h}$. In other words, the centralizer for $\mathfrak{a}$ is either equal to the stabilizing subalgebra
in the spinless case, or contains it in the spinning case. Thus the commutation relations~\eqref{eq:rooteigencondition},~\eqref{eq:rootcommutations} and~\eqref{eq:rootcommutations2} are simply the assumptions that $[\mathfrak{n}_\pm, \mathfrak{n}_\pm] \subset \mathfrak{n}_\pm$ and $[\mathfrak{h},\mathfrak{n}_\pm] \subset \mathfrak{n}_\pm$. The Hermiticity condition $E_{\alpha,p}^\dagger = E_{-\alpha,p}$ is also natural for the root space decomposition. These were the starting points for the proof in section \ref{sec:csg}.

{\bf Conformal algebra}:  Now we will be more explicit about how these abstract ingredients apply to the specific case of the conformal algebra as considered in section~\ref{sec:gdc}. Recall that we are considering the real Euclidean conformal algebra, a semisimple real algebra that can be expressed in terms of its complexification as $\mathfrak{so}(d+1,1) = \left\{X\in \mathfrak{so}(d+2,\mathbb{C})~ |~ X \mbox{ real, } X^T \eta + \eta X = 0\right\}$, where $\eta= \mbox{diag}(-1,1,...,1)$ is the flat metric on $\mathbb{R}^{d+1,1}$. Note that the real matrices obeying these conditions differ from our choice of complex generators in Eq.~\eqref{RgenFund}, however the algebras are isomorphic. The starting point for the Bruhat decomposition is a Cartan decomposition of $\mathfrak{so}(d+1,1)$, which is not unique and can be specified by an involution. A natural choice is
\be \theta(X) = \eta X \eta~.\ee
The Cartan splitting $\theta(X) = X$ for $X\in \mathfrak{s}$, $\theta(X) = -X$ for $x\in \mathfrak{t}$ implies that $\mathfrak{s} = \left\{L_{\mu\nu},P_\mu+K_\mu\right\}$ and $\mathfrak{t} = \left\{D, P_\mu-K_\mu\right\}$. The maximal abelian subalgebra for $\mathfrak{t}$ is $\mathfrak a_0 = \left\{ -D \right\}$, with centralizer $\mathfrak{m}_0 = \left\{L_{\mu\nu}\right\}$ in $\mathfrak{s}$. Note that we have chosen $\mathfrak a_0$ to be generated by $-D$ instead of $D$ to match with the usual conventions for CFT representations.

The positive and negative root spaces with respect to $\mathfrak{a}_0$ are $\mathfrak{n}^- =\left\{P_\mu\right\}$, $\mathfrak{n}^+ =\left\{K_\mu\right\}$, which are related through the involution as $\theta(\mathfrak{n}^\pm) = \mathfrak{n}^\mp$. The positive and negative roots are all degenerate with $\alpha = \pm 1$. The minimal Bruhat decomposition~\eqref{eq:Bruhat} for the Euclidean conformal group is
\begin{equation}
	\label{eq:basisConformalGen}
	\mathfrak{g} = \{P_\mu\} \oplus \{L_{\mu\nu}\} \oplus \{D\} \oplus  \{K_\mu\} ~.
\end{equation}

The usual highest-weight representation for the conformal algebra consists of a conformal primary state $\ket{\Delta}$ annihilated by $K_\mu$, with the remaining descendant states built by acting with $P_\mu$. Both the primary and its descendants are eigenstates of $-D$, with $P_\mu$  acting as a lowering operator since it decreases the $-D$-eigenvalue. But this is precisely a highest-weight representation for the Bruhat decomposition described above. $-D$ generates the abelian algebra $\mathfrak{a}$. The condition~\eqref{eq:extension} that the stabilizer algebra for the highest weight state is the centralizer of $\mathfrak{a}$ is just the statement that the primary state is also an eigenstate under $L_{\mu\nu}$:
\begin{equation}
	D \ket \Delta = \Delta \ket \Delta~, \quad L_{\mu\nu} \ket \Delta = 0~.
\end{equation}

We end by summarizing how our assumptions in section \ref{sec:csg} apply to the case of the conformal algebra. The condition $[\mathfrak{h}, \mathfrak{n}_\pm]\subset \mathfrak{n}_\pm$ is satisfied by the algebra~\eqref{EuclideanConformalAlgebra} and $[\mathfrak{n}_\pm, \mathfrak{n}_\pm]\subset \mathfrak{n}_\pm$ is trivially satisfied since for the conformal algebra, the $P_\mu$'s and $K_\mu$'s commute so $\mathfrak{n}_\pm$ is abelian. With respect to the field theory dagger~\eqref{eq:fundUnitarity}, the generators of $\mathfrak{n}_\pm$ obey the Hermiticity conditions $P_\mu^\dagger = K_\mu$. These are just the conditions $E_{\alpha, p}^\dagger = E_{-\alpha, p}$ taken above for the root space decomposition and used in the proof in section \ref{sec:csg}. 

%% file: sections/holography_appendix.tex
\section{J. Holographic Interpretation}

In this part of the supplemental material, we present further details on the relation of our coherent states in $\mathrm{CFT}_d$ and the trajectories of massive particles in $\mathrm{AdS}_{d+1}$. In particular we explain the  geometric interpretation of the Fubini-Study metric and the complexity of states in terms of the bulk geometry and present an interesting example of geodesics with fixed radius.

\subsection{Background}
Here, we review our conventions for the embedding space and trajectories of massive particles in AdS$_{d+1}$ following \cite{Dorn:2005jt} (up to a modification of the signature). 
 Let us start with the action of a massive particle of mass $m$ in an embedding space description of $\mathrm{AdS}_{d+1}$ of radius $R$ consisting of a hyperbola of radius $R$ in flat space $\Real^{d,2}$ with metric $g_{AB} = \mathrm{diag}(-, -, +, \dots, +)$. The coordinates of the flat space will be denoted $X^A$ with $A \in \{0, 0', 1, \dots, d\}$ and we will use Greek indices to denote the space directions  $\mu \in \{ 1, \dots, d\}$. The action for the massive particle reads
\begin{equation}
	\label{eq:holographyAction}
	S =  - \int \dd \tau \left[ -\dfrac{\dot X(\tau)^2}{2 e(\tau)} + \dfrac{e(\tau) m^2}{2} - \dfrac{\mu(\tau)}{2} \left( X(\tau)^2 + R^2\right)\right]~.
\end{equation}
%- \int \dd t L(\tau) =
 In this action, $e(\tau)$ is the einbein and $\mu(\tau)$ is a Lagrange multiplier enforcing the hyperbola condition $X(\tau)^2 =- R^2$. 
 The equations of motion derived from this action fix
\begin{equation}
 	e(\tau)^2 =- \dot X(\tau)^2/m^2~, \quad \mu(\tau) = -\dfrac{m}{R^2} \sqrt{-\dot X(\tau)^2} ~.
\end{equation}
Solutions will be timelike geodesics in $\mathrm{AdS}_{d+1}$ with
\begin{equation}
	X_{0}(t) = r(t) \cos(t/R)~, \quad X_{0'}(t) = r(t) \sin(t/R)~,
\end{equation}
where $t$ is the $\mathrm{AdS}_{d+1}$ time coordinate and the restriction to the hyperbola fixes
\begin{equation}
	-R^2 = X^2(t) = -r(t)^2 + X_\mu(t) X^\mu(t)~.
\end{equation}
The action \eqref{eq:holographyAction} is SO$(d,2)$ invariant and the associated conserved charges read
\begin{equation}\label{JAB}
	J_{AB} = P_{B}(t) X_{A}(t) - P_{A}(t) X_{B}(t) = P_{B}(0) X_{A}(0) - P_{A}(0) X_{B}(0)~,
\end{equation}
where $P_A(\tau) \equiv \dot X_A(\tau) / e(\tau) = m \dot X^A(\tau) / \sqrt{- \dot X^2(\tau)}$ is the canonical momentum associated with $X_A(\tau)$. The charges corresponding to the compact subgroup $SO(2) \times SO(d)$ are the energy $E \equiv J_{0,0'}$ and angular momentum $J_{\mu\nu}$, with $J^2 = J_{\mu\nu} J_{\mu\nu}/2$  the squared angular momentum of the trajectory. We selected our geodesics to be future oriented, \ie $X_0\dot X_{0'}-X_{0'}\dot X_{0}>0$ and this simply tells us that the energy is positive $E > 0$. The remaining  conserved charges are $J_{0,\mu}$ and $J_{0', \mu}$ which can be reorganized into a pair of complex coordinates describing (without redundancy)  the phase space of timelike geodesics
\begin{equation}
	z_\mu \equiv J_{0', \mu} - i J_{0,\mu}~, \quad z^*_\mu \equiv J_{0', \mu} + i J_{0,\mu}~.
\end{equation}
Using \eqref{JAB}, we can write $d$ equations for the coordinates $X_\mu$
%dynamical integrals
\begin{equation}
 	E X_\mu = J_{0,\mu} X_{0'} - J_{0',\mu} X_{0}
\end{equation}
which yield
\begin{equation}
	X_\mu(t) = - \dfrac{r(t)}{2 E} \left(z_\mu^* e^{i t / R} + z_\mu e^{- i t / R}\right) \text{, ~~~with } r(t)= \dfrac{2 E R}{\sqrt{4 E^2 - z^{*2} e^{2 i t / R} - z^{2} e^{-2 i t / R} - 2 z \cdot z^*}}~.
\end{equation}
$(z, z^*)$ are complex coordinates on the phase space of the particle which can be used to evaluate its symplectic form
\begin{equation}
	\omega_{\mathrm{bulk}} = \dd P_A \wedge \dd X^A ~.
\end{equation}
The energy is minimal when the particle is at rest and equal to $E_0 = mR$. Quantizing the classical system canonically by promoting the observables $\{E, J_{\mu\nu},z_\mu,z^*_\mu\}$ to operators leads to a unitary reducible representation of $\mathrm{SO}(d,2)$. We introduce the following change of coordinates
\begin{equation}
	z_\mu = 2 E_0 \dfrac{\alpha_\mu^* - \alpha^{*2} \alpha_\mu}{A(\alpha, \alpha^*)}~, \quad z^*_\mu = 2 E_0 \dfrac{\alpha_\mu - \alpha^{2} \alpha^*_\mu}{A(\alpha, \alpha^*)}~.
\end{equation}
As shown by \cite{Dorn:2005jt}, an irreducible representation can be obtained upon imposing that $\alpha$ is in the domain \eqref{domain} together with the condition $\alpha\cdot \alpha^*<1$. Furthermore, we can find that
\begin{equation}
	\omega_{\mathrm{bulk}} = \omega_{\mathrm{CFT}_d}~,
\end{equation}
provided that we identify $\Delta = E_0$, where the symplectic form of the CFT can be easily read from the metric \eqref{eq:FSGen} via the relations
\eqref{sym124} and \eqref{eq:appMetricCoset}. Starting with the action \eqref{eq:holographyAction}, the classical geodesic trajectory in the bulk is based on a saddle point approximation with large mass. The mass of the dual quantum particle is related to the field theory operator dimension as $m^2 R^2 = \Delta(\Delta-d)$ which can be inverted as 
\begin{equation}
	\Delta = m R \left(\sqrt{1 + \dfrac{d^2}{4 m^2 R^2}} + \dfrac{d}{2 m R}\right)~.
\end{equation}
In the limit of large mass we simply have $m R = \Delta$. 
Therefore, we have an exact duality between $\mathrm{CFT}_d$ states and timelike geodesics in $\mathrm{AdS}_{d+1}$. It is interesting to notice that due to canonical quantization, the conserved charges associated with the isometries in the bulk map to the expectation values of algebra elements in the $\mathrm{CFT}_d$. For example, we find that $z_\mu = \ev{i P_\mu}{\alpha}$ and the energy and angular momentum of the particle are given by
\begin{equation}
	\label{eq:EJalphasInv}
	 E = \Delta \dfrac{1 - \alpha^2 \alpha^{*2}}{A(\alpha,\alpha^*)} = \ev{D}{\alpha}~, \quad J = 2 \Delta \dfrac{\sqrt{(\alpha\cdot\alpha^*)^2 - \alpha^2 \alpha^{*2}}}{A(\alpha, \alpha^*)} = \sqrt{\dfrac{1}{2}\ev{i L_{\mu\nu}}{\alpha} \ev{i L_{\mu\nu}}{\alpha}} ~,
\end{equation}
which can be inverted as
\begin{equation}
   \label{eq:EJalphas}
   \alpha \cdot \alpha^* = \dfrac{E^2 - J^2 - \Delta^2}{(E + \Delta)^2 - J^2}~, \quad \alpha^2 \alpha^{*2} = \dfrac{(E - \Delta)^2 - J^2}{(E + \Delta)^2 - J^2}~.
\end{equation}
In this coordinate system, $r(t)$ can be simplified into
\begin{equation}
	r(t) = \dfrac{R\, E}{\Delta} \dfrac{\sqrt{A(\alpha,\alpha^*)} }{|B(t,\alpha)|}, \quad B(t,\alpha) \equiv e^{ i t/R} \alpha^2-e^{- i t/R} ~.
\end{equation}

\subsection{Complexity in Holography}

We have established an explicit connection between states in the $\mathrm{CFT}_d$ and timelike geodesics in $\mathrm{AdS}_{d+1}$ (or particle states). We can now re-express the complexity \eqref{eq:TargetComplexityRel}-\eqref{eq:generalComplexityGeodesics} of a target state $\ket{\alpha_T}$ in the field theory in terms of the energy $E_T$ and angular momentum $J_T$ of the associated target particle state in $\mathrm{AdS}_{d+1}$ as
\begin{equation}
	\label{eq:complexityHolography}
   \mathcal C[E_T, J_T] = \sqrt{\Delta} \sqrt{\left(\atanh\sqrt{\dfrac{E_T + J_T - \Delta}{E_T + J_T + \Delta}}\right)^2 + \left(\atanh\sqrt{\dfrac{E_T - J_T - \Delta}{E_T - J_T + \Delta}}\right)^2}~,
\end{equation}
where we used that
\begin{equation}
	\Omega^S_T = \sqrt{\dfrac{E_T + J_T - \Delta}{E_T + J_T + \Delta}}~, \quad \Omega^A_T = \sqrt{\dfrac{E_T - J_T - \Delta}{E_T - J_T + \Delta}}~.
\end{equation}

\subsection{Geometric Interpretation of the Fubini-Study Metric}

We have argued that states $\ket{\alpha}$ are related to timelike geodesics in $\mathrm{AdS}_{d+1}$ according to \eqref{eq:geodesicCorrespondence}-\eqref{eq:geodesicCorrespondence2}. The question that remains to be answered is what is the gravitational dual of the Fubini-Study line element between two infinitesimally close states.

Consider two nearby geodesics $X^\mu(t)$ and $X^\mu(t) + \delta X^\mu(t)$ corresponding to two nearby states $\ket \alpha$ and $\ket {\alpha+\delta \alpha}$, respectively. We can expand $\delta X^\mu(t)$ as follows  
\begin{equation}
	\delta X^\mu(t) = \dd \alpha \cdot \partial_{\alpha} X^\mu(t) + \dd \alpha^* \cdot \partial_{\alpha^*} X^\mu(t) + \dd t \dot X^\mu(t) ~.
\end{equation}
The relation \eqref{eq:geodesicCorrespondence} implies that both $X^\mu$ and $X^\mu + \delta X^\mu$ are geodesics in $\mathrm{AdS}_{d+1}$ corresponding to the parameters $\alpha,\alpha^*$ and $\alpha+d\alpha,\alpha^*+d\alpha^*$. 
We further impose a requirement that $\delta X \cdot \dot X = 0$ which can be used to solve for $\dd t$ as follows
\begin{equation}
	\dd t = -\dfrac{1}{\dot X^2(t)} \left(\dot X_\mu(t) \,
	\dd \alpha^\nu  \frac{\partial X^\mu(t)}{\partial \alpha^\nu} + 
	\dot X_\mu(t) \, \dd \alpha^{*\nu}  \frac{\partial X^\mu(t)}{\partial \alpha^{*\nu}}  \right)~.
\end{equation}
This requirement can be understood as looking at the hyperplane orthogonal to $\dot X^\mu(t)$ and finding the intersection of this hyperplane with the second geodesic at every time $t$.  We will measure the perpendicular distance between the two geodesics along this hyperplane. 
The perpendicular distance $\delta X^2_{\text{perp}}(t)$ can be related to $\dd s^2_{FS}$ in~\eqref{eq:FSGen} as follows. We begin by separating $\delta X^2_{\text{perp}}(t)$ into three contributions
\begin{equation}
	\label{eq:separationBartekProposal}
	\delta X^2_{\text{perp}}(t) = \dfrac{R^2}{\Delta} \dd s^2_{\mathrm{FS}} +  g_{\alpha\alpha}(t)\, \dd \alpha^2 + g_{\alpha\alpha}^*(t) \,\dd \alpha^{*2}~, \qquad \text{with} \qquad g_{\alpha\alpha}(t) = \dfrac{R^2}{A(\alpha, \alpha^*)} 
	\dfrac{e^{2 i t/R} -  \alpha^{*2}}{1 - e^{2 i t/R}\,\alpha^{2}} ~.
\end{equation}
Note that only the last two terms in $\delta X^2_{\text{perp}}(t)$ are time dependent. Next, we look for times at which $\delta X^2_{\text{perp}}(t)$ is extremal, \ie $\partial_t \delta X^2_{\text{perp}}(t) = 0$. This leads to two solutions $t_+$ and $t_-$ corresponding to a maximum (with $\partial_t^2 \delta X^2_{\text{perp}}(t_+)<0$) and a minimum (with $\partial_t^2 \delta X^2_{\text{perp}}(t_-)>0$) perpendicular distance, respectively, which are given by
\begin{equation}
	e^{2 i t_\pm/R} = \dfrac{\alpha^{*2} \,\sqrt{\dd\alpha\cdot \dd\alpha} \pm \sqrt{\dd\alpha^* \cdot\dd\alpha^*}}{\sqrt{\dd\alpha\cdot \dd\alpha} \pm \alpha^2\, \sqrt{\dd\alpha^*\cdot \dd\alpha^*}} ,
\end{equation}
where
\begin{equation}
	\delta X^2_{\text{perp},\mathrm{max/min}} \equiv \delta X^2(t_\pm) = \dfrac{R^2}{\Delta} \dd s^2_{\mathrm{FS}} \pm \dfrac{2 R^2 \sqrt{\dd\alpha^2\,\, \dd \alpha^{*2}}}{A(\alpha,\alpha^*)}~.
\end{equation}
Using the above relations, we finally recover the equivalence \eqref{eq:geometricArgumentProp}
\begin{equation}
	\dd s^2_{\mathrm{FS}} = \dfrac{\Delta}{2 R^2} \left(\delta X^2_{\text{perp},\mathrm{min}} + \delta X^2_{\text{perp},\mathrm{max}} \right)~.
\end{equation}

\subsection{Example: Timelike Geodesics with Fixed Radius}

We will now work out an explicit case for which much of the previous calculations simplify. The goal of this presentation is to give an intuition of how to use our holographic results. We will therefore focus on states for which we have $\alpha(\sigma)^2 = 0$. The condition \eqref{domain} then implies $\alpha(\sigma) \cdot \alpha^*(\sigma) < 1/2$ and for these values, we have, from Eq.~\eqref{eq:geodesicCorrespondence}-\eqref{eq:geodesicCorrespondence2}
\begin{equation}
	\label{eq:energyRadiusCirc}
	E(\sigma) = \dfrac{\Delta}{1 - 2 \alpha(\sigma) \cdot \alpha^*(\sigma)}~, \quad r(t; \sigma) = \dfrac{R}{\sqrt{1 - 2 \alpha(\sigma) \cdot \alpha^*(\sigma)}} \equiv r_0(\sigma) ~,
\end{equation}
where $\rho(t) = R ~ \mathrm{arccosh}(r(t)/R)$ is the global AdS radial coordinate. The fact that $r(t;\sigma)$ does not depend on $t$ means that the associated timelike geodesics have fixed radius which we will define as circular geodesics. This choice of $\alpha(\sigma)$ leads to a path within the subset of fixed radius geodesics for which \eqref{eq:generalComplexityGeodesics22} simplifies into
\begin{equation}
	\begin{gathered}
		\Omega_T^S = \sqrt{2 ~ \alpha_T \cdot \alpha_T^*}~, \quad \Omega_T^A = 0~,\\
		\mathcal C[\alpha_T] = \sqrt{\Delta} ~ \atanh\left(\sqrt{2 ~ \alpha_T \cdot \alpha_T^*}\right)~.
	\end{gathered}
\end{equation}
In the holographic formulation, $J_T = E_T - \Delta$ for such states so the complexity can be expressed using the energy of the particle
\begin{equation}
	\mathcal C_{\mathrm{circular}}[E_T] = \sqrt{\Delta} ~ \atanh \left(\sqrt{1 - \dfrac{\Delta}{E_T}}\right) 
 = \sqrt{\Delta} ~ \atanh \left(\sqrt{1 - \dfrac{R^2}{r_T^2}}\right) 	
	~,
\end{equation} 
where $r_T \equiv r_0(\sigma = 1)$ is the radius of the outermost geodesic in the circuit.  Considering the AdS metric in a Fefferman-Graham expansion $\dd s^2 = \dfrac{1}{z^2}\left(\dd z^2 + \dd x_\mu \dd x^\mu\right)$, we have near the boundary ($\rho \to \infty$) the relation $z \simeq R/\sinh(\rho/R) \simeq \dfrac{R^2}{r}$.  
Close to the boundary $z \sim \delta$ and this means $r_T \sim R^2/\delta$ such that
\begin{equation}
	\mathcal C_{\mathrm{circular}}[\delta] \sim \sqrt{\Delta} \log \left[2 R / \delta\right] ~.
\end{equation}
This result captures the divergent behavior of complexity as we go to states (represented by circular geodesics) which are very far from our reference state (very close to the boundary in AdS$_{d+1}$). Note that this is different  from asking what is the vacuum divergence of complexity evaluated using the holographic proposals as in, \eg \cite{Carmi:2016wjl,Reynolds:2016rvl}.

%% file: ComplexityCFTD.bbl
\begin{thebibliography}{95}%
\makeatletter
\providecommand \@ifxundefined [1]{%
 \@ifx{#1\undefined}
}%
\providecommand \@ifnum [1]{%
 \ifnum #1\expandafter \@firstoftwo
 \else \expandafter \@secondoftwo
 \fi
}%
\providecommand \@ifx [1]{%
 \ifx #1\expandafter \@firstoftwo
 \else \expandafter \@secondoftwo
 \fi
}%
\providecommand \natexlab [1]{#1}%
\providecommand \enquote  [1]{``#1''}%
\providecommand \bibnamefont  [1]{#1}%
\providecommand \bibfnamefont [1]{#1}%
\providecommand \citenamefont [1]{#1}%
\providecommand \href@noop [0]{\@secondoftwo}%
\providecommand \href [0]{\begingroup \@sanitize@url \@href}%
\providecommand \@href[1]{\@@startlink{#1}\@@href}%
\providecommand \@@href[1]{\endgroup#1\@@endlink}%
\providecommand \@sanitize@url [0]{\catcode `\\12\catcode `\$12\catcode
  `\&12\catcode `\#12\catcode `\^12\catcode `\_12\catcode `\%12\relax}%
\providecommand \@@startlink[1]{}%
\providecommand \@@endlink[0]{}%
\providecommand \url  [0]{\begingroup\@sanitize@url \@url }%
\providecommand \@url [1]{\endgroup\@href {#1}{\urlprefix }}%
\providecommand \urlprefix  [0]{URL }%
\providecommand \Eprint [0]{\href }%
\providecommand \doibase [0]{https://doi.org/}%
\providecommand \selectlanguage [0]{\@gobble}%
\providecommand \bibinfo  [0]{\@secondoftwo}%
\providecommand \bibfield  [0]{\@secondoftwo}%
\providecommand \translation [1]{[#1]}%
\providecommand \BibitemOpen [0]{}%
\providecommand \bibitemStop [0]{}%
\providecommand \bibitemNoStop [0]{.\EOS\space}%
\providecommand \EOS [0]{\spacefactor3000\relax}%
\providecommand \BibitemShut  [1]{\csname bibitem#1\endcsname}%
\let\auto@bib@innerbib\@empty
%</preamble>
\bibitem [{\citenamefont {Watrous}(2008)}]{watrous2008quantum}%
  \BibitemOpen
  \bibfield  {author} {\bibinfo {author} {\bibfnamefont {J.}~\bibnamefont
  {Watrous}},\ }\href@noop {} {\bibinfo {title} {Quantum computational
  complexity}} (\bibinfo {year} {2008}),\ \Eprint
  {https://arxiv.org/abs/0804.3401} {arXiv:0804.3401 [quant-ph]} \BibitemShut
  {NoStop}%
\bibitem [{\citenamefont {Aaronson}(2016)}]{Aaronson:2016vto}%
  \BibitemOpen
  \bibfield  {author} {\bibinfo {author} {\bibfnamefont {S.}~\bibnamefont
  {Aaronson}},\ }\bibfield  {title} {\bibinfo {title} {{The Complexity of
  Quantum States and Transformations: From Quantum Money to Black Holes}}\
  }(\bibinfo {year} {2016})\ \Eprint {https://arxiv.org/abs/1607.05256}
  {arXiv:1607.05256 [quant-ph]} \BibitemShut {NoStop}%
\bibitem [{\citenamefont {Nielsen}\ and\ \citenamefont
  {Chuang}(2000)}]{nielsen2002quantum}%
  \BibitemOpen
  \bibfield  {author} {\bibinfo {author} {\bibfnamefont {M.}~\bibnamefont
  {Nielsen}}\ and\ \bibinfo {author} {\bibfnamefont {I.}~\bibnamefont
  {Chuang}},\ }\href@noop {} {\emph {\bibinfo {title} {Quantum Computation and
  Quantum Information}}}\ (\bibinfo  {publisher} {Cambridge University Press},\
  \bibinfo {year} {2000})\BibitemShut {NoStop}%
\bibitem [{\citenamefont {Susskind}(2018)}]{Susskind:2018pmk}%
  \BibitemOpen
  \bibfield  {author} {\bibinfo {author} {\bibfnamefont {L.}~\bibnamefont
  {Susskind}},\ }\bibfield  {title} {\bibinfo {title} {{Three Lectures on
  Complexity and Black Holes}}\ }(\bibinfo  {publisher} {Springer},\ \bibinfo
  {year} {2018})\ \Eprint {https://arxiv.org/abs/1810.11563} {arXiv:1810.11563
  [hep-th]} \BibitemShut {NoStop}%
\bibitem [{\citenamefont {Susskind}(2016{\natexlab{a}})}]{Susskind:2014moa}%
  \BibitemOpen
  \bibfield  {author} {\bibinfo {author} {\bibfnamefont {L.}~\bibnamefont
  {Susskind}},\ }\bibfield  {title} {\bibinfo {title} {{Entanglement is not
  enough}},\ }\href {https://doi.org/10.1002/prop.201500095} {\bibfield
  {journal} {\bibinfo  {journal} {Fortsch. Phys.}\ }\textbf {\bibinfo {volume}
  {64}},\ \bibinfo {pages} {49} (\bibinfo {year} {2016}{\natexlab{a}})},\
  \Eprint {https://arxiv.org/abs/1411.0690} {arXiv:1411.0690 [hep-th]}
  \BibitemShut {NoStop}%
\bibitem [{\citenamefont {Aharony}\ \emph {et~al.}(2000)\citenamefont
  {Aharony}, \citenamefont {Gubser}, \citenamefont {Maldacena}, \citenamefont
  {Ooguri},\ and\ \citenamefont {Oz}}]{Aharony:1999ti}%
  \BibitemOpen
  \bibfield  {author} {\bibinfo {author} {\bibfnamefont {O.}~\bibnamefont
  {Aharony}}, \bibinfo {author} {\bibfnamefont {S.~S.}\ \bibnamefont {Gubser}},
  \bibinfo {author} {\bibfnamefont {J.~M.}\ \bibnamefont {Maldacena}}, \bibinfo
  {author} {\bibfnamefont {H.}~\bibnamefont {Ooguri}},\ and\ \bibinfo {author}
  {\bibfnamefont {Y.}~\bibnamefont {Oz}},\ }\bibfield  {title} {\bibinfo
  {title} {{Large N field theories, string theory and gravity}},\ }\href
  {https://doi.org/10.1016/S0370-1573(99)00083-6} {\bibfield  {journal}
  {\bibinfo  {journal} {Phys. Rept.}\ }\textbf {\bibinfo {volume} {323}},\
  \bibinfo {pages} {183} (\bibinfo {year} {2000})},\ \Eprint
  {https://arxiv.org/abs/hep-th/9905111} {arXiv:hep-th/9905111} \BibitemShut
  {NoStop}%
\bibitem [{\citenamefont {Susskind}(2016{\natexlab{b}})}]{Susskind:2014rva}%
  \BibitemOpen
  \bibfield  {author} {\bibinfo {author} {\bibfnamefont {L.}~\bibnamefont
  {Susskind}},\ }\bibfield  {title} {\bibinfo {title} {{Computational
  Complexity and Black Hole Horizons}},\ }\href
  {https://doi.org/10.1002/prop.201500092} {\bibfield  {journal} {\bibinfo
  {journal} {Fortsch. Phys.}\ }\textbf {\bibinfo {volume} {64}},\ \bibinfo
  {pages} {24} (\bibinfo {year} {2016}{\natexlab{b}})},\ \Eprint
  {https://arxiv.org/abs/1402.5674} {arXiv:1402.5674 [hep-th]} \BibitemShut
  {NoStop}%
\bibitem [{\citenamefont {Brown}\ \emph
  {et~al.}(2016{\natexlab{a}})\citenamefont {Brown}, \citenamefont {Roberts},
  \citenamefont {Susskind}, \citenamefont {Swingle},\ and\ \citenamefont
  {Zhao}}]{Brown:2015lvg}%
  \BibitemOpen
  \bibfield  {author} {\bibinfo {author} {\bibfnamefont {A.~R.}\ \bibnamefont
  {Brown}}, \bibinfo {author} {\bibfnamefont {D.~A.}\ \bibnamefont {Roberts}},
  \bibinfo {author} {\bibfnamefont {L.}~\bibnamefont {Susskind}}, \bibinfo
  {author} {\bibfnamefont {B.}~\bibnamefont {Swingle}},\ and\ \bibinfo {author}
  {\bibfnamefont {Y.}~\bibnamefont {Zhao}},\ }\bibfield  {title} {\bibinfo
  {title} {{Complexity, action, and black holes}},\ }\href
  {https://doi.org/10.1103/PhysRevD.93.086006} {\bibfield  {journal} {\bibinfo
  {journal} {Phys. Rev. D}\ }\textbf {\bibinfo {volume} {93}},\ \bibinfo
  {pages} {086006} (\bibinfo {year} {2016}{\natexlab{a}})},\ \Eprint
  {https://arxiv.org/abs/1512.04993} {arXiv:1512.04993 [hep-th]} \BibitemShut
  {NoStop}%
\bibitem [{\citenamefont {Brown}\ \emph
  {et~al.}(2016{\natexlab{b}})\citenamefont {Brown}, \citenamefont {Roberts},
  \citenamefont {Susskind}, \citenamefont {Swingle},\ and\ \citenamefont
  {Zhao}}]{brown2016holographic}%
  \BibitemOpen
  \bibfield  {author} {\bibinfo {author} {\bibfnamefont {A.~R.}\ \bibnamefont
  {Brown}}, \bibinfo {author} {\bibfnamefont {D.~A.}\ \bibnamefont {Roberts}},
  \bibinfo {author} {\bibfnamefont {L.}~\bibnamefont {Susskind}}, \bibinfo
  {author} {\bibfnamefont {B.}~\bibnamefont {Swingle}},\ and\ \bibinfo {author}
  {\bibfnamefont {Y.}~\bibnamefont {Zhao}},\ }\bibfield  {title} {\bibinfo
  {title} {Holographic complexity equals bulk action?},\ }\href
  {https://doi.org/10.1103/PhysRevLett.116.191301} {\bibfield  {journal}
  {\bibinfo  {journal} {Phys. Rev. Lett.}\ }\textbf {\bibinfo {volume} {116}},\
  \bibinfo {pages} {191301} (\bibinfo {year} {2016}{\natexlab{b}})}\BibitemShut
  {NoStop}%
\bibitem [{\citenamefont {Carmi}\ \emph
  {et~al.}(2017{\natexlab{a}})\citenamefont {Carmi}, \citenamefont {Chapman},
  \citenamefont {Marrochio}, \citenamefont {Myers},\ and\ \citenamefont
  {Sugishita}}]{Carmi:2017jqz}%
  \BibitemOpen
  \bibfield  {author} {\bibinfo {author} {\bibfnamefont {D.}~\bibnamefont
  {Carmi}}, \bibinfo {author} {\bibfnamefont {S.}~\bibnamefont {Chapman}},
  \bibinfo {author} {\bibfnamefont {H.}~\bibnamefont {Marrochio}}, \bibinfo
  {author} {\bibfnamefont {R.~C.}\ \bibnamefont {Myers}},\ and\ \bibinfo
  {author} {\bibfnamefont {S.}~\bibnamefont {Sugishita}},\ }\bibfield  {title}
  {\bibinfo {title} {{On the Time Dependence of Holographic Complexity}},\
  }\href {https://doi.org/10.1007/JHEP11(2017)188} {\bibfield  {journal}
  {\bibinfo  {journal} {JHEP}\ }\textbf {\bibinfo {volume} {11}},\ \bibinfo
  {pages} {188}},\ \Eprint {https://arxiv.org/abs/1709.10184} {arXiv:1709.10184
  [hep-th]} \BibitemShut {NoStop}%
\bibitem [{\citenamefont {Stanford}\ and\ \citenamefont
  {Susskind}(2014)}]{Stanford:2014jda}%
  \BibitemOpen
  \bibfield  {author} {\bibinfo {author} {\bibfnamefont {D.}~\bibnamefont
  {Stanford}}\ and\ \bibinfo {author} {\bibfnamefont {L.}~\bibnamefont
  {Susskind}},\ }\bibfield  {title} {\bibinfo {title} {{Complexity and Shock
  Wave Geometries}},\ }\href {https://doi.org/10.1103/PhysRevD.90.126007}
  {\bibfield  {journal} {\bibinfo  {journal} {Phys. Rev. D}\ }\textbf {\bibinfo
  {volume} {90}},\ \bibinfo {pages} {126007} (\bibinfo {year} {2014})},\
  \Eprint {https://arxiv.org/abs/1406.2678} {arXiv:1406.2678 [hep-th]}
  \BibitemShut {NoStop}%
\bibitem [{\citenamefont {Chapman}\ \emph
  {et~al.}(2018{\natexlab{a}})\citenamefont {Chapman}, \citenamefont
  {Marrochio},\ and\ \citenamefont {Myers}}]{Chapman:2018dem}%
  \BibitemOpen
  \bibfield  {author} {\bibinfo {author} {\bibfnamefont {S.}~\bibnamefont
  {Chapman}}, \bibinfo {author} {\bibfnamefont {H.}~\bibnamefont {Marrochio}},\
  and\ \bibinfo {author} {\bibfnamefont {R.~C.}\ \bibnamefont {Myers}},\
  }\bibfield  {title} {\bibinfo {title} {{Holographic complexity in Vaidya
  spacetimes. Part I}},\ }\href {https://doi.org/10.1007/JHEP06(2018)046}
  {\bibfield  {journal} {\bibinfo  {journal} {JHEP}\ }\textbf {\bibinfo
  {volume} {06}},\ \bibinfo {pages} {046}},\ \Eprint
  {https://arxiv.org/abs/1804.07410} {arXiv:1804.07410 [hep-th]} \BibitemShut
  {NoStop}%
\bibitem [{\citenamefont {Chapman}\ \emph
  {et~al.}(2018{\natexlab{b}})\citenamefont {Chapman}, \citenamefont
  {Marrochio},\ and\ \citenamefont {Myers}}]{Chapman:2018lsv}%
  \BibitemOpen
  \bibfield  {author} {\bibinfo {author} {\bibfnamefont {S.}~\bibnamefont
  {Chapman}}, \bibinfo {author} {\bibfnamefont {H.}~\bibnamefont {Marrochio}},\
  and\ \bibinfo {author} {\bibfnamefont {R.~C.}\ \bibnamefont {Myers}},\
  }\bibfield  {title} {\bibinfo {title} {{Holographic complexity in Vaidya
  spacetimes. Part II}},\ }\href {https://doi.org/10.1007/JHEP06(2018)114}
  {\bibfield  {journal} {\bibinfo  {journal} {JHEP}\ }\textbf {\bibinfo
  {volume} {06}},\ \bibinfo {pages} {114}},\ \Eprint
  {https://arxiv.org/abs/1805.07262} {arXiv:1805.07262 [hep-th]} \BibitemShut
  {NoStop}%
\bibitem [{\citenamefont {Jefferson}\ and\ \citenamefont
  {Myers}(2017)}]{Jefferson:2017sdb}%
  \BibitemOpen
  \bibfield  {author} {\bibinfo {author} {\bibfnamefont {R.}~\bibnamefont
  {Jefferson}}\ and\ \bibinfo {author} {\bibfnamefont {R.~C.}\ \bibnamefont
  {Myers}},\ }\bibfield  {title} {\bibinfo {title} {{Circuit complexity in
  quantum field theory}},\ }\href {https://doi.org/10.1007/JHEP10(2017)107}
  {\bibfield  {journal} {\bibinfo  {journal} {JHEP}\ }\textbf {\bibinfo
  {volume} {10}},\ \bibinfo {pages} {107}},\ \Eprint
  {https://arxiv.org/abs/1707.08570} {arXiv:1707.08570 [hep-th]} \BibitemShut
  {NoStop}%
\bibitem [{\citenamefont {Chapman}\ \emph
  {et~al.}(2018{\natexlab{c}})\citenamefont {Chapman}, \citenamefont {Heller},
  \citenamefont {Marrochio},\ and\ \citenamefont {Pastawski}}]{chapman2018}%
  \BibitemOpen
  \bibfield  {author} {\bibinfo {author} {\bibfnamefont {S.}~\bibnamefont
  {Chapman}}, \bibinfo {author} {\bibfnamefont {M.~P.}\ \bibnamefont {Heller}},
  \bibinfo {author} {\bibfnamefont {H.}~\bibnamefont {Marrochio}},\ and\
  \bibinfo {author} {\bibfnamefont {F.}~\bibnamefont {Pastawski}},\ }\bibfield
  {title} {\bibinfo {title} {{Toward a Definition of Complexity for Quantum
  Field Theory States}},\ }\href
  {https://doi.org/10.1103/PhysRevLett.120.121602} {\bibfield  {journal}
  {\bibinfo  {journal} {Phys. Rev. Lett.}\ }\textbf {\bibinfo {volume} {120}},\
  \bibinfo {pages} {121602} (\bibinfo {year} {2018}{\natexlab{c}})},\ \Eprint
  {https://arxiv.org/abs/1707.08582} {arXiv:1707.08582 [hep-th]} \BibitemShut
  {NoStop}%
\bibitem [{\citenamefont {Khan}\ \emph {et~al.}(2018)\citenamefont {Khan},
  \citenamefont {Krishnan},\ and\ \citenamefont {Sharma}}]{Khan:2018rzm}%
  \BibitemOpen
  \bibfield  {author} {\bibinfo {author} {\bibfnamefont {R.}~\bibnamefont
  {Khan}}, \bibinfo {author} {\bibfnamefont {C.}~\bibnamefont {Krishnan}},\
  and\ \bibinfo {author} {\bibfnamefont {S.}~\bibnamefont {Sharma}},\
  }\bibfield  {title} {\bibinfo {title} {{Circuit Complexity in Fermionic Field
  Theory}},\ }\href {https://doi.org/10.1103/PhysRevD.98.126001} {\bibfield
  {journal} {\bibinfo  {journal} {Phys. Rev. D}\ }\textbf {\bibinfo {volume}
  {98}},\ \bibinfo {pages} {126001} (\bibinfo {year} {2018})},\ \Eprint
  {https://arxiv.org/abs/1801.07620} {arXiv:1801.07620 [hep-th]} \BibitemShut
  {NoStop}%
\bibitem [{\citenamefont {Hackl}\ and\ \citenamefont
  {Myers}(2018)}]{Hackl:2018ptj}%
  \BibitemOpen
  \bibfield  {author} {\bibinfo {author} {\bibfnamefont {L.}~\bibnamefont
  {Hackl}}\ and\ \bibinfo {author} {\bibfnamefont {R.~C.}\ \bibnamefont
  {Myers}},\ }\bibfield  {title} {\bibinfo {title} {{Circuit complexity for
  free fermions}},\ }\href {https://doi.org/10.1007/JHEP07(2018)139} {\bibfield
   {journal} {\bibinfo  {journal} {JHEP}\ }\textbf {\bibinfo {volume} {07}},\
  \bibinfo {pages} {139}},\ \Eprint {https://arxiv.org/abs/1803.10638}
  {arXiv:1803.10638 [hep-th]} \BibitemShut {NoStop}%
\bibitem [{\citenamefont {Chapman}\ \emph
  {et~al.}(2019{\natexlab{a}})\citenamefont {Chapman}, \citenamefont {Eisert},
  \citenamefont {Hackl}, \citenamefont {Heller}, \citenamefont {Jefferson},
  \citenamefont {Marrochio},\ and\ \citenamefont {Myers}}]{chapman2019}%
  \BibitemOpen
  \bibfield  {author} {\bibinfo {author} {\bibfnamefont {S.}~\bibnamefont
  {Chapman}}, \bibinfo {author} {\bibfnamefont {J.}~\bibnamefont {Eisert}},
  \bibinfo {author} {\bibfnamefont {L.}~\bibnamefont {Hackl}}, \bibinfo
  {author} {\bibfnamefont {M.~P.}\ \bibnamefont {Heller}}, \bibinfo {author}
  {\bibfnamefont {R.}~\bibnamefont {Jefferson}}, \bibinfo {author}
  {\bibfnamefont {H.}~\bibnamefont {Marrochio}},\ and\ \bibinfo {author}
  {\bibfnamefont {R.~C.}\ \bibnamefont {Myers}},\ }\bibfield  {title} {\bibinfo
  {title} {{Complexity and entanglement for thermofield double states}},\
  }\href {https://doi.org/10.21468/SciPostPhys.6.3.034} {\bibfield  {journal}
  {\bibinfo  {journal} {SciPost Phys.}\ }\textbf {\bibinfo {volume} {6}},\
  \bibinfo {pages} {034} (\bibinfo {year} {2019}{\natexlab{a}})},\ \Eprint
  {https://arxiv.org/abs/1810.05151} {arXiv:1810.05151 [hep-th]} \BibitemShut
  {NoStop}%
\bibitem [{\citenamefont {Chapman}\ and\ \citenamefont
  {Chen}(2019)}]{Chapman:2019clq}%
  \BibitemOpen
  \bibfield  {author} {\bibinfo {author} {\bibfnamefont {S.}~\bibnamefont
  {Chapman}}\ and\ \bibinfo {author} {\bibfnamefont {H.~Z.}\ \bibnamefont
  {Chen}},\ }\bibfield  {title} {\bibinfo {title} {{Complexity for Charged
  Thermofield Double States}},\ }\href@noop {} {\  (\bibinfo {year} {2019})},\
  \Eprint {https://arxiv.org/abs/1910.07508} {arXiv:1910.07508 [hep-th]}
  \BibitemShut {NoStop}%
\bibitem [{\citenamefont {Bhattacharyya}\ \emph {et~al.}(2018)\citenamefont
  {Bhattacharyya}, \citenamefont {Shekar},\ and\ \citenamefont
  {Sinha}}]{Bhattacharyya:2018bbv}%
  \BibitemOpen
  \bibfield  {author} {\bibinfo {author} {\bibfnamefont {A.}~\bibnamefont
  {Bhattacharyya}}, \bibinfo {author} {\bibfnamefont {A.}~\bibnamefont
  {Shekar}},\ and\ \bibinfo {author} {\bibfnamefont {A.}~\bibnamefont
  {Sinha}},\ }\bibfield  {title} {\bibinfo {title} {{Circuit complexity in
  interacting QFTs and RG flows}},\ }\href
  {https://doi.org/10.1007/JHEP10(2018)140} {\bibfield  {journal} {\bibinfo
  {journal} {JHEP}\ }\textbf {\bibinfo {volume} {10}},\ \bibinfo {pages}
  {140}},\ \Eprint {https://arxiv.org/abs/1808.03105} {arXiv:1808.03105
  [hep-th]} \BibitemShut {NoStop}%
\bibitem [{\citenamefont {Caputa}\ and\ \citenamefont
  {Magan}(2019)}]{caputa2019}%
  \BibitemOpen
  \bibfield  {author} {\bibinfo {author} {\bibfnamefont {P.}~\bibnamefont
  {Caputa}}\ and\ \bibinfo {author} {\bibfnamefont {J.~M.}\ \bibnamefont
  {Magan}},\ }\bibfield  {title} {\bibinfo {title} {{Quantum Computation as
  Gravity}},\ }\href {https://doi.org/10.1103/PhysRevLett.122.231302}
  {\bibfield  {journal} {\bibinfo  {journal} {Phys. Rev. Lett.}\ }\textbf
  {\bibinfo {volume} {122}},\ \bibinfo {pages} {231302} (\bibinfo {year}
  {2019})},\ \Eprint {https://arxiv.org/abs/1807.04422} {arXiv:1807.04422
  [hep-th]} \BibitemShut {NoStop}%
\bibitem [{\citenamefont {Erdmenger}\ \emph {et~al.}(2020)\citenamefont
  {Erdmenger}, \citenamefont {Gerbershagen},\ and\ \citenamefont
  {Weigel}}]{Erdmenger:2020sup}%
  \BibitemOpen
  \bibfield  {author} {\bibinfo {author} {\bibfnamefont {J.}~\bibnamefont
  {Erdmenger}}, \bibinfo {author} {\bibfnamefont {M.}~\bibnamefont
  {Gerbershagen}},\ and\ \bibinfo {author} {\bibfnamefont {A.-L.}\ \bibnamefont
  {Weigel}},\ }\bibfield  {title} {\bibinfo {title} {{Complexity measures from
  geometric actions on Virasoro and Kac-Moody orbits}},\ }\href
  {https://doi.org/10.1007/JHEP11(2020)003} {\bibfield  {journal} {\bibinfo
  {journal} {JHEP}\ }\textbf {\bibinfo {volume} {11}},\ \bibinfo {pages}
  {003}},\ \Eprint {https://arxiv.org/abs/2004.03619} {arXiv:2004.03619
  [hep-th]} \BibitemShut {NoStop}%
\bibitem [{\citenamefont {Flory}\ and\ \citenamefont
  {Heller}(2020{\natexlab{a}})}]{flory2020}%
  \BibitemOpen
  \bibfield  {author} {\bibinfo {author} {\bibfnamefont {M.}~\bibnamefont
  {Flory}}\ and\ \bibinfo {author} {\bibfnamefont {M.~P.}\ \bibnamefont
  {Heller}},\ }\bibfield  {title} {\bibinfo {title} {{Geometry of Complexity in
  Conformal Field Theory}},\ }\href
  {https://doi.org/10.1103/PhysRevResearch.2.043438} {\bibfield  {journal}
  {\bibinfo  {journal} {Phys. Rev. Res.}\ }\textbf {\bibinfo {volume} {2}},\
  \bibinfo {pages} {043438} (\bibinfo {year} {2020}{\natexlab{a}})},\ \Eprint
  {https://arxiv.org/abs/2005.02415} {arXiv:2005.02415 [hep-th]} \BibitemShut
  {NoStop}%
\bibitem [{\citenamefont {Flory}\ and\ \citenamefont
  {Heller}(2020{\natexlab{b}})}]{flory2020a}%
  \BibitemOpen
  \bibfield  {author} {\bibinfo {author} {\bibfnamefont {M.}~\bibnamefont
  {Flory}}\ and\ \bibinfo {author} {\bibfnamefont {M.~P.}\ \bibnamefont
  {Heller}},\ }\bibfield  {title} {\bibinfo {title} {{Conformal field theory
  complexity from Euler-Arnold equations}},\ }\href
  {https://doi.org/10.1007/JHEP12(2020)091} {\bibfield  {journal} {\bibinfo
  {journal} {JHEP}\ }\textbf {\bibinfo {volume} {12}},\ \bibinfo {pages}
  {091}},\ \Eprint {https://arxiv.org/abs/2007.11555} {arXiv:2007.11555
  [hep-th]} \BibitemShut {NoStop}%
\bibitem [{\citenamefont {Bueno}\ \emph {et~al.}(2019)\citenamefont {Bueno},
  \citenamefont {Magan},\ and\ \citenamefont {Shahbazi}}]{Bueno:2019ajd}%
  \BibitemOpen
  \bibfield  {author} {\bibinfo {author} {\bibfnamefont {P.}~\bibnamefont
  {Bueno}}, \bibinfo {author} {\bibfnamefont {J.~M.}\ \bibnamefont {Magan}},\
  and\ \bibinfo {author} {\bibfnamefont {C.~S.}\ \bibnamefont {Shahbazi}},\
  }\bibfield  {title} {\bibinfo {title} {{Complexity measures in QFT and
  constrained geometric actions}},\ }\href@noop {} {\  (\bibinfo {year}
  {2019})},\ \Eprint {https://arxiv.org/abs/1908.03577} {arXiv:1908.03577
  [hep-th]} \BibitemShut {NoStop}%
\bibitem [{\citenamefont {Chapman}\ \emph {et~al.}(2017)\citenamefont
  {Chapman}, \citenamefont {Marrochio},\ and\ \citenamefont
  {Myers}}]{chapman2017}%
  \BibitemOpen
  \bibfield  {author} {\bibinfo {author} {\bibfnamefont {S.}~\bibnamefont
  {Chapman}}, \bibinfo {author} {\bibfnamefont {H.}~\bibnamefont {Marrochio}},\
  and\ \bibinfo {author} {\bibfnamefont {R.~C.}\ \bibnamefont {Myers}},\
  }\bibfield  {title} {\bibinfo {title} {{Complexity of Formation in
  Holography}},\ }\href {https://doi.org/10.1007/JHEP01(2017)062} {\bibfield
  {journal} {\bibinfo  {journal} {JHEP}\ }\textbf {\bibinfo {volume} {01}},\
  \bibinfo {pages} {062}},\ \Eprint {https://arxiv.org/abs/1610.08063}
  {arXiv:1610.08063 [hep-th]} \BibitemShut {NoStop}%
\bibitem [{\citenamefont {Chapman}\ \emph
  {et~al.}(2019{\natexlab{b}})\citenamefont {Chapman}, \citenamefont {Ge},\
  and\ \citenamefont {Policastro}}]{Chapman:2018bqj}%
  \BibitemOpen
  \bibfield  {author} {\bibinfo {author} {\bibfnamefont {S.}~\bibnamefont
  {Chapman}}, \bibinfo {author} {\bibfnamefont {D.}~\bibnamefont {Ge}},\ and\
  \bibinfo {author} {\bibfnamefont {G.}~\bibnamefont {Policastro}},\ }\bibfield
   {title} {\bibinfo {title} {{Holographic Complexity for Defects Distinguishes
  Action from Volume}},\ }\href {https://doi.org/10.1007/JHEP05(2019)049}
  {\bibfield  {journal} {\bibinfo  {journal} {JHEP}\ }\textbf {\bibinfo
  {volume} {05}},\ \bibinfo {pages} {049}},\ \Eprint
  {https://arxiv.org/abs/1811.12549} {arXiv:1811.12549 [hep-th]} \BibitemShut
  {NoStop}%
\bibitem [{\citenamefont {Sato}\ and\ \citenamefont
  {Watanabe}(2019)}]{Sato:2019kik}%
  \BibitemOpen
  \bibfield  {author} {\bibinfo {author} {\bibfnamefont {Y.}~\bibnamefont
  {Sato}}\ and\ \bibinfo {author} {\bibfnamefont {K.}~\bibnamefont
  {Watanabe}},\ }\bibfield  {title} {\bibinfo {title} {{Does Boundary
  Distinguish Complexities?}},\ }\href
  {https://doi.org/10.1007/JHEP11(2019)132} {\bibfield  {journal} {\bibinfo
  {journal} {JHEP}\ }\textbf {\bibinfo {volume} {11}},\ \bibinfo {pages}
  {132}},\ \Eprint {https://arxiv.org/abs/1908.11094} {arXiv:1908.11094
  [hep-th]} \BibitemShut {NoStop}%
\bibitem [{\citenamefont {Nielsen}\ \emph {et~al.}(2006)\citenamefont
  {Nielsen}, \citenamefont {Dowling}, \citenamefont {Gu},\ and\ \citenamefont
  {Doherty}}]{nielsen2006quantum}%
  \BibitemOpen
  \bibfield  {author} {\bibinfo {author} {\bibfnamefont {M.~A.}\ \bibnamefont
  {Nielsen}}, \bibinfo {author} {\bibfnamefont {M.~R.}\ \bibnamefont
  {Dowling}}, \bibinfo {author} {\bibfnamefont {M.}~\bibnamefont {Gu}},\ and\
  \bibinfo {author} {\bibfnamefont {A.~C.}\ \bibnamefont {Doherty}},\
  }\bibfield  {title} {\bibinfo {title} {Quantum computation as geometry},\
  }\href {https://doi.org/10.1126/science.1121541} {\bibfield  {journal}
  {\bibinfo  {journal} {Science}\ }\textbf {\bibinfo {volume} {311}},\ \bibinfo
  {pages} {1133} (\bibinfo {year} {2006})}\BibitemShut {NoStop}%
\bibitem [{\citenamefont {Nielsen}(2006)}]{nielsen2005}%
  \BibitemOpen
  \bibfield  {author} {\bibinfo {author} {\bibfnamefont {M.}~\bibnamefont
  {Nielsen}},\ }\bibfield  {title} {\bibinfo {title} {A geometric approach to
  quantum circuit lower bounds},\ }\href@noop {} {\bibfield  {journal}
  {\bibinfo  {journal} {Quantum Inf. Comput.}\ }\textbf {\bibinfo {volume}
  {6}},\ \bibinfo {pages} {213} (\bibinfo {year} {2006})},\ \Eprint
  {https://arxiv.org/abs/quant-ph/0502070} {arXiv:quant-ph/0502070 [quant-ph]}
  \BibitemShut {NoStop}%
\bibitem [{\citenamefont {Dowling}\ and\ \citenamefont
  {Nielsen}(2008)}]{dowling2008geometry}%
  \BibitemOpen
  \bibfield  {author} {\bibinfo {author} {\bibfnamefont {M.~R.}\ \bibnamefont
  {Dowling}}\ and\ \bibinfo {author} {\bibfnamefont {M.~A.}\ \bibnamefont
  {Nielsen}},\ }\bibfield  {title} {\bibinfo {title} {The geometry of quantum
  computation},\ }\href {https://doi.org/10.26421/QIC8.10-1} {\bibfield
  {journal} {\bibinfo  {journal} {Quantum Information \& Computation}\ }\textbf
  {\bibinfo {volume} {8}},\ \bibinfo {pages} {861} (\bibinfo {year} {2008})},\
  \Eprint {https://arxiv.org/abs/quant-ph/0701004} {arXiv:quant-ph/0701004
  [quant-ph]} \BibitemShut {NoStop}%
\bibitem [{\citenamefont {Anandan}\ and\ \citenamefont
  {Aharonov}(1990)}]{Anandan:1990fq}%
  \BibitemOpen
  \bibfield  {author} {\bibinfo {author} {\bibfnamefont {J.}~\bibnamefont
  {Anandan}}\ and\ \bibinfo {author} {\bibfnamefont {Y.}~\bibnamefont
  {Aharonov}},\ }\bibfield  {title} {\bibinfo {title} {{Geometry of Quantum
  Evolution}},\ }\href {https://doi.org/10.1103/PhysRevLett.65.1697} {\bibfield
   {journal} {\bibinfo  {journal} {Phys. Rev. Lett.}\ }\textbf {\bibinfo
  {volume} {65}},\ \bibinfo {pages} {1697} (\bibinfo {year}
  {1990})}\BibitemShut {NoStop}%
\bibitem [{\citenamefont {Lloyd}(2000)}]{lloyd2000ultimate}%
  \BibitemOpen
  \bibfield  {author} {\bibinfo {author} {\bibfnamefont {S.}~\bibnamefont
  {Lloyd}},\ }\bibfield  {title} {\bibinfo {title} {Ultimate physical limits to
  computation},\ }\href@noop {} {\bibfield  {journal} {\bibinfo  {journal}
  {Nature}\ }\textbf {\bibinfo {volume} {406}},\ \bibinfo {pages} {1047}
  (\bibinfo {year} {2000})}\BibitemShut {NoStop}%
\bibitem [{Sup()}]{SupMat}%
  \BibitemOpen
  \href@noop {} {}\bibinfo {note} {Supplementary material, composed of ten
  sections. In section A, we provide the explicit relation between the
  Euclidean and Lorentzian generators of the conformal group. In section B, we
  provide various additional details for the derivation in section
  \ref{sec:gdc}. In sections C-D, we discuss geodesics in the complexity
  metric. In section E, we discuss bounds on the complexity and its growth. In
  section F, we focus on the explicit $d = 2$ case and compare the results
  derived using our techniques to previous results in the literature. In
  section G, we comment on the spinning case in $d=2$. In section H, we present
  the example of the metric and geometric action on coadjoint orbits of the
  conformal group in the fundamental representation. In section I, we justify
  the decomposition used in section \ref{sec:csg}. In section J, we provide
  additional details about the relation to holography.}\BibitemShut {Stop}%
\bibitem [{\citenamefont {Gibbons}(2000)}]{gibbons2000}%
  \BibitemOpen
  \bibfield  {author} {\bibinfo {author} {\bibfnamefont {G.}~\bibnamefont
  {Gibbons}},\ }\bibfield  {title} {\bibinfo {title} {{Holography and the
  future tube}},\ }\href {https://doi.org/10.1088/0264-9381/17/5/316}
  {\bibfield  {journal} {\bibinfo  {journal} {Class. Quant. Grav.}\ }\textbf
  {\bibinfo {volume} {17}},\ \bibinfo {pages} {1071} (\bibinfo {year}
  {2000})},\ \Eprint {https://arxiv.org/abs/hep-th/9911027}
  {arXiv:hep-th/9911027} \BibitemShut {NoStop}%
\bibitem [{\citenamefont {Andrianopoli}\ \emph {et~al.}(2005)\citenamefont
  {Andrianopoli}, \citenamefont {Ferrara}, \citenamefont {Lledo},\ and\
  \citenamefont {Macia}}]{Andrianopoli:2005de}%
  \BibitemOpen
  \bibfield  {author} {\bibinfo {author} {\bibfnamefont {L.}~\bibnamefont
  {Andrianopoli}}, \bibinfo {author} {\bibfnamefont {S.}~\bibnamefont
  {Ferrara}}, \bibinfo {author} {\bibfnamefont {M.~A.}\ \bibnamefont {Lledo}},\
  and\ \bibinfo {author} {\bibfnamefont {O.}~\bibnamefont {Macia}},\ }\bibfield
   {title} {\bibinfo {title} {{Integration of massive states as contractions of
  non linear sigma-models}},\ }\href {https://doi.org/10.1063/1.1960719}
  {\bibfield  {journal} {\bibinfo  {journal} {J. Math. Phys.}\ }\textbf
  {\bibinfo {volume} {46}},\ \bibinfo {pages} {072307} (\bibinfo {year}
  {2005})},\ \Eprint {https://arxiv.org/abs/hep-th/0503196}
  {arXiv:hep-th/0503196} \BibitemShut {NoStop}%
\bibitem [{\citenamefont {Czech}\ \emph {et~al.}(2015)\citenamefont {Czech},
  \citenamefont {Lamprou}, \citenamefont {McCandlish},\ and\ \citenamefont
  {Sully}}]{Czech:2015qta}%
  \BibitemOpen
  \bibfield  {author} {\bibinfo {author} {\bibfnamefont {B.}~\bibnamefont
  {Czech}}, \bibinfo {author} {\bibfnamefont {L.}~\bibnamefont {Lamprou}},
  \bibinfo {author} {\bibfnamefont {S.}~\bibnamefont {McCandlish}},\ and\
  \bibinfo {author} {\bibfnamefont {J.}~\bibnamefont {Sully}},\ }\bibfield
  {title} {\bibinfo {title} {{Integral Geometry and Holography}},\ }\href
  {https://doi.org/10.1007/JHEP10(2015)175} {\bibfield  {journal} {\bibinfo
  {journal} {JHEP}\ }\textbf {\bibinfo {volume} {10}},\ \bibinfo {pages}
  {175}},\ \Eprint {https://arxiv.org/abs/1505.05515} {arXiv:1505.05515
  [hep-th]} \BibitemShut {NoStop}%
%%CITATION = ARXIV:1505.05515;%%
\bibitem [{\citenamefont {de~Boer}\ \emph
  {et~al.}(2016{\natexlab{a}})\citenamefont {de~Boer}, \citenamefont {Heller},
  \citenamefont {Myers},\ and\ \citenamefont {Neiman}}]{deBoer:2015kda}%
  \BibitemOpen
  \bibfield  {author} {\bibinfo {author} {\bibfnamefont {J.}~\bibnamefont
  {de~Boer}}, \bibinfo {author} {\bibfnamefont {M.~P.}\ \bibnamefont {Heller}},
  \bibinfo {author} {\bibfnamefont {R.~C.}\ \bibnamefont {Myers}},\ and\
  \bibinfo {author} {\bibfnamefont {Y.}~\bibnamefont {Neiman}},\ }\bibfield
  {title} {\bibinfo {title} {{Holographic de Sitter Geometry from Entanglement
  in Conformal Field Theory}},\ }\href
  {https://doi.org/10.1103/PhysRevLett.116.061602} {\bibfield  {journal}
  {\bibinfo  {journal} {Phys. Rev. Lett.}\ }\textbf {\bibinfo {volume} {116}},\
  \bibinfo {pages} {061602} (\bibinfo {year} {2016}{\natexlab{a}})},\ \Eprint
  {https://arxiv.org/abs/1509.00113} {arXiv:1509.00113 [hep-th]} \BibitemShut
  {NoStop}%
%%CITATION = ARXIV:1509.00113;%%
\bibitem [{\citenamefont {Czech}\ \emph {et~al.}(2016)\citenamefont {Czech},
  \citenamefont {Lamprou}, \citenamefont {McCandlish}, \citenamefont {Mosk},\
  and\ \citenamefont {Sully}}]{Czech:2016xec}%
  \BibitemOpen
  \bibfield  {author} {\bibinfo {author} {\bibfnamefont {B.}~\bibnamefont
  {Czech}}, \bibinfo {author} {\bibfnamefont {L.}~\bibnamefont {Lamprou}},
  \bibinfo {author} {\bibfnamefont {S.}~\bibnamefont {McCandlish}}, \bibinfo
  {author} {\bibfnamefont {B.}~\bibnamefont {Mosk}},\ and\ \bibinfo {author}
  {\bibfnamefont {J.}~\bibnamefont {Sully}},\ }\bibfield  {title} {\bibinfo
  {title} {{A Stereoscopic Look into the Bulk}},\ }\href
  {https://doi.org/10.1007/JHEP07(2016)129} {\bibfield  {journal} {\bibinfo
  {journal} {JHEP}\ }\textbf {\bibinfo {volume} {07}},\ \bibinfo {pages}
  {129}},\ \Eprint {https://arxiv.org/abs/1604.03110} {arXiv:1604.03110
  [hep-th]} \BibitemShut {NoStop}%
\bibitem [{\citenamefont {de~Boer}\ \emph
  {et~al.}(2016{\natexlab{b}})\citenamefont {de~Boer}, \citenamefont {Haehl},
  \citenamefont {Heller},\ and\ \citenamefont {Myers}}]{deBoer:2016pqk}%
  \BibitemOpen
  \bibfield  {author} {\bibinfo {author} {\bibfnamefont {J.}~\bibnamefont
  {de~Boer}}, \bibinfo {author} {\bibfnamefont {F.~M.}\ \bibnamefont {Haehl}},
  \bibinfo {author} {\bibfnamefont {M.~P.}\ \bibnamefont {Heller}},\ and\
  \bibinfo {author} {\bibfnamefont {R.~C.}\ \bibnamefont {Myers}},\ }\bibfield
  {title} {\bibinfo {title} {{Entanglement, holography and causal diamonds}},\
  }\href {https://doi.org/10.1007/JHEP08(2016)162} {\bibfield  {journal}
  {\bibinfo  {journal} {JHEP}\ }\textbf {\bibinfo {volume} {08}},\ \bibinfo
  {pages} {162}},\ \Eprint {https://arxiv.org/abs/1606.03307} {arXiv:1606.03307
  [hep-th]} \BibitemShut {NoStop}%
%%CITATION = ARXIV:1606.03307;%%
\bibitem [{\citenamefont {Penna}\ and\ \citenamefont
  {Zukowski}(2019)}]{penna2019}%
  \BibitemOpen
  \bibfield  {author} {\bibinfo {author} {\bibfnamefont {R.~F.}\ \bibnamefont
  {Penna}}\ and\ \bibinfo {author} {\bibfnamefont {C.}~\bibnamefont
  {Zukowski}},\ }\bibfield  {title} {\bibinfo {title} {{Kinematic space and the
  orbit method}},\ }\href {https://doi.org/10.1007/JHEP07(2019)045} {\bibfield
  {journal} {\bibinfo  {journal} {JHEP}\ }\textbf {\bibinfo {volume} {07}},\
  \bibinfo {pages} {045}},\ \Eprint {https://arxiv.org/abs/1812.02176}
  {arXiv:1812.02176 [hep-th]} \BibitemShut {NoStop}%
\bibitem [{\citenamefont {Gallier}\ and\ \citenamefont
  {Quaintance}(2019)}]{gallier2019differential}%
  \BibitemOpen
  \bibfield  {author} {\bibinfo {author} {\bibfnamefont {J.}~\bibnamefont
  {Gallier}}\ and\ \bibinfo {author} {\bibfnamefont {J.}~\bibnamefont
  {Quaintance}},\ }\href@noop {} {\emph {\bibinfo {title} {Differential
  Geometry and Lie Groups}}}\ (\bibinfo  {publisher} {Springer},\ \bibinfo
  {year} {2019})\BibitemShut {NoStop}%
\bibitem [{\citenamefont {Helgason}(1979)}]{helgason}%
  \BibitemOpen
  \bibfield  {author} {\bibinfo {author} {\bibfnamefont {S.}~\bibnamefont
  {Helgason}},\ }\href@noop {} {\emph {\bibinfo {title} {Differential Geometry,
  Lie Groups, and Symmetric Spaces}}}\ (\bibinfo  {publisher} {Elsevier
  Science},\ \bibinfo {year} {1979})\BibitemShut {NoStop}%
\bibitem [{\citenamefont {O'Neill}(1983)}]{Barrett1983}%
  \BibitemOpen
  \bibfield  {author} {\bibinfo {author} {\bibfnamefont {B.}~\bibnamefont
  {O'Neill}},\ }\href {https://books.google.com/books?id=CGk1eRSjFIIC} {\emph
  {\bibinfo {title} {Semi-Riemannian Geometry With Applications to
  Relativity}}}\ (\bibinfo  {publisher} {Elsevier Science},\ \bibinfo {year}
  {1983})\BibitemShut {NoStop}%
\bibitem [{\citenamefont {Oblak}(2017)}]{Oblak:2017ect}%
  \BibitemOpen
  \bibfield  {author} {\bibinfo {author} {\bibfnamefont {B.}~\bibnamefont
  {Oblak}},\ }\bibfield  {title} {\bibinfo {title} {{Berry Phases on Virasoro
  Orbits}},\ }\href {https://doi.org/10.1007/JHEP10(2017)114} {\bibfield
  {journal} {\bibinfo  {journal} {JHEP}\ }\textbf {\bibinfo {volume} {10}},\
  \bibinfo {pages} {114}},\ \Eprint {https://arxiv.org/abs/1703.06142}
  {arXiv:1703.06142 [hep-th]} \BibitemShut {NoStop}%
\bibitem [{\citenamefont {Akal}(2019)}]{Akal:2019hxa}%
  \BibitemOpen
  \bibfield  {author} {\bibinfo {author} {\bibfnamefont {I.}~\bibnamefont
  {Akal}},\ }\bibfield  {title} {\bibinfo {title} {{Reflections on Virasoro
  circuit complexity and Berry phase}},\ }\href@noop {} {\  (\bibinfo {year}
  {2019})},\ \Eprint {https://arxiv.org/abs/1908.08514} {arXiv:1908.08514
  [hep-th]} \BibitemShut {NoStop}%
\bibitem [{\citenamefont {Rychkov}(2016)}]{rychkov2017}%
  \BibitemOpen
  \bibfield  {author} {\bibinfo {author} {\bibfnamefont {S.}~\bibnamefont
  {Rychkov}},\ }\bibfield  {title} {\bibinfo {title} {{EPFL Lectures on
  Conformal Field Theory in D $\geq 3$ Dimensions}},\ }\href@noop {} {\
  \bibinfo {series} {SpringerBriefs in Physics} (\bibinfo {year} {2016})},\
  \Eprint {https://arxiv.org/abs/1601.05000} {arXiv:1601.05000 [hep-th]}
  \BibitemShut {NoStop}%
\bibitem [{\citenamefont {Witten}(1988)}]{witten1988}%
  \BibitemOpen
  \bibfield  {author} {\bibinfo {author} {\bibfnamefont {E.}~\bibnamefont
  {Witten}},\ }\bibfield  {title} {\bibinfo {title} {{Coadjoint Orbits of the
  Virasoro Group}},\ }\href {https://doi.org/10.1007/BF01218287} {\bibfield
  {journal} {\bibinfo  {journal} {Commun. Math. Phys.}\ }\textbf {\bibinfo
  {volume} {114}},\ \bibinfo {pages} {1} (\bibinfo {year} {1988})}\BibitemShut
  {NoStop}%
\bibitem [{\citenamefont {Kirillov}(2004)}]{kirillov2004}%
  \BibitemOpen
  \bibfield  {author} {\bibinfo {author} {\bibfnamefont {A.}~\bibnamefont
  {Kirillov}},\ }\href {https://doi.org/http://dx.doi.org/10.1090/gsm/064}
  {\emph {\bibinfo {title} {Lectures on the {{Orbit Method}}}}},\ \bibinfo
  {series} {Graduate {{Studies}} in {{Mathematics}}}, Vol.~\bibinfo {volume}
  {64}\ (\bibinfo  {publisher} {{American Mathematical Society}},\ \bibinfo
  {year} {2004})\BibitemShut {NoStop}%
\bibitem [{\citenamefont {Alekseev}\ and\ \citenamefont
  {Shatashvili}(1989)}]{alekseev1988}%
  \BibitemOpen
  \bibfield  {author} {\bibinfo {author} {\bibfnamefont {A.}~\bibnamefont
  {Alekseev}}\ and\ \bibinfo {author} {\bibfnamefont {S.~L.}\ \bibnamefont
  {Shatashvili}},\ }\bibfield  {title} {\bibinfo {title} {{Path Integral
  Quantization of the Coadjoint Orbits of the Virasoro Group and 2D Gravity}},\
  }\href {https://doi.org/10.1016/0550-3213(89)90130-2} {\bibfield  {journal}
  {\bibinfo  {journal} {Nucl. Phys. B}\ }\textbf {\bibinfo {volume} {323}},\
  \bibinfo {pages} {719} (\bibinfo {year} {1989})}\BibitemShut {NoStop}%
\bibitem [{\citenamefont {Alekseev}\ and\ \citenamefont
  {Shatashvili}(2018)}]{alekseev2018}%
  \BibitemOpen
  \bibfield  {author} {\bibinfo {author} {\bibfnamefont {A.}~\bibnamefont
  {Alekseev}}\ and\ \bibinfo {author} {\bibfnamefont {S.~L.}\ \bibnamefont
  {Shatashvili}},\ }\bibfield  {title} {\bibinfo {title} {{Coadjoint Orbits,
  Cocycles and Gravitational Wess\textendash{}Zumino}},\ }\href
  {https://doi.org/10.1142/9789813233867_0007} {\bibfield  {journal} {\bibinfo
  {journal} {Rev. Math. Phys.}\ }\textbf {\bibinfo {volume} {30}},\ \bibinfo
  {pages} {1840001} (\bibinfo {year} {2018})},\ \Eprint
  {https://arxiv.org/abs/1801.07963} {arXiv:1801.07963 [hep-th]} \BibitemShut
  {NoStop}%
\bibitem [{\citenamefont {Taylor}(1993)}]{Taylor:1993zp}%
  \BibitemOpen
  \bibfield  {author} {\bibinfo {author} {\bibfnamefont {W.}~\bibnamefont
  {Taylor}},\ }\href@noop {} {\bibinfo {title} {{Coadjoint orbits and conformal
  field theory}}} (\bibinfo {year} {1993}),\ \Eprint
  {https://arxiv.org/abs/hep-th/9310040} {arXiv:hep-th/9310040} \BibitemShut
  {NoStop}%
\bibitem [{\citenamefont {Perelomov}(1972)}]{perelomov1972}%
  \BibitemOpen
  \bibfield  {author} {\bibinfo {author} {\bibfnamefont {A.~M.}\ \bibnamefont
  {Perelomov}},\ }\bibfield  {title} {\bibinfo {title} {Coherent states for
  arbitrary lie group},\ }\href
  {https://projecteuclid.org:443/euclid.cmp/1103858078} {\bibfield  {journal}
  {\bibinfo  {journal} {Comm. Math. Phys.}\ }\textbf {\bibinfo {volume} {26}},\
  \bibinfo {pages} {222} (\bibinfo {year} {1972})}\BibitemShut {NoStop}%
\bibitem [{\citenamefont {Gilmore}(1974)}]{gilmore}%
  \BibitemOpen
  \bibfield  {author} {\bibinfo {author} {\bibfnamefont {R.}~\bibnamefont
  {Gilmore}},\ }\bibfield  {title} {\bibinfo {title} {On the properties of
  coherent states},\ }\href@noop {} {\bibfield  {journal} {\bibinfo  {journal}
  {Rev. Mex. de Fisica}\ }\textbf {\bibinfo {volume} {23}},\ \bibinfo {pages}
  {143} (\bibinfo {year} {1974})}\BibitemShut {NoStop}%
\bibitem [{\citenamefont {Yaffe}(1982)}]{RevModPhys.54.407}%
  \BibitemOpen
  \bibfield  {author} {\bibinfo {author} {\bibfnamefont {L.~G.}\ \bibnamefont
  {Yaffe}},\ }\bibfield  {title} {\bibinfo {title} {Large $n$ limits as
  classical mechanics},\ }\href {https://doi.org/10.1103/RevModPhys.54.407}
  {\bibfield  {journal} {\bibinfo  {journal} {Rev. Mod. Phys.}\ }\textbf
  {\bibinfo {volume} {54}},\ \bibinfo {pages} {407} (\bibinfo {year}
  {1982})}\BibitemShut {NoStop}%
\bibitem [{\citenamefont {Provost}\ and\ \citenamefont
  {Vallee}(1980)}]{Provost:1980nc}%
  \BibitemOpen
  \bibfield  {author} {\bibinfo {author} {\bibfnamefont {J.~P.}\ \bibnamefont
  {Provost}}\ and\ \bibinfo {author} {\bibfnamefont {G.}~\bibnamefont
  {Vallee}},\ }\bibfield  {title} {\bibinfo {title} {{Riemannian Structure on
  Manifolds of Quantum States}},\ }\href {https://doi.org/10.1007/BF02193559}
  {\bibfield  {journal} {\bibinfo  {journal} {Commun. Math. Phys.}\ }\textbf
  {\bibinfo {volume} {76}},\ \bibinfo {pages} {289} (\bibinfo {year}
  {1980})}\BibitemShut {NoStop}%
\bibitem [{\citenamefont {Rowe}\ \emph {et~al.}(1985)\citenamefont {Rowe},
  \citenamefont {Rosensteel},\ and\ \citenamefont {Gilmore}}]{Rowe1985}%
  \BibitemOpen
  \bibfield  {author} {\bibinfo {author} {\bibfnamefont {D.~J.}\ \bibnamefont
  {Rowe}}, \bibinfo {author} {\bibfnamefont {G.}~\bibnamefont {Rosensteel}},\
  and\ \bibinfo {author} {\bibfnamefont {R.}~\bibnamefont {Gilmore}},\
  }\bibfield  {title} {\bibinfo {title} {Vector coherent state representation
  theory},\ }\href {https://doi.org/10.1063/1.526702} {\bibfield  {journal}
  {\bibinfo  {journal} {Journal of Mathematical Physics}\ }\textbf {\bibinfo
  {volume} {26}},\ \bibinfo {pages} {2787} (\bibinfo {year}
  {1985})}\BibitemShut {NoStop}%
\bibitem [{\citenamefont {Rowe}\ \emph {et~al.}(1988)\citenamefont {Rowe},
  \citenamefont {Blanc},\ and\ \citenamefont {Hecht}}]{Rowe1988}%
  \BibitemOpen
  \bibfield  {author} {\bibinfo {author} {\bibfnamefont {D.~J.}\ \bibnamefont
  {Rowe}}, \bibinfo {author} {\bibfnamefont {R.~L.}\ \bibnamefont {Blanc}},\
  and\ \bibinfo {author} {\bibfnamefont {K.~T.}\ \bibnamefont {Hecht}},\
  }\bibfield  {title} {\bibinfo {title} {Vector coherent state theory and its
  application to the orthogonal groups},\ }\href
  {https://doi.org/10.1063/1.528066} {\bibfield  {journal} {\bibinfo  {journal}
  {Journal of Mathematical Physics}\ }\textbf {\bibinfo {volume} {29}},\
  \bibinfo {pages} {287} (\bibinfo {year} {1988})}\BibitemShut {NoStop}%
\bibitem [{\citenamefont {Rowe}(2012)}]{Rowe:2012yi}%
  \BibitemOpen
  \bibfield  {author} {\bibinfo {author} {\bibfnamefont {D.~J.}\ \bibnamefont
  {Rowe}},\ }\bibfield  {title} {\bibinfo {title} {{Vector coherent state
  representations and their inner products}},\ }\href
  {https://doi.org/10.1088/1751-8113/45/24/244003} {\bibfield  {journal}
  {\bibinfo  {journal} {J. Phys. A}\ }\textbf {\bibinfo {volume} {45}},\
  \bibinfo {pages} {24003} (\bibinfo {year} {2012})},\ \Eprint
  {https://arxiv.org/abs/1207.0126} {arXiv:1207.0126 [math-ph]} \BibitemShut
  {NoStop}%
\bibitem [{\citenamefont {Bartlett}\ \emph {et~al.}(2002)\citenamefont
  {Bartlett}, \citenamefont {Rowe},\ and\ \citenamefont
  {Repka}}]{Bartlett2002}%
  \BibitemOpen
  \bibfield  {author} {\bibinfo {author} {\bibfnamefont {S.~D.}\ \bibnamefont
  {Bartlett}}, \bibinfo {author} {\bibfnamefont {D.~J.}\ \bibnamefont {Rowe}},\
  and\ \bibinfo {author} {\bibfnamefont {J.}~\bibnamefont {Repka}},\ }\bibfield
   {title} {\bibinfo {title} {Vector coherent state representations, induced
  representations and geometric quantization: {II}. vector coherent state
  representations},\ }\href {https://doi.org/10.1088/0305-4470/35/27/307}
  {\bibfield  {journal} {\bibinfo  {journal} {Journal of Physics A:
  Mathematical and General}\ }\textbf {\bibinfo {volume} {35}},\ \bibinfo
  {pages} {5625} (\bibinfo {year} {2002})}\BibitemShut {NoStop}%
\bibitem [{\citenamefont {Dorn}\ and\ \citenamefont
  {Jorjadze}(2005)}]{Dorn:2005jt}%
  \BibitemOpen
  \bibfield  {author} {\bibinfo {author} {\bibfnamefont {H.}~\bibnamefont
  {Dorn}}\ and\ \bibinfo {author} {\bibfnamefont {G.}~\bibnamefont
  {Jorjadze}},\ }\bibfield  {title} {\bibinfo {title} {{On particle dynamics in
  AdS(N+1) space-time}},\ }\href {https://doi.org/10.1002/prop.200510208}
  {\bibfield  {journal} {\bibinfo  {journal} {Fortsch. Phys.}\ }\textbf
  {\bibinfo {volume} {53}},\ \bibinfo {pages} {486} (\bibinfo {year} {2005})},\
  \Eprint {https://arxiv.org/abs/hep-th/0502081} {arXiv:hep-th/0502081}
  \BibitemShut {NoStop}%
\bibitem [{\citenamefont {Miyaji}\ \emph
  {et~al.}(2015{\natexlab{a}})\citenamefont {Miyaji}, \citenamefont {Ryu},
  \citenamefont {Takayanagi},\ and\ \citenamefont {Wen}}]{Miyaji:2014mca}%
  \BibitemOpen
  \bibfield  {author} {\bibinfo {author} {\bibfnamefont {M.}~\bibnamefont
  {Miyaji}}, \bibinfo {author} {\bibfnamefont {S.}~\bibnamefont {Ryu}},
  \bibinfo {author} {\bibfnamefont {T.}~\bibnamefont {Takayanagi}},\ and\
  \bibinfo {author} {\bibfnamefont {X.}~\bibnamefont {Wen}},\ }\bibfield
  {title} {\bibinfo {title} {{Boundary States as Holographic Duals of Trivial
  Spacetimes}},\ }\href {https://doi.org/10.1007/JHEP05(2015)152} {\bibfield
  {journal} {\bibinfo  {journal} {JHEP}\ }\textbf {\bibinfo {volume} {05}},\
  \bibinfo {pages} {152}},\ \Eprint {https://arxiv.org/abs/1412.6226}
  {arXiv:1412.6226 [hep-th]} \BibitemShut {NoStop}%
\bibitem [{\citenamefont {Fu}\ \emph {et~al.}(2018)\citenamefont {Fu},
  \citenamefont {Maloney}, \citenamefont {Marolf}, \citenamefont {Maxfield},\
  and\ \citenamefont {Wang}}]{Fu:2018kcp}%
  \BibitemOpen
  \bibfield  {author} {\bibinfo {author} {\bibfnamefont {Z.}~\bibnamefont
  {Fu}}, \bibinfo {author} {\bibfnamefont {A.}~\bibnamefont {Maloney}},
  \bibinfo {author} {\bibfnamefont {D.}~\bibnamefont {Marolf}}, \bibinfo
  {author} {\bibfnamefont {H.}~\bibnamefont {Maxfield}},\ and\ \bibinfo
  {author} {\bibfnamefont {Z.}~\bibnamefont {Wang}},\ }\bibfield  {title}
  {\bibinfo {title} {{Holographic complexity is nonlocal}},\ }\href
  {https://doi.org/10.1007/JHEP02(2018)072} {\bibfield  {journal} {\bibinfo
  {journal} {JHEP}\ }\textbf {\bibinfo {volume} {02}},\ \bibinfo {pages}
  {072}},\ \Eprint {https://arxiv.org/abs/1801.01137} {arXiv:1801.01137
  [hep-th]} \BibitemShut {NoStop}%
\bibitem [{\citenamefont {Ag\'on}\ \emph {et~al.}(2019)\citenamefont {Ag\'on},
  \citenamefont {Headrick},\ and\ \citenamefont {Swingle}}]{Agon:2018zso}%
  \BibitemOpen
  \bibfield  {author} {\bibinfo {author} {\bibfnamefont {C.~A.}\ \bibnamefont
  {Ag\'on}}, \bibinfo {author} {\bibfnamefont {M.}~\bibnamefont {Headrick}},\
  and\ \bibinfo {author} {\bibfnamefont {B.}~\bibnamefont {Swingle}},\
  }\bibfield  {title} {\bibinfo {title} {{Subsystem Complexity and
  Holography}},\ }\href {https://doi.org/10.1007/JHEP02(2019)145} {\bibfield
  {journal} {\bibinfo  {journal} {JHEP}\ }\textbf {\bibinfo {volume} {02}},\
  \bibinfo {pages} {145}},\ \Eprint {https://arxiv.org/abs/1804.01561}
  {arXiv:1804.01561 [hep-th]} \BibitemShut {NoStop}%
\bibitem [{\citenamefont {Braccia}\ \emph {et~al.}(2020)\citenamefont
  {Braccia}, \citenamefont {Cotrone},\ and\ \citenamefont
  {Tonni}}]{Braccia:2019xxi}%
  \BibitemOpen
  \bibfield  {author} {\bibinfo {author} {\bibfnamefont {P.}~\bibnamefont
  {Braccia}}, \bibinfo {author} {\bibfnamefont {A.~L.}\ \bibnamefont
  {Cotrone}},\ and\ \bibinfo {author} {\bibfnamefont {E.}~\bibnamefont
  {Tonni}},\ }\bibfield  {title} {\bibinfo {title} {{Complexity in the presence
  of a boundary}},\ }\href {https://doi.org/10.1007/JHEP02(2020)051} {\bibfield
   {journal} {\bibinfo  {journal} {JHEP}\ }\textbf {\bibinfo {volume} {02}},\
  \bibinfo {pages} {051}},\ \Eprint {https://arxiv.org/abs/1910.03489}
  {arXiv:1910.03489 [hep-th]} \BibitemShut {NoStop}%
\bibitem [{\citenamefont {Caputa}\ \emph {et~al.}(2017)\citenamefont {Caputa},
  \citenamefont {Kundu}, \citenamefont {Miyaji}, \citenamefont {Takayanagi},\
  and\ \citenamefont {Watanabe}}]{caputa2017}%
  \BibitemOpen
  \bibfield  {author} {\bibinfo {author} {\bibfnamefont {P.}~\bibnamefont
  {Caputa}}, \bibinfo {author} {\bibfnamefont {N.}~\bibnamefont {Kundu}},
  \bibinfo {author} {\bibfnamefont {M.}~\bibnamefont {Miyaji}}, \bibinfo
  {author} {\bibfnamefont {T.}~\bibnamefont {Takayanagi}},\ and\ \bibinfo
  {author} {\bibfnamefont {K.}~\bibnamefont {Watanabe}},\ }\bibfield  {title}
  {\bibinfo {title} {{Liouville Action as Path-Integral Complexity: From
  Continuous Tensor Networks to AdS/CFT}},\ }\href
  {https://doi.org/10.1007/JHEP11(2017)097} {\bibfield  {journal} {\bibinfo
  {journal} {JHEP}\ }\textbf {\bibinfo {volume} {11}},\ \bibinfo {pages}
  {097}},\ \Eprint {https://arxiv.org/abs/1706.07056} {arXiv:1706.07056
  [hep-th]} \BibitemShut {NoStop}%
\bibitem [{\citenamefont {Camargo}\ \emph {et~al.}(2019)\citenamefont
  {Camargo}, \citenamefont {Heller}, \citenamefont {Jefferson},\ and\
  \citenamefont {Knaute}}]{Camargo:2019isp}%
  \BibitemOpen
  \bibfield  {author} {\bibinfo {author} {\bibfnamefont {H.~A.}\ \bibnamefont
  {Camargo}}, \bibinfo {author} {\bibfnamefont {M.~P.}\ \bibnamefont {Heller}},
  \bibinfo {author} {\bibfnamefont {R.}~\bibnamefont {Jefferson}},\ and\
  \bibinfo {author} {\bibfnamefont {J.}~\bibnamefont {Knaute}},\ }\bibfield
  {title} {\bibinfo {title} {{Path integral optimization as circuit
  complexity}},\ }\href {https://doi.org/10.1103/PhysRevLett.123.011601}
  {\bibfield  {journal} {\bibinfo  {journal} {Phys. Rev. Lett.}\ }\textbf
  {\bibinfo {volume} {123}},\ \bibinfo {pages} {011601} (\bibinfo {year}
  {2019})},\ \Eprint {https://arxiv.org/abs/1904.02713} {arXiv:1904.02713
  [hep-th]} \BibitemShut {NoStop}%
\bibitem [{\citenamefont {Milsted}\ and\ \citenamefont
  {Vidal}(2018{\natexlab{a}})}]{Milsted:2018yur}%
  \BibitemOpen
  \bibfield  {author} {\bibinfo {author} {\bibfnamefont {A.}~\bibnamefont
  {Milsted}}\ and\ \bibinfo {author} {\bibfnamefont {G.}~\bibnamefont
  {Vidal}},\ }\bibfield  {title} {\bibinfo {title} {{Tensor networks as path
  integral geometry}},\ }\href@noop {} {\  (\bibinfo {year}
  {2018}{\natexlab{a}})},\ \Eprint {https://arxiv.org/abs/1807.02501}
  {arXiv:1807.02501 [cond-mat.str-el]} \BibitemShut {NoStop}%
\bibitem [{\citenamefont {Milsted}\ and\ \citenamefont
  {Vidal}(2018{\natexlab{b}})}]{Milsted:2018san}%
  \BibitemOpen
  \bibfield  {author} {\bibinfo {author} {\bibfnamefont {A.}~\bibnamefont
  {Milsted}}\ and\ \bibinfo {author} {\bibfnamefont {G.}~\bibnamefont
  {Vidal}},\ }\bibfield  {title} {\bibinfo {title} {{Geometric interpretation
  of the multi-scale entanglement renormalization ansatz}},\ }\href@noop {} {\
  (\bibinfo {year} {2018}{\natexlab{b}})},\ \Eprint
  {https://arxiv.org/abs/1812.00529} {arXiv:1812.00529 [hep-th]} \BibitemShut
  {NoStop}%
\bibitem [{\citenamefont {Czech}(2018)}]{Czech:2017ryf}%
  \BibitemOpen
  \bibfield  {author} {\bibinfo {author} {\bibfnamefont {B.}~\bibnamefont
  {Czech}},\ }\bibfield  {title} {\bibinfo {title} {{Einstein Equations from
  Varying Complexity}},\ }\href
  {https://doi.org/10.1103/PhysRevLett.120.031601} {\bibfield  {journal}
  {\bibinfo  {journal} {Phys. Rev. Lett.}\ }\textbf {\bibinfo {volume} {120}},\
  \bibinfo {pages} {031601} (\bibinfo {year} {2018})},\ \Eprint
  {https://arxiv.org/abs/1706.00965} {arXiv:1706.00965 [hep-th]} \BibitemShut
  {NoStop}%
\bibitem [{\citenamefont {Chandra}\ \emph {et~al.}(2021)\citenamefont
  {Chandra}, \citenamefont {de~Boer}, \citenamefont {Flory}, \citenamefont
  {Heller}, \citenamefont {H\"ortner},\ and\ \citenamefont
  {Rolph}}]{Chandra:2021kdv}%
  \BibitemOpen
  \bibfield  {author} {\bibinfo {author} {\bibfnamefont {A.~R.}\ \bibnamefont
  {Chandra}}, \bibinfo {author} {\bibfnamefont {J.}~\bibnamefont {de~Boer}},
  \bibinfo {author} {\bibfnamefont {M.}~\bibnamefont {Flory}}, \bibinfo
  {author} {\bibfnamefont {M.~P.}\ \bibnamefont {Heller}}, \bibinfo {author}
  {\bibfnamefont {S.}~\bibnamefont {H\"ortner}},\ and\ \bibinfo {author}
  {\bibfnamefont {A.}~\bibnamefont {Rolph}},\ }\bibfield  {title} {\bibinfo
  {title} {{Spacetime as a quantum circuit}},\ }\href
  {https://doi.org/10.1007/JHEP04(2021)207} {\bibfield  {journal} {\bibinfo
  {journal} {JHEP}\ }\textbf {\bibinfo {volume} {21}},\ \bibinfo {pages}
  {207}},\ \Eprint {https://arxiv.org/abs/2101.01185} {arXiv:2101.01185
  [hep-th]} \BibitemShut {NoStop}%
\bibitem [{\citenamefont {Levy}\ and\ \citenamefont {Oz}(2018)}]{Levy:2018bdc}%
  \BibitemOpen
  \bibfield  {author} {\bibinfo {author} {\bibfnamefont {T.}~\bibnamefont
  {Levy}}\ and\ \bibinfo {author} {\bibfnamefont {Y.}~\bibnamefont {Oz}},\
  }\bibfield  {title} {\bibinfo {title} {{Liouville Conformal Field Theories in
  Higher Dimensions}},\ }\href {https://doi.org/10.1007/JHEP06(2018)119}
  {\bibfield  {journal} {\bibinfo  {journal} {JHEP}\ }\textbf {\bibinfo
  {volume} {06}},\ \bibinfo {pages} {119}},\ \Eprint
  {https://arxiv.org/abs/1804.02283} {arXiv:1804.02283 [hep-th]} \BibitemShut
  {NoStop}%
\bibitem [{\citenamefont {Marolf}\ \emph {et~al.}(2018)\citenamefont {Marolf},
  \citenamefont {Parrikar}, \citenamefont {Rabideau}, \citenamefont
  {Izadi~Rad},\ and\ \citenamefont {Van~Raamsdonk}}]{Marolf:2017kvq}%
  \BibitemOpen
  \bibfield  {author} {\bibinfo {author} {\bibfnamefont {D.}~\bibnamefont
  {Marolf}}, \bibinfo {author} {\bibfnamefont {O.}~\bibnamefont {Parrikar}},
  \bibinfo {author} {\bibfnamefont {C.}~\bibnamefont {Rabideau}}, \bibinfo
  {author} {\bibfnamefont {A.}~\bibnamefont {Izadi~Rad}},\ and\ \bibinfo
  {author} {\bibfnamefont {M.}~\bibnamefont {Van~Raamsdonk}},\ }\bibfield
  {title} {\bibinfo {title} {{From Euclidean Sources to Lorentzian Spacetimes
  in Holographic Conformal Field Theories}},\ }\href
  {https://doi.org/10.1007/JHEP06(2018)077} {\bibfield  {journal} {\bibinfo
  {journal} {JHEP}\ }\textbf {\bibinfo {volume} {06}},\ \bibinfo {pages}
  {077}},\ \Eprint {https://arxiv.org/abs/1709.10101} {arXiv:1709.10101
  [hep-th]} \BibitemShut {NoStop}%
\bibitem [{\citenamefont {Belin}\ \emph
  {et~al.}(2019{\natexlab{a}})\citenamefont {Belin}, \citenamefont
  {Lewkowycz},\ and\ \citenamefont {S\'arosi}}]{Belin:2018fxe}%
  \BibitemOpen
  \bibfield  {author} {\bibinfo {author} {\bibfnamefont {A.}~\bibnamefont
  {Belin}}, \bibinfo {author} {\bibfnamefont {A.}~\bibnamefont {Lewkowycz}},\
  and\ \bibinfo {author} {\bibfnamefont {G.}~\bibnamefont {S\'arosi}},\
  }\bibfield  {title} {\bibinfo {title} {{The boundary dual of the bulk
  symplectic form}},\ }\href {https://doi.org/10.1016/j.physletb.2018.10.071}
  {\bibfield  {journal} {\bibinfo  {journal} {Phys. Lett. B}\ }\textbf
  {\bibinfo {volume} {789}},\ \bibinfo {pages} {71} (\bibinfo {year}
  {2019}{\natexlab{a}})},\ \Eprint {https://arxiv.org/abs/1806.10144}
  {arXiv:1806.10144 [hep-th]} \BibitemShut {NoStop}%
\bibitem [{\citenamefont {Belin}\ \emph
  {et~al.}(2019{\natexlab{b}})\citenamefont {Belin}, \citenamefont
  {Lewkowycz},\ and\ \citenamefont {S\'arosi}}]{Belin:2018bpg}%
  \BibitemOpen
  \bibfield  {author} {\bibinfo {author} {\bibfnamefont {A.}~\bibnamefont
  {Belin}}, \bibinfo {author} {\bibfnamefont {A.}~\bibnamefont {Lewkowycz}},\
  and\ \bibinfo {author} {\bibfnamefont {G.}~\bibnamefont {S\'arosi}},\
  }\bibfield  {title} {\bibinfo {title} {{Complexity and the bulk volume, a new
  York time story}},\ }\href {https://doi.org/10.1007/JHEP03(2019)044}
  {\bibfield  {journal} {\bibinfo  {journal} {JHEP}\ }\textbf {\bibinfo
  {volume} {03}},\ \bibinfo {pages} {044}},\ \Eprint
  {https://arxiv.org/abs/1811.03097} {arXiv:1811.03097 [hep-th]} \BibitemShut
  {NoStop}%
\bibitem [{\citenamefont {Miyaji}\ \emph
  {et~al.}(2015{\natexlab{b}})\citenamefont {Miyaji}, \citenamefont {Numasawa},
  \citenamefont {Shiba}, \citenamefont {Takayanagi},\ and\ \citenamefont
  {Watanabe}}]{Miyaji:2015woj}%
  \BibitemOpen
  \bibfield  {author} {\bibinfo {author} {\bibfnamefont {M.}~\bibnamefont
  {Miyaji}}, \bibinfo {author} {\bibfnamefont {T.}~\bibnamefont {Numasawa}},
  \bibinfo {author} {\bibfnamefont {N.}~\bibnamefont {Shiba}}, \bibinfo
  {author} {\bibfnamefont {T.}~\bibnamefont {Takayanagi}},\ and\ \bibinfo
  {author} {\bibfnamefont {K.}~\bibnamefont {Watanabe}},\ }\bibfield  {title}
  {\bibinfo {title} {{Distance between Quantum States and Gauge-Gravity
  Duality}},\ }\href {https://doi.org/10.1103/PhysRevLett.115.261602}
  {\bibfield  {journal} {\bibinfo  {journal} {Phys. Rev. Lett.}\ }\textbf
  {\bibinfo {volume} {115}},\ \bibinfo {pages} {261602} (\bibinfo {year}
  {2015}{\natexlab{b}})},\ \Eprint {https://arxiv.org/abs/1507.07555}
  {arXiv:1507.07555 [hep-th]} \BibitemShut {NoStop}%
\bibitem [{\citenamefont {Abt}\ \emph {et~al.}(2019)\citenamefont {Abt},
  \citenamefont {Erdmenger}, \citenamefont {Gerbershagen}, \citenamefont
  {Melby-Thompson},\ and\ \citenamefont {Northe}}]{Abt:2018ywl}%
  \BibitemOpen
  \bibfield  {author} {\bibinfo {author} {\bibfnamefont {R.}~\bibnamefont
  {Abt}}, \bibinfo {author} {\bibfnamefont {J.}~\bibnamefont {Erdmenger}},
  \bibinfo {author} {\bibfnamefont {M.}~\bibnamefont {Gerbershagen}}, \bibinfo
  {author} {\bibfnamefont {C.~M.}\ \bibnamefont {Melby-Thompson}},\ and\
  \bibinfo {author} {\bibfnamefont {C.}~\bibnamefont {Northe}},\ }\bibfield
  {title} {\bibinfo {title} {{Holographic Subregion Complexity from Kinematic
  Space}},\ }\href {https://doi.org/10.1007/JHEP01(2019)012} {\bibfield
  {journal} {\bibinfo  {journal} {JHEP}\ }\textbf {\bibinfo {volume} {01}},\
  \bibinfo {pages} {012}},\ \Eprint {https://arxiv.org/abs/1805.10298}
  {arXiv:1805.10298 [hep-th]} \BibitemShut {NoStop}%
\bibitem [{\citenamefont {Czech}\ \emph {et~al.}(2018)\citenamefont {Czech},
  \citenamefont {Lamprou}, \citenamefont {McCandlish},\ and\ \citenamefont
  {Sully}}]{Czech:2017zfq}%
  \BibitemOpen
  \bibfield  {author} {\bibinfo {author} {\bibfnamefont {B.}~\bibnamefont
  {Czech}}, \bibinfo {author} {\bibfnamefont {L.}~\bibnamefont {Lamprou}},
  \bibinfo {author} {\bibfnamefont {S.}~\bibnamefont {McCandlish}},\ and\
  \bibinfo {author} {\bibfnamefont {J.}~\bibnamefont {Sully}},\ }\bibfield
  {title} {\bibinfo {title} {{Modular Berry Connection for Entangled Subregions
  in AdS/CFT}},\ }\href {https://doi.org/10.1103/PhysRevLett.120.091601}
  {\bibfield  {journal} {\bibinfo  {journal} {Phys. Rev. Lett.}\ }\textbf
  {\bibinfo {volume} {120}},\ \bibinfo {pages} {091601} (\bibinfo {year}
  {2018})},\ \Eprint {https://arxiv.org/abs/1712.07123} {arXiv:1712.07123
  [hep-th]} \BibitemShut {NoStop}%
\bibitem [{\citenamefont {Czech}\ \emph {et~al.}(2019)\citenamefont {Czech},
  \citenamefont {De~Boer}, \citenamefont {Ge},\ and\ \citenamefont
  {Lamprou}}]{Czech:2019vih}%
  \BibitemOpen
  \bibfield  {author} {\bibinfo {author} {\bibfnamefont {B.}~\bibnamefont
  {Czech}}, \bibinfo {author} {\bibfnamefont {J.}~\bibnamefont {De~Boer}},
  \bibinfo {author} {\bibfnamefont {D.}~\bibnamefont {Ge}},\ and\ \bibinfo
  {author} {\bibfnamefont {L.}~\bibnamefont {Lamprou}},\ }\bibfield  {title}
  {\bibinfo {title} {{A modular sewing kit for entanglement wedges}},\ }\href
  {https://doi.org/10.1007/JHEP11(2019)094} {\bibfield  {journal} {\bibinfo
  {journal} {JHEP}\ }\textbf {\bibinfo {volume} {11}},\ \bibinfo {pages}
  {094}},\ \Eprint {https://arxiv.org/abs/1903.04493} {arXiv:1903.04493
  [hep-th]} \BibitemShut {NoStop}%
\bibitem [{\citenamefont {Campos~Venuti}\ and\ \citenamefont
  {Zanardi}(2007)}]{PhysRevLett.99.095701}%
  \BibitemOpen
  \bibfield  {author} {\bibinfo {author} {\bibfnamefont {L.}~\bibnamefont
  {Campos~Venuti}}\ and\ \bibinfo {author} {\bibfnamefont {P.}~\bibnamefont
  {Zanardi}},\ }\bibfield  {title} {\bibinfo {title} {Quantum critical scaling
  of the geometric tensors},\ }\href
  {https://doi.org/10.1103/PhysRevLett.99.095701} {\bibfield  {journal}
  {\bibinfo  {journal} {Phys. Rev. Lett.}\ }\textbf {\bibinfo {volume} {99}},\
  \bibinfo {pages} {095701} (\bibinfo {year} {2007})}\BibitemShut {NoStop}%
\bibitem [{\citenamefont {Kolodrubetz}\ \emph {et~al.}(2013)\citenamefont
  {Kolodrubetz}, \citenamefont {Gritsev},\ and\ \citenamefont
  {Polkovnikov}}]{PhysRevB.88.064304}%
  \BibitemOpen
  \bibfield  {author} {\bibinfo {author} {\bibfnamefont {M.}~\bibnamefont
  {Kolodrubetz}}, \bibinfo {author} {\bibfnamefont {V.}~\bibnamefont
  {Gritsev}},\ and\ \bibinfo {author} {\bibfnamefont {A.}~\bibnamefont
  {Polkovnikov}},\ }\bibfield  {title} {\bibinfo {title} {{Classifying and
  measuring geometry of a quantum ground state manifold}},\ }\href
  {https://doi.org/10.1103/PhysRevB.88.064304} {\bibfield  {journal} {\bibinfo
  {journal} {Phys. Rev. B}\ }\textbf {\bibinfo {volume} {88}},\ \bibinfo
  {pages} {064304} (\bibinfo {year} {2013})},\ \Eprint
  {https://arxiv.org/abs/1305.0568} {arXiv:1305.0568 [cond-mat.stat-mech]}
  \BibitemShut {NoStop}%
\bibitem [{\citenamefont {Gritsev}\ and\ \citenamefont
  {Polkovnikov}(2017)}]{SciPostPhys.2.3.021}%
  \BibitemOpen
  \bibfield  {author} {\bibinfo {author} {\bibfnamefont {V.}~\bibnamefont
  {Gritsev}}\ and\ \bibinfo {author} {\bibfnamefont {A.}~\bibnamefont
  {Polkovnikov}},\ }\bibfield  {title} {\bibinfo {title} {{Integrable Floquet
  dynamics}},\ }\href {https://doi.org/10.21468/SciPostPhys.2.3.021} {\bibfield
   {journal} {\bibinfo  {journal} {SciPost Phys.}\ }\textbf {\bibinfo {volume}
  {2}},\ \bibinfo {pages} {021} (\bibinfo {year} {2017})},\ \Eprint
  {https://arxiv.org/abs/1701.05276} {arXiv:1701.05276 [cond-mat.stat-mech]}
  \BibitemShut {NoStop}%
\bibitem [{\citenamefont {Camilo}\ and\ \citenamefont
  {Teixeira}(2020)}]{PhysRevB.102.174304}%
  \BibitemOpen
  \bibfield  {author} {\bibinfo {author} {\bibfnamefont {G.}~\bibnamefont
  {Camilo}}\ and\ \bibinfo {author} {\bibfnamefont {D.}~\bibnamefont
  {Teixeira}},\ }\bibfield  {title} {\bibinfo {title} {Complexity and floquet
  dynamics: Nonequilibrium ising phase transitions},\ }\href
  {https://doi.org/10.1103/PhysRevB.102.174304} {\bibfield  {journal} {\bibinfo
   {journal} {Phys. Rev. B}\ }\textbf {\bibinfo {volume} {102}},\ \bibinfo
  {pages} {174304} (\bibinfo {year} {2020})}\BibitemShut {NoStop}%
\bibitem [{\citenamefont {Liu}\ \emph {et~al.}(2020)\citenamefont {Liu},
  \citenamefont {Whitsitt}, \citenamefont {Curtis}, \citenamefont {Lundgren},
  \citenamefont {Titum}, \citenamefont {Yang}, \citenamefont {Garrison},\ and\
  \citenamefont {Gorshkov}}]{PhysRevResearch.2.013323}%
  \BibitemOpen
  \bibfield  {author} {\bibinfo {author} {\bibfnamefont {F.}~\bibnamefont
  {Liu}}, \bibinfo {author} {\bibfnamefont {S.}~\bibnamefont {Whitsitt}},
  \bibinfo {author} {\bibfnamefont {J.~B.}\ \bibnamefont {Curtis}}, \bibinfo
  {author} {\bibfnamefont {R.}~\bibnamefont {Lundgren}}, \bibinfo {author}
  {\bibfnamefont {P.}~\bibnamefont {Titum}}, \bibinfo {author} {\bibfnamefont
  {Z.-C.}\ \bibnamefont {Yang}}, \bibinfo {author} {\bibfnamefont {J.~R.}\
  \bibnamefont {Garrison}},\ and\ \bibinfo {author} {\bibfnamefont {A.~V.}\
  \bibnamefont {Gorshkov}},\ }\bibfield  {title} {\bibinfo {title} {Circuit
  complexity across a topological phase transition},\ }\href
  {https://doi.org/10.1103/PhysRevResearch.2.013323} {\bibfield  {journal}
  {\bibinfo  {journal} {Phys. Rev. Research}\ }\textbf {\bibinfo {volume}
  {2}},\ \bibinfo {pages} {013323} (\bibinfo {year} {2020})}\BibitemShut
  {NoStop}%
\bibitem [{\citenamefont {Mazac}(2018)}]{Dalimil}%
  \BibitemOpen
  \bibfield  {author} {\bibinfo {author} {\bibfnamefont {D.}~\bibnamefont
  {Mazac}},\ }\bibfield  {title} {\bibinfo {title} {Bootstrap introduction}}
  (\bibinfo {year} {2018}),\ \bibinfo {note} {2018 Bootstrap School
  Lecture}\BibitemShut {NoStop}%
\bibitem [{\citenamefont {Minwalla}(1998)}]{Minwalla1998}%
  \BibitemOpen
  \bibfield  {author} {\bibinfo {author} {\bibfnamefont {S.}~\bibnamefont
  {Minwalla}},\ }\bibfield  {title} {\bibinfo {title} {{Restrictions imposed by
  superconformal invariance on quantum field theories}},\ }\href
  {https://doi.org/10.4310/ATMP.1998.v2.n4.a4} {\bibfield  {journal} {\bibinfo
  {journal} {Adv. Theor. Math. Phys.}\ }\textbf {\bibinfo {volume} {2}},\
  \bibinfo {pages} {783} (\bibinfo {year} {1998})},\ \Eprint
  {https://arxiv.org/abs/hep-th/9712074} {arXiv:hep-th/9712074} \BibitemShut
  {NoStop}%
\bibitem [{\citenamefont {Luscher}\ and\ \citenamefont
  {Mack}(1975)}]{luscher1975global}%
  \BibitemOpen
  \bibfield  {author} {\bibinfo {author} {\bibfnamefont {M.}~\bibnamefont
  {Luscher}}\ and\ \bibinfo {author} {\bibfnamefont {G.}~\bibnamefont {Mack}},\
  }\bibfield  {title} {\bibinfo {title} {{Global Conformal Invariance in
  Quantum Field Theory}},\ }\href {https://doi.org/10.1007/BF01608988}
  {\bibfield  {journal} {\bibinfo  {journal} {Commun. Math. Phys.}\ }\textbf
  {\bibinfo {volume} {41}},\ \bibinfo {pages} {203} (\bibinfo {year}
  {1975})}\BibitemShut {NoStop}%
\bibitem [{\citenamefont {Zhang}\ \emph {et~al.}(1990)\citenamefont {Zhang},
  \citenamefont {Feng},\ and\ \citenamefont {Gilmore}}]{Zhang:1990fy}%
  \BibitemOpen
  \bibfield  {author} {\bibinfo {author} {\bibfnamefont {W.-M.}\ \bibnamefont
  {Zhang}}, \bibinfo {author} {\bibfnamefont {D.~H.}\ \bibnamefont {Feng}},\
  and\ \bibinfo {author} {\bibfnamefont {R.}~\bibnamefont {Gilmore}},\
  }\bibfield  {title} {\bibinfo {title} {{Coherent States: Theory and Some
  Applications}},\ }\href {https://doi.org/10.1103/RevModPhys.62.867}
  {\bibfield  {journal} {\bibinfo  {journal} {Rev. Mod. Phys.}\ }\textbf
  {\bibinfo {volume} {62}},\ \bibinfo {pages} {867} (\bibinfo {year}
  {1990})}\BibitemShut {NoStop}%
\bibitem [{\citenamefont {Klimov}\ and\ \citenamefont
  {Chumakov}(2009)}]{klimovchumakov}%
  \BibitemOpen
  \bibfield  {author} {\bibinfo {author} {\bibfnamefont {A.}~\bibnamefont
  {Klimov}}\ and\ \bibinfo {author} {\bibfnamefont {S.}~\bibnamefont
  {Chumakov}},\ }\href@noop {} {\emph {\bibinfo {title} {A Group-Theoretical
  Approach to Quantum Optics: Models of Atom-Field Interactions}}}\ (\bibinfo
  {publisher} {Wiley},\ \bibinfo {year} {2009})\BibitemShut {NoStop}%
\bibitem [{\citenamefont {Feinsilver}\ \emph {et~al.}(2001)\citenamefont
  {Feinsilver}, \citenamefont {Kocik},\ and\ \citenamefont
  {Giering}}]{feinsilver2001canonical}%
  \BibitemOpen
  \bibfield  {author} {\bibinfo {author} {\bibfnamefont {P.}~\bibnamefont
  {Feinsilver}}, \bibinfo {author} {\bibfnamefont {J.}~\bibnamefont {Kocik}},\
  and\ \bibinfo {author} {\bibfnamefont {M.}~\bibnamefont {Giering}},\
  }\bibfield  {title} {\bibinfo {title} {Canonical variables and analysis on
  $so(n, 2)$},\ }\href@noop {} {\bibfield  {journal} {\bibinfo  {journal}
  {Journal of Physics A: Mathematical and General}\ }\textbf {\bibinfo {volume}
  {34}},\ \bibinfo {pages} {2367} (\bibinfo {year} {2001})}\BibitemShut
  {NoStop}%
\bibitem [{\citenamefont {Dobrev}(2016)}]{Dobrev}%
  \BibitemOpen
  \bibfield  {author} {\bibinfo {author} {\bibfnamefont {V.~K.}\ \bibnamefont
  {Dobrev}},\ }\href@noop {} {\emph {\bibinfo {title} {Noncompact Semisimple
  Lie Algebras and Groups}}}\ (\bibinfo  {publisher} {De Gruyter},\ \bibinfo
  {year} {2016})\BibitemShut {NoStop}%
\bibitem [{\citenamefont {Penedones}\ \emph {et~al.}(2016)\citenamefont
  {Penedones}, \citenamefont {Trevisani},\ and\ \citenamefont
  {Yamazaki}}]{Penedones:2015aga}%
  \BibitemOpen
  \bibfield  {author} {\bibinfo {author} {\bibfnamefont {J.}~\bibnamefont
  {Penedones}}, \bibinfo {author} {\bibfnamefont {E.}~\bibnamefont
  {Trevisani}},\ and\ \bibinfo {author} {\bibfnamefont {M.}~\bibnamefont
  {Yamazaki}},\ }\bibfield  {title} {\bibinfo {title} {{Recursion Relations for
  Conformal Blocks}},\ }\href {https://doi.org/10.1007/JHEP09(2016)070}
  {\bibfield  {journal} {\bibinfo  {journal} {JHEP}\ }\textbf {\bibinfo
  {volume} {09}},\ \bibinfo {pages} {070}},\ \Eprint
  {https://arxiv.org/abs/1509.00428} {arXiv:1509.00428 [hep-th]} \BibitemShut
  {NoStop}%
\bibitem [{\citenamefont {Yamazaki}(2016)}]{Yamazaki:2016vqi}%
  \BibitemOpen
  \bibfield  {author} {\bibinfo {author} {\bibfnamefont {M.}~\bibnamefont
  {Yamazaki}},\ }\bibfield  {title} {\bibinfo {title} {{Comments on Determinant
  Formulas for General CFTs}},\ }\href
  {https://doi.org/10.1007/JHEP10(2016)035} {\bibfield  {journal} {\bibinfo
  {journal} {JHEP}\ }\textbf {\bibinfo {volume} {10}},\ \bibinfo {pages}
  {035}},\ \Eprint {https://arxiv.org/abs/1601.04072} {arXiv:1601.04072
  [hep-th]} \BibitemShut {NoStop}%
\bibitem [{\citenamefont {Carmi}\ \emph
  {et~al.}(2017{\natexlab{b}})\citenamefont {Carmi}, \citenamefont {Myers},\
  and\ \citenamefont {Rath}}]{Carmi:2016wjl}%
  \BibitemOpen
  \bibfield  {author} {\bibinfo {author} {\bibfnamefont {D.}~\bibnamefont
  {Carmi}}, \bibinfo {author} {\bibfnamefont {R.~C.}\ \bibnamefont {Myers}},\
  and\ \bibinfo {author} {\bibfnamefont {P.}~\bibnamefont {Rath}},\ }\bibfield
  {title} {\bibinfo {title} {{Comments on Holographic Complexity}},\ }\href
  {https://doi.org/10.1007/JHEP03(2017)118} {\bibfield  {journal} {\bibinfo
  {journal} {JHEP}\ }\textbf {\bibinfo {volume} {03}},\ \bibinfo {pages}
  {118}},\ \Eprint {https://arxiv.org/abs/1612.00433} {arXiv:1612.00433
  [hep-th]} \BibitemShut {NoStop}%
\bibitem [{\citenamefont {Reynolds}\ and\ \citenamefont
  {Ross}(2017)}]{Reynolds:2016rvl}%
  \BibitemOpen
  \bibfield  {author} {\bibinfo {author} {\bibfnamefont {A.}~\bibnamefont
  {Reynolds}}\ and\ \bibinfo {author} {\bibfnamefont {S.~F.}\ \bibnamefont
  {Ross}},\ }\bibfield  {title} {\bibinfo {title} {{Divergences in Holographic
  Complexity}},\ }\href {https://doi.org/10.1088/1361-6382/aa6925} {\bibfield
  {journal} {\bibinfo  {journal} {Class. Quant. Grav.}\ }\textbf {\bibinfo
  {volume} {34}},\ \bibinfo {pages} {105004} (\bibinfo {year} {2017})},\
  \Eprint {https://arxiv.org/abs/1612.05439} {arXiv:1612.05439 [hep-th]}
  \BibitemShut {NoStop}%
\end{thebibliography}%
